\documentstyle[amssymb,aps,multicol,floats,psfig,eqsecnum]{revtex}
\input epsf
\def\ourlim{+}  
\def\bdg{BDG}   
\def\volV{{\cal V}}             
\def\notV{{\overline{\cal V}}}  
\def\surS{{\cal \partial V}}             

\def\fms{\phantom{-}}           

\def\ket#1{\vert#1\rangle}

\def\gf{{\bf G}}

\def\romE{{\rm e}}              
\def\psm#1{{\bbox{\sigma}_{#1}}}        
\def\s1{{\psm{1}}}
\def\s2{{\psm{2}}}
\def\s3{{\psm{3}}}

\def\ourhat{}           
\def\Dop{{\ourhat{\Bbb D}}} 
\def\Oop{{\ourhat{\Bbb O}}} 
\def\Iop{{\ourhat{\Bbb I}}} 
\def\bbone{{BB-I}}
\def\bbtwo{{BB-II}}
\def\bbthr{{BB-III}}
\def\dis{\displaystyle}
\def\amp{{\cal M}}       
\def\genco{{\cal A}}     
\def\phfunc{{\cal S}}    
\def\proman{{\frak M}}
\def\smde{{f}}          
\def\Kf{k_{\rm F}}
\def\sudel{\Delta}
\def\Ef{\mu}
\def\SC{{$\Sigma$}}
\def\SCm{\Sigma}
\def\ab{Andreev billiard}
\def\surel{d{\sigma}} 
\newcommand{\eqbreak}{
\end{multicols}
\widetext
\noindent
\rule{.48\linewidth}{.1mm}\rule{.1mm}{.1cm}
}
\newcommand{\eqresume}{
\noindent
\rule{.52\linewidth}{.0mm}\rule[-.1cm]{.1mm}{.1cm}\rule{.48\linewidth}{.1mm}
\begin{multicols}{2}
\narrowtext
}
\begin{document}
\draft
\title{Quantal Andreev billiards: Semiclassical approach
to mesoscale oscillations in the density of states}
\author{\.{I}nan\c{c}~Adagideli\cite{REF:IAnote}
and Paul M.~Goldbart\cite{REF:PGnote}}
\address{Department of Physics
and Materials Research Laboratory,\\
University of Illinois at Urbana-Champaign,
Urbana, Illinois 61801, U.S.A.}
 \date{\today}
\maketitle
\begin{abstract}
Andreev billiards are finite, arbitrarily-shaped, normal-state regions,
surrounded by superconductor.  At energies below the superconducting gap,
single-quasiparticle excitations are confined to the normal region and
its vicinity, the essential mechanism for this confinement being Andreev
reflection.
This Paper develops and implements a theoretical framework for the
investigation of the short-wave quantal properties of these
single-quasiparticle excitations.
The focus is primarily on the relationship between the quasiparticle
energy eigenvalue spectrum and the geometrical shape of the normal-state
region, i.e., the question of spectral geometry in the novel setting of
excitations confined by a superconducting pair-potential.
Among the central results of this investigation are
two semiclassical trace formulas for the density of states.
The first, a lower-resolution formula, corresponds to the
well-known quasiclassical approximation, conventionally invoked in
settings involving superconductivity.
The second, a higher-resolution formula, allows the density of states
to be expressed in terms of:
(i)~An explicit formula for the level density, valid in the short-wave
limit, for billiards of arbitrary shape and dimensionality.  This level
density depends on the billiard shape only through the set of
stationary-length chords of the billiard and the curvature of the
boundary at the endpoints of these chords; and
(ii)~Higher-resolution corrections to the level density, expressed as a
sum over periodic orbits that creep around the billiard boundary.
Owing to the fact that these creeping orbits are much longer than the
stationary chords, one can, inter alia, \lq\lq hear\rq\rq\ the
stationary chords of Andreev billiards.
\end{abstract}
\widetext
\section{Introduction and overview}
\label{SEC:Intro}
The purpose of this paper is to consider the quantal dynamics of
elementary electron and hole quasiparticle excitations existing
within and in the vicinity of a normal-state region of matter that
is completely surrounded by an essentially infinite region of
conventional superconductor.  The entire system---normal-state region
and superconducting surround---may be envisaged as three-dimensional,
although the approach that we shall be developing is applicable in
any number of dimensions.  Owing to the inability of the surrounding
superconductor to support propagating quasiparticle excitations at
sufficiently low energies, electron and hole quasiparticle excitations
at such energies are bound to the normal-state region and its vicinity,
and it is on the properties of such bound states that we shall be
focusing our attention.

The process responsible for the confinement of these excitations
to the normal-state region and its vicinity is {\it Andreev
reflection\/}~\cite{REF:AFAndreev} from the surrounding
superconducting condensate; we shall therefore refer to such
structures, near to which quasiparticles are confined, as {\it
Andreev billiards\/}.  Andreev billiards were introduced and
certain simple aspects of their classical dynamics were discussed
in Ref.~\cite{REF:Kosztin}.  A very brief account of the approach
and results contained in the present Paper were reported in
Ref.~\cite{REF:QABlett}.  Certain quantum-mechanical properties of
\ab s were studied in Refs.~\cite{REF:Matrix}.

As we explore the quantal dynamics of quasiparticle excitations
of Andreev billiards, our primary focus will be on the relationship
between the quasiparticle energy eigenvalue spectrum and the
geometrical shape of the normal-state region, i.e., the question of
{\it spectral geometry\/} in this novel setting of excitations
confined by a superconducting pair-potential.  In the setting of
{\it conventional billiard\/} systems~\cite{REF:ConBillRev} confinement
is, by contrast, accomplished by an infinite (or occasionally
finite) single-particle potential~\cite{REF:AGraysplit}.
As mentioned above, we shall primarily be concerned with {\it confined\/}
quasiparticle states, and therefore in energy eigenvalues lying within
the quasiparticle gap of the surrounding superconductor (although our
approach is also suited to the study of scattering states).
The strategy that we shall develop is inspired by the beautiful work
of Balian and Bloch, in which, {\it inter alia\/}, the eigenvalue spectrum of
the Laplace operator was investigated for generically-shaped spatial regions and
various types of boundary conditions~\cite{REF:BaBoOne,REF:BaBoTwo,REF:BaBoThree}.
The central theme of the work of Balian and Bloch is the relationship between the
boundary shape, the type of boundary conditions, and the the eigenvalue spectrum.
We shall refer to Refs.~\cite{REF:BaBoOne,REF:BaBoTwo,REF:BaBoThree} respectively
as\ \bbone, \bbtwo, and \bbthr.

As it is so central to the properties of Andreev billiards, let us pause
to review the core qualitative features of the Andreev reflection
process: to a high degree of accuracy it
(i)~interconverts electron and hole excitations; and
(ii)~reverses the velocity of excitations.
It is this latter, {\it retro-reflective\/}, character of the Andreev
reflection process that endows Andreev billiards with dynamical characteristics
quite distinct from those of conventional billiards, in which confinement is
caused by {\it specular reflection\/} from a single-particle potential.

We shall take as a model for the normal-state region of an Andreev billiard
a Fermi gas, parametrized by the Fermi energy.  Thus, in the normal-state region
we shall be neglecting the effects of band structure, impurity scattering,
and quasiparticle interactions.  We shall account for the
superconducting nature of the matter surrounding the normal-state region by
asserting that there is a superconducting pair-potential $\Delta$ that
varies discontinuously: inside the normal-state region $\Delta=0$; outside
the normal-state region $\Delta$ takes on the constant value $\Delta_{0}$
($\ne 0$). We shall refer to the surface on which $\Delta$ changes
discontinuously as the shape of the Andreev billiard.
Thus, we shall not be working self-consistently, but
shall benefit from being in a position to develop an interface-scattering
approach to the quasiparticle dynamics, in which we are able to focus on
processes occurring at the interface.  Hence, we can incorporate in a
direct and natural manner the impact of the shape of the billiard
(i.e.~the shape of the interface) on the spectrum of energy eigenvalues
of the confined electron-hole quasiparticles.

We see four principal sources of motivation for the present work.
First, Andreev billiards provide a novel setting for the exploration of
spectrum-shape relationships, a branch of mathematics with a distinguished
history~\cite{REF:MarKKac,REF:Baltes}.  The novelty is fed in by the Andreev
reflection process occuring at the normal-to-superconductor boundary.
Second, the usual spectral-geometric scenario [in which deviations of the
density of modes from its large-system limit become appreciable as the
wavelength become comparable to the characteristic linear size of the system]
is not the whole story for the case of Andreev billiards.
Instead, owing to the presence of a second, much larger, lengthscale, set
by the difference between the momenta of incident electrons and the holes
they become upon Andreev reflection (and {\it vice versa\/}).  As this momenutm
difference is small (on the scale of the Fermi momentum), the corresponding
lengthscale is much larger that the Fermi wavelength.  Through this new
lengthscale the eigenvalue spectrum can be sensitive to the shape of the
billiard even when the characteristic size of the billiard is much larger
than the underlying wave length associated with quasiparticle motion.
This relevance of the new lengthscale has long been appreciated, showing
up, e.g., in the effectively one-dimensional settings of
Tomasch~\cite{REF:Tomasch}
and McMillan-Rowell~\cite{REF:McMillan}
oscillations in the tunneling density of states above the superconducting
gap, and in de~Gennes--Saint-James
bound states~\cite{REF:StJames} below the superconducting gap
Third, as we shall see when we develop a trace formula for the (oscillatory
part of the) density of quasiparticle eigenstates (DOS), there turns out to
be a novel and useful separation in the scale of periods of the two
dominating classes of (primitive, classical, periodic) trajectories that feature.
As a consequence, the DOS will comprise:
(i)~a relatively smooth contribution due to retracings of geometrical
chords across the billiard of stationary lengths, dressed by
(ii)~a more rapidly varying contribution arising from orbits located near
the boundary and involving charge-preserving as well as charge-interconverting
reflection processes.  Thus, from the oscillatory part of the DOS
one can \lq\lq hear\rq\rq\ aspects of the shape of the billiard such as
the stationary values of the lengths of the chord ).  We are not aware of
any other spectral-geometric contexts that feature this type of
information.
Fourth, the quasiparticle energy eigenvalue spectrum, and its sensitivity to
the shape of the billiard, should be experimentally accessible, e.g., via
tunneling spectroscopy on hybrid superconducting/normal-state structures. The
current state of microfabrication technology makes such experiments
realizable~\cite{REF:antidot}.

We see the following as the principal results of the present work.
First, we provide the machinery for computing the Green function for
Andreev billiards of arbitrary shape in terms of a multiple scattering
expansion that focuses on the influence of the billiard shape.
Second, we implement this machinery to construct two semiclassical schemes
(resulting in two semiclassical trace formulas)
for computing the oscillatory component of the DOS.
One, which we shall refer to as Scheme~A, simply amounts to an elaboration
(to billiards of arbitrary shape) of Andreev's approximation.
Thus, it gives a DOS that takes the form of an integral over the
chords of the normal-state region with an appropriate weight
function.  Hence, it realizes the intuitively natural notion that the
chords, being the periodic orbits at the Andreev level, determine the
energy eigenvalue spectrum, according to Bohr-Sommerfeld quantization
conditions.  The other, Scheme~B, captures certain physical effects that are
inaccessible to Scheme~A, such as  mesoscale oscillations in the DOS.

This paper is organized as follows.
Following the present introduction and overview
we define, in Sec.~\ref{SEC:BDAB}, Andreev billiards and present the
corresponding Bogoliubov--De~Gennes (\bdg) eigenproblem.  In this section
we also introduce the
Green function for the \bdg\ eigenproblem and provide its connection to
the DOS, and review the standard {\it quasiclassical\/} approach to the
\bdg\ eigenproblem, due to Andreev. In Sec.~\ref{SEC:MuScEx} we formulate
the computation of the Green
function in terms of an expansion in which the basic processes are
reflection from and transmission through the interface separating the normal(N) and
superconducting(S) regions.  We then make a physically-motivated reorganization
of this expansion, which will later allow us to integrate out states in the S region, thus
obtaining a description solely in terms of states within the billiard.
(Section~\ref{SEC:MuScEx} is the technical basis of the Paper; however, it may
safely be omitted by readers wishing to focus on results rather than
methods.)\thinspace\
After this exact reformulation we proceed, in Sec.~\ref{SEC:AsMuScEx},
to integrate out the states in the S region within an approximation scheme
valid for short waves.  In this way, we obtain an effective Multiple
Reflection (rather than Scattering) Expansion.
We continue, in Sec.~\ref{SEC:DenOfStates}, by following certain approximation
strategies that allow us to compute the DOS at various
levels of energy-resolution, in each case obtaining the corresponding trace
formula.  This section is the heart of the Paper.
In Sec.~\ref{SEC:Summ} we make some concluding remarks and hint at some
possible applications of the ideas we have presented.
We have relegated to appendices some background material, including a
derivation of the \bdg\ wave equation and the rudiments of boundary
integral methods, as well as some technical and parenthetical passages.
\section{Andreev billiards}
\label{SEC:BDAB}
\subsection{Idealization of the physical system}
\label{SEC:Idealize}
\begin{figure}[hbt]
 \epsfxsize=10cm
\centerline{\epsfbox{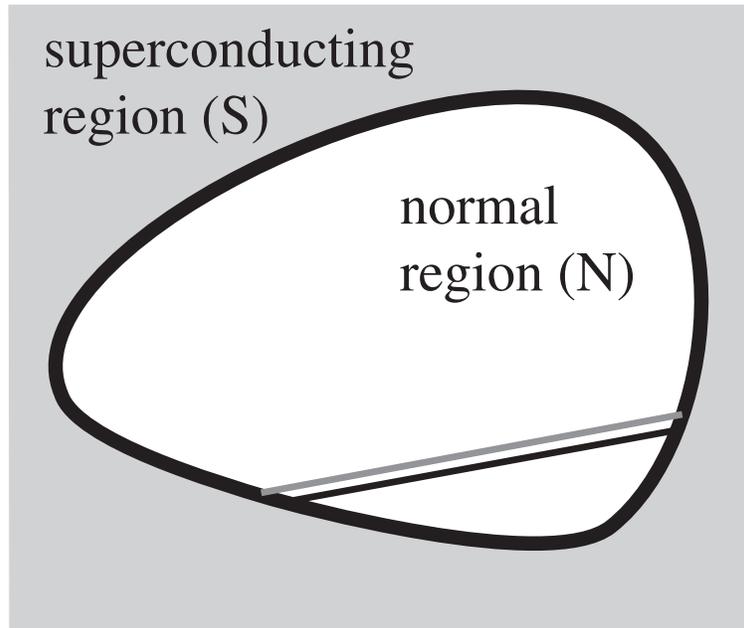}} \vskip+0.4truecm
\caption{Two-dimensional example of an Andreev billiard, showing a
normal region (N) surrounded by a superconducting region (R). In
this example the billiard is convex.} \label{FIG:and_bill}
\end{figure}

The physical system of interest in the present Paper is an Andreev
billiard of arbitrary shape, i.e., a normal-state region embedded
inside an infinite superconducting region, as depicted in
Fig.~\ref{FIG:and_bill}.  Following Gor'kov's mean-field approach
to superconductivity~\cite{REF:LPG,REF:AGD}, we describe the system
by the (variable particle-number) Hamiltonian $H$, given by
\begin{equation}
H=\sum_{\alpha=\pm}\int d^{d}x\,
\psi_\alpha^{\dag }({\bf x})\left\{
-{\frac{\hbar^{2}}{2m}}\nabla^{2}-\mu\right\}
 \psi_{\alpha}^{\phantom\dag}({\bf x})
+\int d^{d}x\,
        \left\{
 \Delta ({\bf x})^{*}\,
 \psi_{+}^{\phantom\dag}({\bf x})\,
 \psi_{-}^{\phantom\dag}({\bf x})
+\Delta({\bf x})\,
 \psi_{+}^{\dag}({\bf x})\,
 \psi_{-}^{\dag}({\bf x})
        \right\} .
\label{EQ:Gorkov}
\end{equation}
Here, $m$ is the effective electron mass, $\mu$ is the electron
chemical potential (which we take to be uniform throughout the
system), $\Delta ({\bf x})$ is a given superconducting pair-potential
which characterizes the superconducting condensate,
$\psi_{\alpha}^{\dag}({\bf x})$ and
$\psi_{\alpha}^{\phantom\dag}({\bf x})$ are creation and
annihilation field operators for electron quasiparticles having
position ${\bf x}$ and spin-projection ${\alpha}=\pm $, and $d$ is
the dimension of space (typically two or three).  We shall
not go beyond the picture of quasiparticle
excitations propagating in the presence of a superconducting condensate implied by this
description.
Within the Gor'kov description of the consequences of the
electron-electron interaction, any (Heisenberg-representation)
excited state that is arrived at by the addition of a single spin-up electron
quasiparticle to the (Heisenberg-representation) ground state
$\vert\Phi_{0}\rangle$
evolves into a coherent superposition of such a state and a state
arrived at by the removal of a spin-down electron quasiparticle
from the ground state\cite{FT:spinflip},
the Hamiltonian~(\ref{EQ:Gorkov}) maintaining the
system in this sector of Fock space.
Thus, it is adequate to address states of the form
\begin{equation}
\vert\Phi_{1}\rangle\equiv\int d^{d}x\,
\left(
u({\bf x})\,\psi_{\uparrow}^{\dag}({\bf x})+
v({\bf x})\,\psi_{\downarrow}     ({\bf x})\right)
\ket{\Phi_{0}},
\label{EQ:BareState}
\end{equation}
which are described by the two-component complex-valued amplitudes
$\big(u({\bf x}),v({\bf x})\big)$,
i.e., the family of one-quasiparticle excited states.

To derive the equation of motion for these one-quasiparticle states
(i.e.~the so-called time-dependent \bdg\ equation~\cite{REF:PGdGbook}), we follow
Andreev~\cite{REF:AFAndreev} and suppose that the system is
in some (Heisenberg-representation) one-quasiparticle excited state
$\vert\Phi_{1}\rangle $.
Then the amplitude $u({\bf x},t)$ for finding the state at time $t$
to be the ground state with an up-spin electron quasiparticle added at
position ${\bf x}$ (i.e.~the Heisenberg-representation state
$\psi_{+}^{\dag}({\bf x},t)\vert\Phi_{0}\rangle$)
is given by
\begin{mathletters}
\begin{equation}
\langle\Phi_{0}\vert\psi_{+}^{\phantom\dag}({\bf x},t)
\vert\Phi_{1}\rangle.
\end{equation}
Similarly, the amplitude $v({\bf x},t)$ for finding the state at time
$t$ to be the ground state with a down spin electron quasiparticle removed at
position ${\bf x}$ (i.e.~the Heisenberg-representation state
$\psi_{-}^{\phantom\dag}({\bf x},t)\vert \Phi_{0}\rangle$)
is given by
\begin{equation}
\langle\Phi_{0}\vert \psi_{-}^{\dag}({\bf x},t)\vert \Phi_{1}\rangle.
\end{equation}%
\end{mathletters}%
Here,
$\psi_{+}^{\phantom\dag}({\bf x},t)
\equiv
{\romE}^{iHt/\hbar}\,
{\psi_{+}^{\phantom\dag}({\bf x})}\,
{\romE}^{-iHt/\hbar}$
and
$\psi_{-}^{\dag}({\bf x},t)
\equiv
{\romE}^{iHt/\hbar}\,
\psi_{-}^{\dag}({\bf x})\,
{\romE}^{-iHt/\hbar}$
are, respectively, Heisenberg-representation field operators.
Thus, the wave functions $u({\bf x},t)$ and $v({\bf x},t)$ serve as
amplitudes for the present up-spin electron quasiparticle and
the absent down-spin electron (i.e.~up-spin hole) quasiparticle.
Then, by virtue of the Heisenberg equation of motion for the field
operators (see, e.g., Ref.~\cite{REF:AGD}, Sec.~6), together with
the Hamiltonian~(\ref{EQ:Gorkov}), it is a straightforward exercise
in computing commutators of field operators to show that the amplitudes
$u({\bf x},t)$ and $v({\bf x},t)$ evolve according to the appropriate
time-dependent Schr\"odinger equation, i.e., the time-dependent \bdg\
equation:
\begin{equation}
i\hbar{\frac{\partial}{{\partial t}}}
\left(
\matrix{
u({\bf x},t)\cr
\noalign{\bigskip}
v({\bf x},t)}
\right)
=
\pmatrix{
-{\hbar^{2}\over{2m}}\nabla^{2}-\mu
&
\Delta({\bf x})
\cr
\noalign{\medskip}
\Delta^{\ast}({\bf x})
&
-\left(-{\hbar^{2}\over{2m}}\nabla^{2}-\mu\right)}
\left(
\matrix{
u({\bf x},t)
\cr
\noalign{\bigskip}
v({\bf x},t)}
\right).
\label{EQ:TDBdG}
\end{equation}
Analysis of this equation via the separation of the time variable, appropriate
when there is no external time-dependence, leads to the \bdg\ eigenproblem
\begin{equation}
\pmatrix{
-{\hbar^{2}\over{2m}}\nabla^{2}-\mu
&
\Delta({\bf x})
\cr
\noalign{\medskip}
\Delta^{\ast}({\bf x})
&
-\left(-{\hbar^{2}\over{2m}}\nabla^{2}-\mu\right)
}
\left(
\matrix{ u_{n}({\bf x})
\cr
\noalign{\bigskip}
v_{n}({\bf x})}
\right)
=
E_{n}\left(
\matrix{
u_{n}({\bf x})
\cr
\noalign{\bigskip}
v_{n}({\bf x})}
\right),
\label{EQ:TIBdG}
\end{equation}
where $n$ is a (collective) index for all quantum numbers and $\{E_{n}\}$ and
$\left\{\big(u_{n}({\bf x}),v_{n}({\bf x})\big)\right\}$ are the corresponding
energy eigenvalues and (two-component) eigenfunctions.

In general, $\Delta$ may vary spatially.  However, we shall consider the situation
in which deep inside the superconductor $\Delta$
goes to a constant value $\Delta_{0}$, whereas throughout the normal metal it
vanishes. In the intermediate region (i.e.~within a superconducting coherence
length outside of the billiard boundary) $\Delta$ is suppressed to a value
smaller than $\Delta_{0}$, and falls to zero as the N region is entered. We
shall ignore the effects of this suppression and assume that $\Delta$ varies
discontinuously between $0$ to $\Delta_{0}$ at a surface, which we refer to
as the billiard boundary and denote by $\surS$, that divides the system into
two homogeneous regions, the billard interior (denoted $\volV$) and the
billiard exterior (denoted $\notV$).  For the sake of simplicity,
we further assume that there are no metallurgical differences between the normal-state
and superconducting regions, inasmuch as the only difference between them is
the value of the pair potential (the effective mass, e.g., being common).

To ease the notation we shall adopt units in which $\hbar^{2}/2m=1$.
To recover results in terms of the original physical units, one
multiplies the three variables having the dimensions of energy
[viz.~$\mu$, $\Delta({\bf r})$ and $E$] by the factor
$\left(2m/\hbar^{2}\right)$.

The \bdg\ eigenproblem plays the same role for Andreev billiards that the
Schr{\"o}dinger eigenproblem plays for conventional billiards.  If a
conventional billiard is surrounded by a region in which the single-particle
potential is {\it infinite\/} then we refer to the billiard as a {\it hard\/}
 billiard, and the Schr{\"o}dinger equation {\it outside\/} the billiard is
replaced by the homogeneous Dirichlet (i.e.~vanishing) boundary condition on
the Schr{\"o}dinger eigenfunction, this boundary condition leading to the
quantization of the eigenvalue spectrum.
If, on the other hand, a conventional billiard is surrounded by a region
in which the single-particle potential is {\it finite\/} then we refer
to the billiard as a {\it soft\/} billiard, and the solution to the
Schr{\"o}dinger equation {\it outside\/} must be matched on to the the
solution of the Schr{\"o}dinger equation {\it inside\/}, this matching leading
to the quantization of the eigenvalue spectrum.  The Andreev billiard is,
therefore, analogous to a {\it soft\/} Schr{\"o}dinger billiard; its
hard limit seems difficult to realize because, at least in known superconductors,
the pair potential is far smaller than the chemical potential.

The eigenproblem for Andreev billiards, then, is given by
Eq.~(\ref{EQ:TIBdG})
with $\Delta({\bf x})=0$        for ${\bf x}$ inside the billiard,
and  $\Delta({\bf x})=\Delta_0$ for ${\bf x}$ outside the billiard.
Thus, we are faced with the task of addressing the \bdg\ eigenproblem for
the case in which the system comprises two spatially homogeneous regions
(one normal, one superconducting) that meet at a closed surface. Owing to
the spatial homogeneity of these regions, the general solution of the \bdg\
equation can readily be obtained in each region. The quantization of the
eigenvalue spectrum results from the matching of the solutions and their
normal derivatives across this surface, together with the confinement of
the eigenfunctions to the vicinity of the Andeev billiard. The resulting
spectrum depends, therefore, on the shape of this surface. Exploring this
dependence is the central aim of the present work.  It would be straightforward
to extend the present framework to handle issues such as Josephson coupling
between superconducting regions, scattering from Andreev billiards, etc.
\subsection{Green function and density of states}
The spectrum of eigenvalues of the \bdg\ wave equation $\{E_{n}\}$
is assembled into the DOS $\rho(E)$, which is defined as
\begin{equation}
\rho(E)\equiv\sum_{n}\delta(E-E_{n}).
\label{EQ:DOSdef}
\end{equation}
As is commonly the case, it is convenient to approach
$\rho(E)$ via a Green function for the \bdg\ wave equation,
which we now introduce.
\subsubsection{Green function for the \bdg\ equation}
\label{SEC:GFbdg}
The ($2\times2$ matrix)
Green function $\gf({\bf x},{\bf x}^{\prime};z)$
for the \bdg\ wave equation is defined by the following
matrix partial differential equation:
\begin{equation}
\pmatrix{
-\nabla^{2}-\mu -z & \Delta({\bf x})
\cr
\noalign{\medskip}
\Delta^{\ast}({\bf x}) & \nabla^{2}+\mu -z}
\gf({\bf x},{\bf x}^{\prime};z)
=
{\bf I}\,\delta({\bf x}-{\bf x}^{\prime}),
\label{EQ:GBdG}
\end{equation}
where $z$ is the complex energy and ${\bf I}$ is the ($2\times2$)
identity matrix.  Under the far-field boundary condition
${\bf G}({\bf x},{\bf x}^{\prime};z)\rightarrow{\bf 0}$ as
$|{\bf x}-{\bf x}^{\prime}|\rightarrow\infty$
the Green function $\gf({\bf x},{\bf x}^{\prime};z)$ is unique~\cite{FT:bndryinfty}.
For the case of the Andreev billiard we have
$\Delta({\bf x})=0$ for ${\bf x}$ in $\volV$ and
$\Delta({\bf x})=\Delta_{0}$ for ${\bf x}$ in $\notV$.

It is useful to express $\gf({\bf x},{\bf x}^{\prime};z)$ in terms of the
eigenfunctions $\left({u_{n},v_{n}}\right)$ of the \bdg\ Hamiltonian, i.e.,
\begin{equation}
\gf({\bf x},{\bf x}^{\prime};z)
=\sum_{n}\frac{1}{E_{n}-z}
\pmatrix{
u_{n}({\bf x})\,u_{n}^{*}({\bf x}^{\prime})&
u_{n}({\bf x})\,v_{n}^{*}({\bf x}^{\prime})\cr
\noalign{\medskip}
v_{n}({\bf x})\,u_{n}^{*}({\bf x}^{\prime})&
v_{n}({\bf x})\,v_{n}^{*}({\bf x}^{\prime})}.
\label{EQ:GfEvE}
\end{equation}
In order to see that this form does indeed satisfy
Eq.~(\ref{EQ:GBdG}), one may substitute this expression into
Eq.~(\ref{EQ:GBdG}) and make use of Eq.~(\ref{EQ:TIBdG})
and the completeness of the eigenfunctions, i.e.,
\begin{equation}
\sum_{n}
\pmatrix{
u_{n}({\bf x})\,u_{n}^{*}({\bf x}^{\prime})&
u_{n}({\bf x})\,v_{n}^{*}({\bf x}^{\prime})\cr
\noalign{\medskip}
v_{n}({\bf x})\,u_{n}^{*}({\bf x}^{\prime})&
v_{n}({\bf x})\,v_{n}^{*}({\bf x}^{\prime})}
={\bf I}\,\delta({\bf x}-{\bf x}^{\prime}).
\end{equation}
\subsubsection{Connection between \bdg\ Green function and
density of states}
We now follow the standard practice of expressing a DOS
in terms of the corresponding Green function by making
use of the identity
\begin{equation}
\frac{1}{\pi}\,
{\rm Im}\,\lim_{\epsilon\to+0}
{1\over{E_{n}-E-i\epsilon}}=
\delta(E_{n}-E).
\end{equation}
Together with Eqs.~(\ref{EQ:GfEvE}) and (\ref{EQ:DOSdef}),
this allows us to see that
\begin{eqnarray}
\rho (E)
&=&
\frac{1}{\pi}
\int d^{d}x\,
\left\{
{\rm Im}\,{\rm Tr}\,
\gf({\bf x},{\bf x}^{\prime};E+i\epsilon )
\right\}_{{\bf x}^{\prime}={\bf x}}
\label{EQ:DOSGF}
=
\int d^{d}x\,
\sum_{n}\frac{1}{\pi}\,{\rm Im}
\frac{
\vert{u_{n}({\bf x})}\vert^{2}+
\vert{v_{n}({\bf x})}\vert^{2}
      }{E_{n}-E-i\epsilon}
\nonumber\\
&=&
\sum_{n}
\frac{1}{\pi}\,
{\rm Im}\,\frac{1}{E_{n}-E-i\epsilon}
=\sum_{n}\delta (E-E_{n}).
\nonumber
\end{eqnarray}
Here, ${\rm Tr}\,\gf$ denotes the (matrix) trace over the
diagonal components of $\gf$.

Equation~(\ref{EQ:DOSGF}) is an expression for $\rho(E)$ in terms
of $\gf({\bf x},{\bf x}^{\prime};E+i\epsilon )$, and any
approximation to $\gf({\bf x},{\bf x}^{\prime};E+i\epsilon )$ thus
furnishes an approximation to $\rho(E)$.  However, being a sum of delta
functions, $\rho (E)$ is not a smooth function and can, therefore, be
extremely awkward to approximate.  In order to find approximations to
$\rho (E)$ it is preferable to seek a continuous function that carries
essentially the same information as it.  One candidate is the
{\it smoothed\/} DOS $\rho_{\gamma}(E)$, defined via
\begin{equation}
\rho_{\gamma}(E)\equiv
\int_{-\infty}^{\infty}
dE^{\prime}\,
\smde(E-E^{\prime};\gamma)\,\rho(E^{\prime}),
\end{equation}
where $\smde(E-E^{\prime};\gamma)$ is some smoothing function and
$\gamma $ is the (real) smoothing width.   We remark, parenthetically,
that any DOS derived from experiment will be smoothed to
some extent, depending on the resolving power of the apparatus and/or
the lifetime of the single-particle excitations.

There are several possible choices for the smoothing fuction
$\smde(E-E^{\prime};\gamma)$, including, e.g., Lorentzian, Gaussian and
logarithmic-Gaussian.  It is also possible to define a continuous integral
transform of $\rho(E)$ that is, itself, a physical
property, and for which we can derive some approximation.  For example, under the
Lambert transform of $\rho (E)$, which exchanges energy
for temperature, $\rho (E)$ is transformed into the total (equilibrium internal) energy
as a function of temperature~\cite{FT:smthproc}.
Which smoothing procedure is the best choice depends on the
method used to approximate $\rho(E)$.   For Green-function--based methods,
such as the one we shall adopt, the Lorentzian smoothing function,
\begin{equation}
\smde(E-E^{\prime};\gamma)\equiv
\frac{1}{\pi}
\frac{\gamma}{(E-E^{\prime})^{2}+\gamma^{2}},
\end{equation}
is the most appropriate, for reasons that should become clear below.

We now give the analogue of Eq.~(\ref{EQ:DOSGF}) for relating the
Lorentzian-smoothed DOS $\rho_{\gamma}(E)$ and the
Green function at complex energy $\gf({\bf x},{\bf x}^{\prime};z)$.
Following essentially the procedure that lead to
Eq.~(\ref{EQ:DOSGF}), we have the following identity for arbitrary
(real) $\gamma$:
\begin{eqnarray}
\rho_{\gamma}(E)
&=&
\frac{1}{\pi}\int d^{d}x\,
{\rm Im}\,{\rm Tr}\,\gf({\bf x},{\bf x}^{\prime};E+i\gamma)
\Big\vert _{{\bf x}^{\prime}={\bf x}}
=
\int d^{d}x\,\sum_{n}
\frac{1}{\pi}\,{\rm Im}\,
\frac{\left\vert{u_{n}({\bf x})}
\right\vert^{2}+\left\vert{v_{n}({\bf x})}\right\vert^{2}}
{E_n -E-i\gamma}
\nonumber
\\
&=&
\sum_{n}
\frac{1}{\pi}
\frac{\gamma}{(E-E_{n})^{2}+\gamma ^{2}}
=
\int_{-\infty}^{\infty}
dE^{\prime}\,
\frac{1}{\pi}
\frac{\gamma}
{(E-E^{\prime})^{2}+\gamma^{2}}\,\,\sum_{n}
\delta(E^{\prime}-E_{n})
\label{EQ:DOSGF_S}.
\end{eqnarray}%
Naturally, in the limit $\gamma\to 0^{+}$, $\rho_{\gamma}(E)$ passes to
$\rho(E)$.
\subsubsection{Fundamental Green function for a homogeneous normal region}
\label{SEC:FundaGFN}
In this section we derive the fundamental (i.e.~homogeneous) normal-state
Green function, the first of two Green functions that are central to the construction of
the Green function for an Andreev billiard.  Along the way, we
introduce some convenient notation that we shall subsequently make use of.
This fundamental Green function is the translationally-invariant solution
$\gf^{\rm N}({\bf x},{\bf x}^{\prime};z)$ of the equation
\begin{mathletters}
\begin{eqnarray}
&&
({\bf H}-z{\bf I})\,\gf_0({\bf x},{\bf x}^{\prime};z)=
\delta ^{(d)}({\bf x-x}^{\prime})\,{\bf I},
\label{defn_of_Greenf1}
\\
&&
{\bf H}=
\pmatrix{{\bf\hat{p}}^{2} -\mu & 0 \cr 0 &
        -{\bf\hat{p}}^{2} +\mu }
\equiv
(-{\bf \hat p}^2+\mu )\,\psm{3}\;,
\end{eqnarray}
\end{mathletters}
where $\psm{1,2,3}$ are the three Pauli matrices, together with the
boundary condition that the Green function vanishes at infinity:
\begin{equation}
\gf^{\rm N}({\bf x},{\bf x}^{\prime};z)\rightarrow {\bf 0}
\quad \mbox{as}\quad
|{\bf x-x}^{\prime}|\rightarrow \infty\;.
\end{equation}
The solution for this Green function is given by the (spatial) matrix
element of the operator $(z{\bf I}-{\bf H})^{-1}$, i.e.,
\begin{eqnarray}
\gf^{\rm N}({\bf x},{\bf x}^{\prime};z)
&=&
\langle{\bf x}|\frac{1}{z{\bf I}-{\bf H}}|{\bf x}^{\prime}\rangle=
\langle{\bf x}|\frac{z{\bf I}+({\bf\hat p}^2-\mu)\psm{3}}
{z^2-({\bf \hat p}^2-\mu )^2}|{\bf x}^{\prime}\rangle
\nonumber
\\
&=&(z{\bf I}-(\nabla_x^2+\mu )\,\psm{3})
\frac{1}{2E}
\Big(
\langle{\bf x}|(z -{\bf \hat p}^2+\mu )^{-1}|
{\bf x}^{\prime}\rangle+
\langle{\bf x}|
(z+{\bf \hat p}^2-\mu )^{-1}|
{\bf x}^{\prime}\rangle
\Big),
\end{eqnarray}
which is, of course, a matrix in electron/hole space.
The two terms inside the parantheses are quite familiar: they are the usual
Green functions of the Helmholtz wave equation, and can be
evaluated in the standard way.  In three dimensions, e.g., one has
\begin{equation}
\langle{\bf x}|
(z \mp {\bf \hat{p}}^{2} \pm \mu)^{-1}|{\bf x}^{\prime}\rangle
=
\pm \frac{1}{4 \pi}
\frac{{\romE}^{\pm i k_{\pm} |{\bf x - x}^{\prime}|}}
{|{\bf x - x}^{\prime}| } \; ,
\end{equation}
The symbols $k_{\pm }$, which will be used throughout this Paper, denote
particle and hole wave numbers, and are given by the expressions
\begin{equation}
k_{\pm }=\sqrt{\mu \pm z}\;,
\end{equation}
where it is understood that the roots having positive real parts are
the ones that are adopted.  Then, for the three-dimensional case, one
arrives at the the fundamental Green function for the normal state:
\begin{eqnarray}
\gf^{\rm N}({\bf x},{\bf x}^{\prime};z)
&=&
\frac{z{\bf I}-(\nabla ^2+\mu )\psm{3}}{2z}
\frac 1{4\pi }
\left(
 \frac{{\romE}^{ ik_{+}|{\bf x-x}^{\prime}|}}{|{\bf x-x}^{\prime}|}
-\frac{{\romE}^{-ik_{-}|{\bf x-x}^{\prime}|}}{|{\bf x-x}^{\prime}|}
\right)
=
\frac{1}{{4\pi|{\bf x}-{\bf x}^{\prime}|}}
\pmatrix{\ {\romE}^{ik_{+}|{\bf x}-{\bf x}'|}&0 \cr 0
         &-{\romE}^{-ik_{-}|{\bf x}-{\bf x}'|} }.
\label{EQ:GFhnm}
\end{eqnarray}
This Green function has the expected form: an outgoing spherical wave in
the particle component (i.e.~the particle Green function) and an incoming
spherical wave in the hole component (i.e.~the hole Green function), the
latter representing an outgoing hole.  Moreover, as we expect in the normal state,
there is no mixing between the particle and hole components, the off-diagonal
elements being zero.  Next, we introduce some convenient notation for the
components of this fundamental Green function, viz.,
\begin{equation}
g_{\pm }({\bf x},{\bf x}^{\prime})
\equiv
\frac{{\romE}^{\pm ik_{\pm }|{\bf x-x}^{\prime}|}}
{|{\bf x-x}^{\prime}|}\;,
\end{equation}
in terms of which $\gf^{\rm N}$ becomes
\begin{equation}
\gf^{\rm N}({\bf x},{\bf x}^{\prime};z)=
\pmatrix{
g_{+}({\bf x},{\bf x}^{\prime}) & 0 \cr
0 & -g_{-}({\bf x},{\bf x}^{\prime})
}.
\label{EQ:FGFhnr}
\end{equation}
\subsubsection{Fundamental Green function for a homogeneous
superconducting region}
\label{SEC:GS}
The fundamental Green function for a homogeneous superconducuting
region is the translationally invariant solution
$\gf^{\rm S}({\bf x},{\bf x}^{\prime};z)$ of the equation:
\begin{equation}
\left\{
z{\bf I}+({\nabla }^{2}+\mu )\,\psm{3}
+\Delta _{1}\psm{1}+\Delta_{2}{\psm{2}}
\right\}
\,{\bf G}^{{\rm S}}({\bf x}-{\bf x}^{\prime})=
{\bf I}\,\delta^{(d)}({\bf x}-{\bf x^{\prime }}),
\label{EQ:gSGF}
\end{equation}
together with the boundary condition that it vanishes as $\vert{\bf x}-{\bf x}'\vert$
goes to infinity. Here $\Delta _{1}$ and $\Delta _{2}$ represent the (constant) real and
imaginary parts of the complex pair-potential.
Note that we shall henceforth take the complex energy $z$ to be the real energy $E$.
We shall be concerned with
situations in which the quasiparticles are bound to the billiard and shall,
threfore, assume that $\vert{E}\vert<\Delta$, where the magnitude
$\Delta$ of the pair potential obeys
$\Delta^{2}\equiv\Delta_{1}^{2}+\Delta_{2}^{2}$.
In the present work, it is convenient to choose a specific gauge, which
we do by setting $\Delta_{1}=0$ and $\Delta_{2}=\Delta$.  However, for
extensions of the present work to settings, such as SNS junctions, in
which there are physical implications of phase differences it is necessary
to ksow the Green function for arbitrary gauges, and it is therefore for
this case that we provide the Green function.

To obtain the Green function one first formally inverts Eq.~(\ref{EQ:gSGF})
to obtain
\begin{equation}
{\bf G}^{{\rm S}}({\bf x}-{\bf x}^{\prime })=
\left\{
E\,{\bf I}+({\nabla }^{2}+\mu)\,{\psm{3}}
+\Delta_{1}\psm{1}+\Delta _{2}{\psm{2}}
\right\}^{-1}
\delta^{(d)}({\bf x}-{\bf x^{\prime }}).
\end{equation}
By manipulating this equation so as to separate the matrix
and partial-differential aspects of the inversion operation
one then arrives at
\begin{mathletters}
\begin{eqnarray}
{\bf G}^{{\rm S}}({\bf x}-{\bf x}^{\prime })
&=&
\left\{
E\,{\bf I}+({\nabla }^{2}+\mu ){\psm{3}}
+\Delta_{1}\psm{1}+\Delta_{2}{\psm{2}}\right\}^{-1}
\, \delta ^{3}({\bf x}-{\bf x^{\prime }})
\\
&=&
\left\{
E\,{\bf I}-({\nabla}^{2}+\mu ){\psm{3}}
-\Delta _{1}\psm{1}-\Delta_{2}{\psm{2}}
\right\}
\left\{
E^{2}-\Delta^{2}-({\nabla }^{2}+\mu )^{2}
\right\}^{-1}\,
\delta ^{(d)}({\bf x}-{\bf x^{\prime}})
\\
&=&
\left\{
E\,{\bf I}-({\nabla }^{2}+\mu ){\psm{3}}
-\Delta _{1}\psm{1}-\Delta _{2}{\psm{2}}
\right\}
\frac{1}{2i\sqrt{\Delta^{2}-E^{2}}}
\nonumber \\
&&
\quad\times
\left\{
 \left({\nabla}^{2}+\mu+i\sqrt{\Delta^{2}-E^{2}}\right)^{-1}
-\left({\nabla}^{2}+\mu-i\sqrt{\Delta^{2}-E^{2}}\right)^{-1}
\right\}\,
\delta^{(d)}({\bf x}-{\bf x^{\prime }}).
\end{eqnarray}
We identify the wave vectors $k_{+}^{{\rm S}}$ and $k_{-}^{{\rm S}}$
for the electron-like and hole-like components, respectively, as follows:
\end{mathletters}
\begin{equation}
k_{\pm }^{{\rm S}}
\equiv
\sqrt{\mu\pm i\sqrt{\Delta ^{2}-E^{2}}},
\qquad\mathop{\rm Re}k_{\pm }^{{\rm S}}>0.
\end{equation}
As the partial differential operators
$\left( \nabla^{2}+\left( k_{\pm }^{{\rm S}}\right)^{2}\right)$
are partial differential operators of the Helmholtz wave equation
for wave vectors of length $k_{\pm }^{{\rm S}}$, their inversion is well known,
\begin{equation}
g_{\pm }^{{\rm S}}({\bf x}-{\bf x}^{\prime })
\equiv
-\left(\nabla^{2}+\left( k_{\pm }^{{\rm S}}\right)^{2}\right)^{-1}
\delta ^{3}({\bf x}-{\bf x^{\prime }})
=\frac{{\romE}^{\pm ik_{\pm }^{{\rm S}}|{\bf x}-{\bf x^{\prime }}|}}
{4\pi |{\bf x}-{\bf x}^{\prime }|},
\label{EQ:invHelmholtz}
\end{equation}
where the final form on the right hand side holds for the three-dimensional
case, and the minus sign in the exponent for holes ensures the proper decay at large
distances.  Upon using Eq.~(\ref{EQ:invHelmholtz}),
${\bf G}^{{\rm S}}$ becomes
\begin{mathletters}
\begin{eqnarray}
{\bf G}^{{\rm S}}({\bf x}-{\bf x}^{\prime })
&=&
\left\{
E\,{\bf I}-({\nabla}^{2}+\mu ){\psm{3}}
-\Delta _{1}\psm{1}-\Delta _{2}{\psm{2}}
\right\}
\frac{1}{2i\sqrt{\Delta ^{2}-E^{2}}}
\left(
 g_{+}^{{\rm S}}({\bf x}-{\bf x}^{\prime })
-g_{-}^{{\rm S}}({\bf x}-{\bf x}^{\prime })
\right)
\\
&=&
\frac{1}{2}
\left\{
\frac{E}{i\sqrt{\Delta ^{2}-E^{2}}}\,{\bf I}
-\frac{\Delta_{1}}{i\sqrt{\Delta^{2}-E^{2}}}\,\psm{1}
-\frac{\Delta_{2}}{i\sqrt{\Delta^{2}-E^{2}}}\,\psm{2}
\right\}
\left(
 g_{+}^{{\rm S}}({\bf x}-{\bf x}^{\prime })
-g_{-}^{{\rm S}}({\bf x}-{\bf x}^{\prime })
\right)
\nonumber\\
&&
\qquad\qquad\qquad+\frac{1}{2}{\psm{3}}
\left(
 g_{+}^{{\rm S}}({\bf x}-{\bf x}^{\prime})
+g_{-}^{{\rm S}}({\bf x}-{\bf x}^{\prime })
\right).
\label{EQ:GFhsc}
\end{eqnarray}
\end{mathletters}%
To get to the second line, the action of $({\nabla }^{2}+\mu)$ on
$g_{\pm }^{{\rm S}}$ is calculated with the help of
Eq.~(\ref{EQ:invHelmholtz}).
\subsection{Andreev's quasiclassical aproximation scheme}
\label{SEC:Andreev}
%
%
We now review the conventional approximation
scheme for studying N-S hybrid systems, such as Andreev billiards.
This approximation scheme was put forth by Andreev~\cite{REF:AFAndreev}, and we
shall refer to it as Andreev's approximation.
It takes advantage of the fact that for typical N-S systems
the dimensionless parameter $\Delta /\mu$ is much less than unity.  The scheme consists of the
separation of rapid and slow oscillations in the wavefunction.
It can be motivated in the following way: consider an arbitrary N-S structure
(i.e.~arbitrary $\Delta ({\bf x})\ll \mu $). Suppose that the electronic
properties of the system are probed during a short time $\Delta t$, so that
there is  insufficient energy resolution around $E=0$ to resolve $\Delta ({\bf x})$.
Then no measurement obtained via this probe is capable of
distinguishing between the original system and a system described by the same \bdg\
equation but with $\Delta({\bf x})$ set to zero.  An energy resolution of $\Delta E$
around $E=0$ corresponds to a momentum resolution of $\Delta p=m\,\Delta E/\Kf$ which,
via the Heisenberg uncertainty relation, corresponds to
a spatial resolution of $\Delta x=\hbar^{2}\Kf/m\,\Delta E$.  Therefore, in order to
resolve any effects of the pair-potential on the electronic states, the system must be
probed on lengthscales {\em larger\/} than $\xi \equiv\hbar^{2} \Kf/m\Delta$, i.e.,
the superconducting coherence length.  Thus, the single-particle wavefunctions
must have the form of plane-waves at the Fermi momentum with an envelope that
varies on the lengthscale $\xi$, i.e.,
\begin{equation}
\pmatrix{u({\bf x}) \cr
         v({\bf x}) }
={\rm e}^{ik_{F}{\bf n}\cdot {\bf x}}
\pmatrix{\bar{u}({\bf x}) \cr
         \bar{v}({\bf x})},
\label{envelope}
\end{equation}
where the unit vector ${\bf n}$ defines the orientation of the wavevector of the
plane wave and $\bar{u}$ and $\bar{v}$ are the slowly varying envelope amplitudes.
By substituting this form into the \bdg\ equation~(\ref{EQ:TIBdG}) one obtains
\begin{equation}
\pmatrix{-\left(\nabla^{2}+2ik_{F}{\bf n}\cdot\bbox{\nabla}\right)  & \Delta({\bf x}) \cr
         \Delta^{\ast }({\bf x}) & \left(\nabla^{2}+2ik_{F}{\bf n}\cdot\bbox{\nabla}\right)}
\pmatrix{\bar{u}({\bf x}) \cr
         \bar{v}({\bf x})}
=E\pmatrix{\bar{u}({\bf x}) \cr
           \bar{v}({\bf x})}.
\label{EQ:subs}
\end{equation}
As $\bar{u}$ and $\bar{v}$ vary on the lengthscale $\xi $, one has that
\begin{equation}
\frac{\left|
\nabla^{2}\bar{u}\right| }{\left| k_{F}{\bf n}\cdot
\bbox{\nabla} \bar{u}\right| }\sim {\cal O}\left( \frac{1}{k_{F}\xi }\right)
={\cal O}\left( \frac{\sudel}{\mu}\right)\ll 1\,.
\label{EQ:smallparam}
\end{equation}
Thus, to leading order in $\sudel/\mu$ it is permissible to ignore the $\nabla^{2}$
term in Eq.~(\ref{EQ:subs}), a singular approximation because it involves changing
the order of the system of partial differential equations.  This approximation to the
\bdg\ Hamiltonian is Andreev's approximation, and leads to the Andreev eigenproblem
\begin{equation}
\pmatrix{-2ik_{F}{\bf n}\cdot\bbox{\nabla}  & \Delta({\bf x}) \cr
         \Delta^{\ast }({\bf x}) & 2ik_{F}{\bf n}\cdot\bbox{\nabla}}
\pmatrix{\bar{u}({\bf x}) \cr
         \bar{v}({\bf x})}
=E\pmatrix{\bar{u}({\bf x}) \cr
           \bar{v}({\bf x})}.
\label{EQ:Andreev_PDE}
\end{equation}

Let us now apply Eq.~(\ref{EQ:Andreev_PDE}) to Andreev billiards.  First, it is useful
to express the variable ${\bf x}$ in terms of ${\bf b}$ (i.e.~the impact parameter or,
equivalently, the transverse parameter), and $s$ (i.e.~the longitudinal parameter) such
that
\begin{equation}
{\bf x}={\bf b}+ {\bf n}\,s.
\end{equation}
Notice that ${\bf b}$ represents the transverse degree(s) of freedom whereas $s$
represents the longitudinal degree of freedom of the excitation.  As, in
Eq.~(\ref{EQ:Andreev_PDE}), there is no differential operator acting on the variable
${\bf b}$, the wavefunctions take the form
\begin{equation}
\pmatrix{\bar{u}({\bf b},s) \cr
         \bar{v}({\bf b},s)}
=\delta({\bf b}-{\bf b}_0)
\pmatrix{u(s;{\bf b}_0) \cr v(s;{\bf b}_0)},
\end{equation}
where ${\bf b}_0$ can be interpreted as the transverse quantum number.
By substituting this form into Eq.~(\ref{EQ:Andreev_PDE}) one obtains
\begin{equation}
\pmatrix{-2ik_{F}\partial/\partial s  & \Delta({\bf b}_0,s) \cr
         \Delta^{\ast}({\bf b}_0,s) & 2ik_{F}\partial/\partial s}
\pmatrix{u(s;{\bf b}_0) \cr v(s;{\bf b}_0)}
=E\pmatrix{u(s;{\bf b}_0) \cr v(s;{\bf b}_0)}.
\label{EQ:Andreev_ODE}
\end{equation}
Thus, one has reduced the partial differential eigenvalue
equation~(\ref{EQ:TIBdG}) to a family of approximate ordinary differential
eigenvalue equations parametrized by $({\bf n},{\bf b}_0)$.

Now, for Andreev billiards $\sudel$ is real and piecewise constant.  In this
case, the solution of Eq.~(\ref{EQ:Andreev_PDE}) proceeds as follows.
The parameters $({\bf n},{\bf b}_0)$ fix a line, which
determines $\Delta({\bf b}_0,s)$, in the sense that
$\Delta({\bf b}_0,s)=\Delta({\bf b}_0+ {\bf n}\,s)$.
As $\Delta({\bf b}_0,s)$ is piecewise constant, the task of solving
Eq.~(\ref{EQ:Andreev_ODE}) is straightforward. The quantization
condition depends on the chord length $\ell({\bf n},{\bf b}_0)$
[i.e.~the length of the part of the line (specified by $({\bf n},{\bf b}_0)$)
lying inside N]:
\begin{equation}
E\,\ell({\bf n},{\bf b}_0)/\Kf-2\varphi=2\pi m\,,
\end{equation}
where $m=0,\pm 1,\pm 2,\ldots$ and $\varphi\equiv\cos^{-1}(E/\sudel)$.
Thus, for each chord there is a ladder of energy eigenvalues.  In order to
obtain the DOS one must first obtain the DOS for a single chord, and then sum
over all chords.  However, it is not {\it a priori\/} completely clear what weight should
be assigned to each chord in performing the continuous summation. In fact, as we shall see
in Sec.~\ref{SEC:SAandreev}, at the Andreev level of approximation the DOS for
a (convex, $d$-dimensional) Andreev billiard is given by
\begin{equation}
\rho(E)
\approx
\sum_{m=-\infty}^{\infty}
\frac{\Kf^{d-2}}
{2(2\pi)^{d-1}}\,
\int d{\bf n}\,d{\bf b}\,\,\ell({\bf n},{\bf b}_0)\,
\delta\left(\frac{E}{\Kf}\ell({\bf n},{\bf b}_0)-2\varphi-2\pi m\right),
\end{equation}
the two $(d-1)$-dimensional integrations,
one over ${\bf n}$ and one over ${\bf b}$,
implement the summation over chords.
\section{Multiple scattering expansion}
\label{SEC:MuScEx}
We now construct a Multiple Scattering Expansion capable of
yielding the Green function
$\gf({\bf x},{\bf x}^{\prime})$
associated with the \bdg\ equation~(\ref{EQ:GBdG}), in settings in
which $\Delta({\bf x})$ is
piecewise constant (i.e.~takes on certain constant values in various
regions of space that are delineated by $d-1$-dimensional surfaces).  Although the
construction is applicable to a
wider range of settings, we shall have in mind the application to an
Andreev billiard, for which $\Delta({\bf x})$ vanishes inside the
billiard and has a constant nonzero value $\Delta_{0}$ outside it.  The
spirit of our approach very much parallels that of BB-I, although there
there are significant differences arising from (i)~the form of the
eigenproblem (\bdg\ rather than Laplace), and (ii)~our need to employ
matching (rather than boundary) conditions.

In this section we set up the general formalism for the exact
\bdg\ Green function for the case in which the system is divided, via a
change in the pair potential, into at least two distinct regions.
As we shall see, the exact Green function can be expressed as a sum,
generated by an interation scheme, over all possible scatterings from
the boundary dividing these regions.

Readers not familiar with the elements of potential theory that we
shall be using (which are sometimes referred to as boundary integral techniques)
may wish to pause to read App.~\ref{APP:BasicIngredients}, in which we
give a self-contained introduction to this subject and develop the
elaborations necessary for application to the \bdg\ eigenproblem.
Specifically, we shall need to handle two-component wave functions in
the context of matching (rather than boundary) conditions.
\subsection{Matching conditions and boundary integral equations for
the \bdg\ Green function}
We now introduce a convenient parametrization of the Green function
$\gf({\bf x},{\bf x}^{\prime})$.  We do this by decomposing
$\gf({\bf x},{\bf x}^{\prime})$ into
a particular integral and a complementary function.
The particular integral, which yields the delta function
under the action of the \bdg\ operator, is built from
the fundamental Green functions
$\gf^{\rm N}({\bf x}-{\bf x}^{\prime})$ and
$\gf^{\rm S}({\bf x}-{\bf x}^{\prime})$.
The complementary function, which obeys the
\bdg\ wave equation [i.e.~the homogeneous version of
Eq.~(\ref{EQ:GBdG}], is specified in terms of as-yet
undetermined single and double layers
${\bbox{\mu}}^{\rm ii}({\bbox{\alpha}},{\bf x}^{\prime})$,
${\bbox{\nu}}^{\rm oi}({\bbox{\alpha}},{\bf x}^{\prime})$,
${\bbox{\mu}}^{\rm io}({\bbox{\alpha}},{\bf x}^{\prime})$, and
${\bbox{\nu}}^{\rm oo}({\bbox{\alpha}},{\bf x}^{\prime})$.
Thus we write
\begin{equation}
\gf({\bf x},{\bf x}^{\prime})
=\cases{
\gf^{\rm ii}({\bf x},{\bf x}^{\prime})
\equiv
\gf^{\rm N}({\bf x}-{\bf x}^{\prime})+
\int_{\surS}d\sigma_{\alpha}\,\partial_{\alpha}\,
\gf^{\rm N}({\bf x}-{\bbox{\alpha}})\,
        {\bbox{\mu}}^{\rm ii}({\bbox{\alpha}},{\bf x}^{\prime}),&
if ${\bf x}\in\volV$ and ${\bf x}^{\prime}\in\volV$;
\cr\noalign{\medskip}
\gf^{\rm oi}({\bf x},{\bf x}^{\prime})
\equiv
\phantom{\gf^{\rm N}({\bf x}-{\bf x}^{\prime})+}
\int_{\surS}d\sigma_{\alpha}\,
\gf^{\rm S}({\bf x}-{\bbox{\alpha}})\,
        {\bbox{\nu}}^{\rm oi}({\bbox{\alpha}},{\bf x}^{\prime}),&
if ${\bf x}\in\notV$ and ${\bf x}^{\prime}\in\volV$;
\cr\noalign{\medskip}
\gf^{\rm io}({\bf x},{\bf x}^{\prime})
\equiv
\phantom{\gf^{\rm N}({\bf x}-{\bf x}^{\prime})+}
\int_{\surS}d\sigma_{\alpha}\,\partial_{\alpha}\,
\gf^{\rm N}({\bf x}-{\bbox{\alpha}})\,
        {\bbox{\mu}}^{\rm io}({\bbox{\alpha}},{\bf x}^{\prime}),&
if ${\bf x}\in\volV$ and ${\bf x}^{\prime}\in\notV$;
\cr\noalign{\medskip}
\gf^{\rm oo}({\bf x},{\bf x}^{\prime})
\equiv
\gf^{\rm S}({\bf x}-{\bf x}^{\prime})+
\int_{\surS}d\sigma_{\alpha}\,
\gf^{\rm S}({\bf x}-{\bbox{\alpha}})\,
        {\bbox{\nu}}^{\rm oo}({\bbox{\alpha}},{\bf x}^{\prime}),&
if ${\bf x}\in\notV$ and ${\bf x}^{\prime}\in\notV$.
\cr}
\label{EQ:decorate}
\end{equation}
We are employing the notation $\partial_{\alpha}f(\bbox{\alpha})$
to indicate the value of the component of the gradient of $f({\bf x})$
with respect to ${\bf x}$, evaluated at the surface point $\bbox{\alpha}$,
directed along the inward normal direction at $\bbox{\alpha}$.  Similarly,
$d\sigma_{\alpha}$ indicates the (scalar) surface element at the point
$\bbox{\alpha}$.  We shall find it useful to decorate
$\gf({\bf x},{\bf x}^{\prime})$
with labels (such as, e.g., ${\rm ii}$ and ${\rm io}$)
according to the region of space in which the variables
${\bf x}$ and ${\bf x}^{\prime}$ lie.  For example,
if ${\bf x}\in\volV$ and ${\bf x}^{\prime}\in\notV$
then we write
$\gf^{\rm io}({\bf x},{\bf x}^{\prime})$ for
$\gf({\bf x},{\bf x}^{\prime})$,
conveying the idea that
${\bf x}$          lies inside  $\volV$ whereas
${\bf x}^{\prime}$ lies outside $\volV$.

Next, we focus on the surface $\surS$ across which the pair-potential
is discontinuous.  From the Green function equation~(\ref{EQ:GBdG}) it
is evident that both
$\gf({\bf x},{\bf x}^{\prime})$ and its normal derivative
${\bf n}\cdot{\nabla}_{x}\gf({\bf x},{\bf x}^{\prime})$ are continuous
as ${\bf x}$ varies across $\surS$.  What this means is that
\begin{mathletters}
\begin{eqnarray}
\lim\limits_{{\bf x}\in\volV\rightarrow{\bbox{\beta}}\in\surS}
\gf({\bf x},{\bf x}^{\prime})
&=&
\lim\limits_{{\bf x}\in\notV\rightarrow{\bbox{\beta}}\in\surS}
\gf({\bf x},{\bf x}^{\prime});
\\
\lim\limits_{{\bf x}\in\volV\rightarrow{\bbox{\beta}}\in\surS}
{\bf n}_{\beta}\,\nabla_{x}
\gf({\bf x},{\bf x}^{\prime})
&=&
\lim\limits_{{\bf x}\in\notV\rightarrow{\bbox{\beta}}\in\surS}
{\bf n}_{\beta}\,\nabla_{x}
\gf({\bf x},{\bf x}^{\prime}).
\end{eqnarray}
\end{mathletters}
We apply this pair of matching conditions across $\surS$
to the parametrizations~(\ref{EQ:decorate}),
once for ${\bf x}^{\prime}\in\volV$ and
once for ${\bf x}^{\prime}\in\notV$.
To evaluate the necessary limits we make use of the using the continuity
conditions~(\ref{EQ:CCscWF}) and (\ref{EQ:JCnsGR}), as well as the jump conditions~(\ref{EQ:JCnsWF}) and (\ref{EQ:JCscGR}),
thus arriving at the following conditions on
the single layers ${\bbox{\nu}}^{\rm oi}$ and ${\bbox{\nu}}^{\rm oo}$ and
double layers ${\bbox{\mu}}^{\rm ii}$ and ${\bbox{\mu}}^{\rm io}$:
\begin{mathletters}
\begin{eqnarray}
&&
\frac{1}{2}\psm{3}\,
{\bbox{\mu}}^{\rm ii}({\bbox{\beta}},{\bf x}^{\prime})=
\phantom{\partial_{\gamma}}
-\gf^{\rm N}({\bbox{\beta}}-{\bf x}^{\prime})
-\int_{\surS}d\sigma_{\alpha}\,
 \phantom{\partial_{\gamma}\,}
 \partial_{\alpha}\,
 \gf^{\rm N}({\bbox{\beta}}-{\bbox{\alpha}})
 \,{\bbox{\mu}}^{\rm ii}({\bbox{\alpha}},{\bf x}^{\prime})
+\int_{\surS}d\sigma_{\alpha}\,
 \phantom{\partial_{\gamma}\,}
 \phantom{\partial_{\gamma}\,}
 \gf^{\rm S}({\bbox{\beta}}-{\bbox{\alpha}})
 \,{\bbox{\nu}}^{\rm oi}({\bbox{\alpha}},{\bf x}^{\prime});
\label{EQ:MaFuIn}
\\
&&
\frac{1}{2}\psm{3}\,
{\bbox{\nu}}^{\rm oi}({\bbox{\gamma}},{\bf x}^{\prime})=
\phantom{-}
\partial_{\gamma}\,
 \gf^{{\rm N}}({\bbox{\gamma}}-{\bf x}^{\prime})
+\int_{\surS}d\sigma_{\alpha}\,
 \partial_{\gamma}^{\ourlim}\,
 \partial_{\alpha}\,
 \gf^{{\rm N}}({\bbox{\gamma}}-{\bbox{\alpha}})
 \,{\bbox{\mu}}^{\rm ii}({\bbox{\alpha}},{\bf x}^{\prime})
-\int_{\surS}d\sigma_{\alpha}\,
 \phantom{\partial_{\gamma}\,}
 \partial_{\gamma}\,
 \gf^{{\rm S}}({\bbox{\gamma}}-{\bbox{\alpha}})
 \,{\bbox{\nu}}^{\rm oi}({\bbox{\alpha}},{\bf x}^{\prime});
\label{EQ:MaGrIn}
\\
&&
\frac{1}{2}\psm{3}\,
{\bbox{\mu}}^{\rm io}({\bbox{\beta}},{\bf x}^{\prime})=
\phantom{-\partial_{\gamma}\,}
 \gf^{\rm S}({\bbox{\gamma}}-{\bf x}^{\prime})
+\int_{\surS}d\sigma_{\alpha}\,
 \phantom{\partial_{\gamma}\,}
 \phantom{\partial_{\gamma}\,}
 \gf^{\rm S}({\bbox{\beta}}-{\bbox{\alpha}})
 \,{\bbox{\nu}}^{\rm oo}({\bbox{\alpha}},{\bf x}^{\prime})
-\int_{\surS}d\sigma_{\alpha}\,
 \phantom{\partial_{\gamma}\,}
 \partial_{\alpha}\,
 \gf^{\rm N}({\bbox{\beta}}-{\bbox{\alpha}})
 \,{\bbox{\mu}}^{\rm io}({\bbox{\alpha}},{\bf x}^{\prime});
\label{EQ:MaFuOu}
\\
&&
\frac{1}{2}\psm{3}\,
{\bbox{\nu}}^{\rm oo}({\bbox{\gamma}},{\bf x}^{\prime})=
-
 \partial_{\gamma}\,\gf^{\rm S}({\bbox{\gamma}}-{\bf x}^{\prime})
-\int_{\surS}d\sigma_{\alpha}\,
 \phantom{\partial_{\gamma}\,}
 \partial_{\gamma}\,
 \gf^{{\rm S}}({\bbox{\gamma}}-{\bbox{\alpha}})
 \,{\bbox{\nu}}^{\rm oo}({\bbox{\alpha}},{\bf x}^{\prime})
+\int_{\surS}d\sigma_{\alpha}\,
 \partial_{\gamma}^{\ourlim}\,
 \partial_{\alpha}\,
 \gf^{{\rm N}}({\bbox{\gamma}}-{\bbox{\alpha}})
 \,{\bbox{\mu}}^{\rm io}({\bbox{\alpha}},{\bf x}^{\prime}).
\label{EQ:MaGrOu}
\end{eqnarray}
\end{mathletters}
These four matching conditions constitute a system of coupled
integral equations for the single layers
${\bbox{\nu}}^{\rm oi}$ and ${\bbox{\nu}}^{\rm oo}$
and the double layers
${\bbox{\mu}}^{\rm ii}$ and ${\bbox{\mu}}^{\rm io}$.
It is convenient to collect together the single and double layers
into a $4\times4$ matrix ${\Bbb M}({\bbox{\gamma}},{\bf x}^{\prime})$,
defined via
\begin{equation}
{\Bbb M}({\bbox{\gamma}},{\bf x}^{\prime})
\equiv
\pmatrix{
{\bbox{\mu}}^{\rm ii}({\bbox{\gamma}},{\bf x}^{\prime})&
{\bbox{\mu}}^{\rm io}({\bbox{\gamma}},{\bf x}^{\prime})\cr
\noalign{\medskip}
{\bbox{\nu}}^{\rm oi}({\bbox{\gamma}},{\bf x}^{\prime})&
{\bbox{\nu}}^{\rm oo}({\bbox{\gamma}},{\bf x}^{\prime})\cr}.
\end{equation}
In terms of ${\Bbb M}({\bbox{\gamma}},{\bf x}^{\prime})$
the system becomes
\begin{mathletters}
\begin{equation}
  {\Bbb M}({\bbox{\gamma}},{\bf x}^{\prime})
=2{\Bbb M}^{0}({\bbox{\gamma}}-{\bf x}^{\prime})
+2\int_{\surS}d\sigma_{\alpha}\,
 {\Bbb G}({\bbox{\gamma}},{\bbox{\alpha}})\,
 {\Bbb M}({\bbox{\alpha}},{\bf x}^{\prime}),
\label{EQ:MSE}
\end{equation}
where the inhomogeneity ${\Bbb M}^{0}$
and the kernel ${\Bbb G}$ are defined via
\begin{eqnarray}
 {\Bbb M}^0({\bbox{\gamma}}-{\bf x}^{\prime})
&\equiv&
 \pmatrix{
-\psm{3}\,
 \gf^{\rm N}({\bbox{\gamma}}-{\bf x}^{\prime})
\hfill&
 \fms\psm{3}\,
 \gf^{\rm S}({\bbox{\gamma}}-{\bf x}^{\prime})
\hfill\cr
\noalign{\medskip}
\fms\psm{3}\,\partial_{\gamma}\,
 \gf^{\rm N}({\bbox{\gamma}}-{\bf x}^{\prime})
\hfill&
-\psm{3}\,
 \partial_{\gamma}\,
 \gf^{\rm S}({\bbox{\gamma}}-{\bf x}^{\prime})
\hfill};
\\
\noalign{\bigskip}
 {\Bbb G}({\bbox{\gamma}},{\bbox{\alpha}})
&\equiv&
 \pmatrix{
-\psm{3}\,
 \partial_{\alpha}\,
 \gf^{{\rm N}}({\bbox{\gamma}}-{\bbox{\alpha}})
\hfill&
\fms\psm{3}\,
 \gf^{{\rm S}}({\bbox{\gamma}}-{\bbox{\alpha}})
\hfill\cr
\noalign{\medskip}
\psm{3}\,
 \partial_{\gamma}\,
 \partial_{\alpha}\,
 \gf^{{\rm N}}({\bbox{\gamma}}-{\bbox{\alpha}})
\hfill&
-\psm{3}\,
 \partial_{\gamma}\,
 \gf^{{\rm S}}({\bbox{\gamma}}-{\bbox{\alpha}})
\hfill\cr}.
\end{eqnarray}
\end{mathletters}%
Equation~(\ref{EQ:MSE}) is one of the central elements of this Paper.
Its output (${\bbox{\mu}}^{\rm ii}$ etc.) can be fed  into Eq.~(\ref{EQ:decorate})
to obtain the ingredients ($\gf^{\rm ii}$, etc.) of $\gf$
In terms of our $4\times4$ notation, this connection becomes
\begin{equation}
\pmatrix{
\gf^{\rm ii}({\bf x,x}^{\prime})&
\gf^{\rm io}({\bf x,x}^{\prime})\cr
\noalign{\medskip}
\gf^{\rm oi}({\bf x,x}^{\prime})&
\gf^{\rm oo}({\bf x,x}^{\prime})\cr}
=\pmatrix{
\gf^{\rm N}({\bf x}-{\bf x}^{\prime})&
{\bf 0}\cr
\noalign{\medskip}
{\bf 0}&
\gf^{\rm S}({\bf x}-{\bf x}^{\prime})\cr}
+\int_{\surS}d\sigma_{\gamma}
\pmatrix{
\partial_{\gamma}\gf^{{\rm N}}({\bf x}-{\bbox{\gamma}})&
{\bf 0}\cr
\noalign{\medskip}
{\bf 0}&
\gf^{{\rm S}}({\bf x}-{\bbox{\gamma}})\cr}
\,
{\Bbb M}({\bbox{\gamma}},{\bf x}^{\prime}).
\label{EQ:GFExp}
\end{equation}
\begin{figure}[hbt]
\epsfxsize=5.0truecm
\centerline{\epsfbox{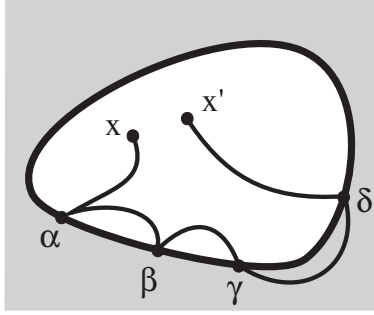}} \vskip+0.4truecm \caption{Typical
term in the MSE for the Green function. Lines running internally
(externally) to the billiard represent homogeneous-region normal
(superconducting) Green functions $\gf^{\rm N}$ ($\gf^{\rm S}$);
each point on the boundary ($\bbox{\alpha}$, $\bbox{\beta}$, etc.)
at which a scattering event occurs is to be integrated over the
complete boundary. \label{FIG:mse}}
\end{figure}%
When the iterative solution of Eq.~(\ref{EQ:MSE}) is fed into
Eq.~(\ref{EQ:GFExp}) for the Green function, the resulting
expansion is called a {\it Multiple Scattering Expansion\/} (MSE)
for the Green function.  A typical term in this expansion is shown
diagrammatically in Fig.~\ref{FIG:mse}.  The physical content of
the MSE is this: the iterations generate terms that correct the
free Green function by accounting for multiple scatterings from the
superconducting condensate surrounding the billiard.  However, the
expansion is not simply a perturbation expansion in powers of the
pair-potential; instead, terms involving $n$ transmissions (i.e.~terms
with $\gf^{\rm N}$ followed by $\gf^{\rm S}$ and {\it vice versa\/}) account
nonperturbatively for all Feynman trajectories that traverse the
boundary $n$ times, thus spending intervals in the superconducting
region.  (The reason such a re-organization of the simple perturbation
expansion in powers of the pair-potential is possible is that one knows
fully the fundamental Green functions that describe propagation in
homogeneous N or S regions.)

\subsection{Reorganized multiple scattering expansion}
\label{SEC:reorganize}
The multiple scattering expansion for the Green function was constructed
by iterating a four-by-four matrix integral equation, viz., Eq.~(\ref{EQ:MSE}).
In fact, this $4\times 4$ structure consists of two substructures:
(i)~the electron-hole structure, which is an essential ingredient in Andreev billiards.
(In fact it is an essential ingredient for any system involving superconductivity.)
(ii)~the inside-outside structure: this structure is specific to Andreev billiards,
as they consist of two distinct regions separated by a boundary.
As for the inside-outside structure, it is possible to diagonalize this (as we
shall soon show), and thus to obtain a $2\times 2$ matrix integral
equation, whose iteration produces exactly the same MSE
as the $4\times 4$ matrix integral equation does.  In this way, we reduce the
problem of determining the full-space Green function to an effective, but
nonetheless exact interior (or, if one wishes, exterior) problem.  (Indeed, in
the following section we shall obtain effective boundary conditions for the
interior Green function problem, the solution of which coincides with the that
for the full-space problem.)\thinspace\

The advantages of this formulation are two-fold:
First it allows us to easily obtain the (Feynman) rules for evaluating a generic
term in the MSE.
Second, as we shall see when we develop approximation schemes for the
Green function, it is especially well suited for the task of integrating
out the outside-propagation processes and, thus, obtaining an effective
Multiple {\it Reflection\/} Expansion.  In this way, the physics of
Andreev reflection as well as the corrections associated with charge-preserving
reflection will become evident.

To make this reorganization, let us introduce the operators
$\Dop$ (diagonal) and $\Oop$ (off-diagonal), defined as follows by their action on
4-component functions:
\begin{mathletters}
\begin{eqnarray}
\Dop\,
{\bbox{\Psi}}({\bbox{\gamma}})
&\equiv&
2\int_{\surS}d\sigma_{\beta}\,
\pmatrix{
-\psm{3}\,\partial_{\alpha}
\gf^{\rm N}({\bbox{\gamma}}-{\bbox{\beta}})
&{\bf 0}
\cr
\noalign{\medskip}
{\bf 0}
&-\psm{3}\,\partial_{\gamma}
\gf^{{\rm S}}({\bbox{\gamma}}-{\bbox{\beta}})
}
{\bbox{\Psi}}({\bbox{\beta}}),
\\
\noalign{\medskip}
\Oop\,{\bbox{\Psi}}({\bbox{\gamma}})
&\equiv&
2\int_{\surS}d\sigma_{\beta}\,
\pmatrix{
{\bf 0}&\psm{3}\,
\gf^{\rm S}({\bbox{\gamma}}-{\bbox{\beta}})
\cr
\noalign{\medskip}
\psm{3}\,\partial_{\gamma}
\partial_{\alpha}
\gf^{{\rm N}}({\bbox{\gamma}}-{\bbox{\beta}} )
&{\bf 0}
}
{\bbox{\Psi}}({\bbox{\beta}}).
\end{eqnarray}
\end{mathletters}%
These operators constitute the (two-by-two block) {\it diagonal\/}
and {\it off-diagonal\/} operator elements of the
kernel ${\Bbb G}$ in the integral equation~(\ref{EQ:MSE}), the iterative solution
of which generates the MSE.  Note that we write the four-by-four identity as
$\Iop$.  In terms of $\Dop$ and $\Oop$ Eq.~(\ref{EQ:MSE}) can be written
symbolically (i.e.~using an obvious condensed notation) as
\begin{equation}
{\Bbb M}=2{\Bbb M}^{0}+2\left(\Dop+\Oop\right){\Bbb M}\,.
\label{EQ:SymMSE}
\end{equation}
To obtain the reorganized MSE, first we rewrite Eq.~(\ref{EQ:SymMSE}) as
\begin{equation}
\left(\Iop-2\Dop\right){\Bbb M}
=2{\Bbb M}^{0}
+2\Oop{\Bbb M}\, ,
\end{equation}
and then, by inverting the operator $\left(\Iop-2\Dop\right)$, we obtain
\begin{equation}
{\Bbb M}=
 \left(\Iop-2\Dop\right)^{-1}2{\Bbb M}^{0}
+\left(\Iop-2\Dop\right)^{-1}2\Oop{\Bbb M}\, ,
\end{equation}
in which the kernel has become (block) off-diagonal.  Next, in order to
obtain a (block) diagonal structure, we iterate this equation once,
thus arriving at
\begin{equation}
{\Bbb M}=
 \left(\Iop-2\Dop\right)^{-1}2{\Bbb M}^{0}
+\left(\Iop-2\Dop\right)^{-1}2\Oop
 \left(\Iop-2\Dop\right)^{-1}2{\Bbb M}^{0}
+\left(\Iop-2\Dop\right)^{-1}2\Oop
 \left(\Iop-2\Dop\right)^{-1}2\Oop{\Bbb M}\,.
\label{EQ:OnceIt}
\end{equation}
At this stage it is useful to define the kernel
${\Bbb K}$ obeying
\begin{equation}
\left(\Iop-2\Dop\right){\Bbb K}=\Iop\,.
\end{equation}
This kernel is block diagonal,
\begin{equation}
{\Bbb K}=
\pmatrix{{\bf K}^{\rm ii}&{\bf 0}\cr
         {\bf 0}&{\bf K}^{\rm oo}\cr},
\end{equation}
and the diagonal (two-by-two) blocks obey
\begin{mathletters}
\begin{eqnarray}
 {\bf K}^{\rm ii}
+2\psm{3}\,\partial{\ourhat{\gf}}^{\rm N}\,{\bf K}^{\rm ii}
&=&{\bf I},
\\
 {\bf K}^{\rm oo}
+2\psm{3}\,\delta{\ourhat{\gf}}^{\rm S}\,{\bf K}^{\rm oo}
&=&{\bf I}.
\label{EQ:ImplDefK}
\end{eqnarray}
\end{mathletters}%
Here and elsewhere we shall use $\delta$ (resp.~$\partial$) without
a subscript to indicate an inward normal derivative with respect to
the first (resp.~second) argument of the succeeding Green function, i.e.,
\begin{mathletters}
\begin{eqnarray}
\delta\gf(\bbox{\alpha},\bbox{\beta})\equiv \partial_{\alpha}\gf(\bbox{\alpha},\bbox{\beta}),
\\
\partial\gf(\bbox{\alpha},\bbox{\beta})\equiv \partial_{\beta}\gf(\bbox{\alpha},\bbox{\beta}).
\end{eqnarray}%
\end{mathletters}%
In terms of ${\Bbb K}\,$, Eq.~(\ref{EQ:OnceIt}), when pre-multiplied by
$\left(\Iop-2\Dop\right)$, becomes
\begin{equation}
\left(\Iop-2\Dop\right){\Bbb M}=
 2{\Bbb M}^{0}
+4\Oop\,{\Bbb K}\,{\Bbb M}^{0}
+4\Oop\,{\Bbb K}\,\Oop\,{\Bbb M},
\end{equation}
the upper-left (two-by-two) block of which can be rearranged to read
\begin{equation}
{\bbox{\mu}}^{\rm ii}= -2\psm{3}\,\gf^{\rm N} +4\psm{3}\,\gf^{\rm
S}\,{\bf K}^{\rm oo}\,\psm{3}\,\delta\gf^{\rm N}
+\left\{
-2\psm{3}\,\partial\gf^{\rm N}
+4\psm{3}\,\gf^{\rm S}\,{\bf K}^{\rm oo}\,
  \psm{3}\,\partial\delta\gf^{\rm N}
\right\}{\bbox{\mu}}^{\rm ii}.
\label{EQ:NewIter}
\end{equation}
What we have accomplished via these transformations is the construction
of a closed ($2\times 2$) system of integral equations for the boundary
layer ${\bbox{\mu}}^{\rm ii}$, which is all that is needed to complete
the computation of the Green function $\gf^{\rm ii}$.  The virtue of this
transformation is that it facilitates the subsequent elimination (via the
integrating out of processes involving virtual propagation in the
superconducting region) of the states located in the superconducting
region.  This elimination can now be made straightforwardly, owing to the
fact that  all superconducting Green functions are now conveniently
located in the kernel in Eq.~(\ref{EQ:NewIter}).

The reorganization just described also allows one to identify the
following rules for the construction of all possible contributions at
any order $n$ ($=1,2,3,\ldots$) to the inside-to-inside Green function
$\gf^{\rm ii}$:
\begin{enumerate}
\item
Write down all possible permutations of $\gf^{\rm N}$ and $\gf^{\rm S}$
(having a total of $n+1$ Green functions), subject to the restriction
that the first and last Green functions are $\gf^{\rm N}$.
\item
Associate to each permutation a factor $(-1)^{i_{\rm N}+1}$, where
$i_{\rm N}$ is the number of $\gf^{\rm N}$ factors.
\item
Furnish all $\gf^{\rm N}$ factors (except the last) with normal
derivatives acting on their {\it second\/} arguments;
all $\gf^{\rm S}$ factors carry no normal derivatives.
\item
Furnish any Green function factor that follows a $\gf^{\rm S}$ factor
with an additional normal derivative acting on its {\it first\/}
argument.
\item
Insert a Pauli-matrix factor $\psm{3}$ before every $\gf^{\rm N}$
and $\gf^{\rm S}$
except the first.
\end{enumerate}
In this way one can construct $\gf^{\rm ii}$.
An example is provided by the process depicted in Fig.~\ref{FIG:mse},
for which the corresponding amplitude is
\begin{eqnarray}
\int_{\surS}
d\sigma_{\alpha}\,
d\sigma_{\beta}\,
d\sigma_{\gamma}\,
d\sigma_{\delta}\,
\partial\gf^{\rm N}({\bf x},{\bbox{\alpha}})\,\psm{3}\,
\partial\gf^{\rm N}({\bbox{\alpha}},{\bbox{\beta}})
\,\psm{3}\,
\partial\gf^{\rm N}({\bbox{\beta}},{\bbox{\gamma}})\,\psm{3}\,
\gf^{\rm S}({\bbox{\gamma}},{\bbox{\delta}})\,\psm{3}\,
\delta\gf^{\rm N}({\bbox{\delta}},{\bf x}^{\prime}).
\label{EQ:FRexample}
\end{eqnarray}

It is worth noting that the resulting series features terms containing
two or more consecutive factors of $\gf^{\rm N}$.  As $\gf^{\rm N}$ is
diagonal, such terms correspond to electron-to-electron and hole-to-hole
reflection processes.  At first sight, the presence of such terms might
be disconcerting, given the charge-interconverting character of Andreev
reflection.  However, it should be recalled that not only does the
superconducting surround interconvert electrons and holes, but also it
confines these quasiparticles to the normal region.  For example, consider
the series of terms that contain {\it no\/} superconducting Green functions
$\gf^{\rm S}$.  This series is precisely the Dirichlet series obtained in
BB-I.  Furthermore, this series can be embedded in any term of the MSE,
which amounts to replacing the fundamental (i.e.~unconfined) Green function
by a suitably confined Green function.  Therefore, terms involving
consecutive factors of $\gf^{\rm N}$, rather than being disconcerting,
are necessary to correct the free-propagation term, doing so by cancelling
the Feynman paths that venture into the superconductor.

\subsection{Effective boundary conditions}
\label{SEC:EffBC}
Before proceeding with our main issues (viz.~the computation of the \bdg\
Green function inside the billiard), we pause to pose and answer two
questions:
(i)~Is there any boundary condition
that can be imposed on $\gf^{\rm ii}$ so that $\gf^{\rm ii}$ can
be computed by solving the \bdg\ Green function equation solely in
$\volV$, i.e., without any reference to the region $\notV$.
And if so, (ii)~what is the precise form of this boundary condition?
(If such an approach turns out to possible then one could dispense with
the cumbersome task of dealing with Green functions having arguments
outside $\volV$, as well as the concomitant need to match Green functions
across the boundary.)

To see that such a boundary condition does indeed exist, and to determine
its explicit form, we substitute into Eq.~(\ref{EQ:NewIter})
the parametrization of $\gf^{\rm ii}$ given in Eq.~(\ref{EQ:decorate})
in terms of ${\bbox{\mu}}^{\rm ii}$.  Then all reference to
${\bbox{\mu}}^{\rm ii}$ cancels, and we arrive at a {\it nonlocal and
billiard-shape--dependent effective boundary condition\/} obeyed by
$\gf^{\rm ii}$, viz.,
\begin{equation}
\gf^{\rm ii}=-2\gf^{\rm S}\,{\bf K}^{\rm oo}\,\psm{3}\,\,\delta\gf^{\rm ii}.
\end{equation}
The reason that the boundary condition is nonlocal is that there exists the
possibility of virtual propagation within the superconductor surrounding the
billiard.  The reason that the boundary condition is shape-dependent is that
this virtual propagation outside the billiard is modified (from the value it
would have in a homogeneous superconductor) due to the presence of the normal
region.

\section{Asymptotics of the multiple scattering expansion}
\label{SEC:AsMuScEx}
Up to the present point, our investigation of the Green function for the
\bdg\ wave equation has been exact, and we have developed the
machinery---the MSE---for computing this Green function in terms of the
fundamental (N and S) Green functions and the shape of the billiard.  The
construction, however, is in terms of an infinite series, each term in
which involves repeated integration over the boundary of the billiard.
Thus, the direct computation of an arbitrary term in this series is
prohibitively difficult, unless the shape of the billiard is
exceptionally simple.  To make progress we therefore need to invoke some
approximation scheme and, as the Fermi wavelength is taken to be much
smaller than the characteristic linear dimension of the billiard $L$,
a very natural one to consider is the semiclassical approximation.
In the present setting, this involves the evaluating of the repeated
boundary integrals via a short-wave asymptotic approximation scheme.
What we mean by this is that we seek an asymptotic approximation for every
term in the MSE, the expansion parameter being $1/k_{\rm F}L$,
where $k_{\rm F}$ is the Fermi wave vector; having invoked such an
approximation, we shall re-sum the MSE.

By following this scheme we shall be
able to obtain, {\it inter alia\/}, the Green function for single-particle
excitations, as well as a {\it trace formula\/} for the oscillatory part of
the density of energy eigenvalues, valid in the semiclassical regime.
What we mean by a {\it trace formula\/} is an explicit formula for the
oscillatory part of the density of energy eigenvalues, expressed in terms
of a sum over all closed semiclassical particle orbits.  As we shall see,
by virtue of the retro-reflective character of Andreev reflection, this sum
over particle orbits is quite distinct from that arising in the setting of
conventional billiards.

Our approach has the virtue of delivering results not only for the DOS
at the {\it coarsest\/} of energy resolutions (i.e.~the Andreev level of
approximation, in which motion is confined to chords traversing the billiard)
but also at the {\it finer\/} level, thus revealing the mesoscale oscillations due to
the quantal particle motion transverse to each chord.
\subsection{Classical dynamics in Andreev billiards}
\label{SEC:ARCL}
The purpose of this subsection is to make a brief intermezzo in which discuss the physics of Andreev reflection
from the point of view of geometrical optics.  To this end, we develop stationary
phase arguments aimed at elucidating the origin of the retro-reflective character
of Andreev reflection.  Along the way, we shall see that owing to the difference
in the wavelengths of the incoming and reflected quasiparticles there is
imperfectness in the this retro-reflection, i.e., the reflected excitation does
not, in general, precisely retrace the path of the incoming one.  Arguments of this
type will be useful, subsequently, when we come to incorporate quantum fluctuations
around the classical trajectories associated with Andreev reflection.

\begin{figure}[hbt]
\epsfxsize=10.0cm \centerline{\epsfbox{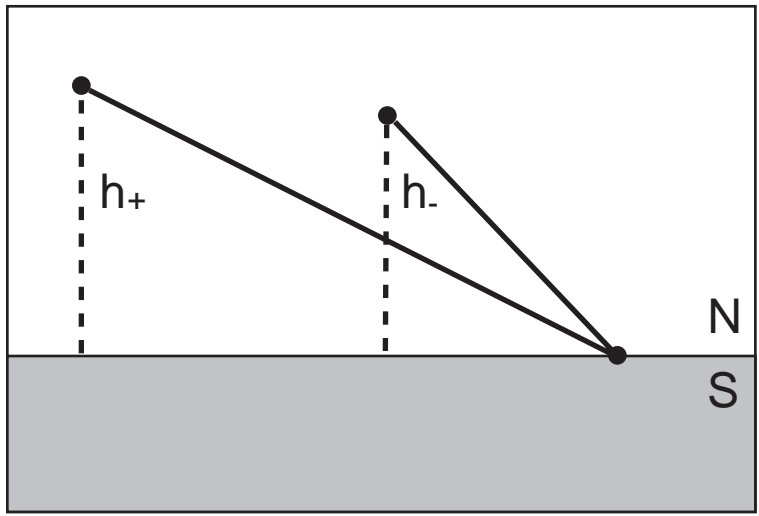}}
\vskip+0.4truecm \caption{Geometry for Andreev reflection from a
planar surface. Initial and final points (i.e.~the off-boundary
points) are kept fixed, and the point of reflection from the
boundary is then determined via the stationarity of the
corresponding phase.} \label{FIG:lloyd}
\end{figure}%
It is well known that Eq.~(\ref{EQ:TIBdG}) gives rise to the Andreev reflection
phenomenon, in which the electrons arriving from the normal metal are converted
into holes (and {\it vice versa\/}) at the superconductor boundary, and are retro-reflected
(i.e.~have the excitation velocity reversed).  In the present section we discuss
the classical limit of this reflection process by making use of the method of
stationary phase (i.e.~via the principle of least action).  Throughout this
section we make the (physical-optics--type) assumption that an electron wave
having energy $E$ and traveling a distance $r$ acquires a phase
\begin{equation}
{\romE}^{ik_{+}r},
\label{EQ:elph}
\end{equation}
whereas a hole traveling the same direction acquires the phase
\begin{equation}
{\romE}^{-ik_{-}r},
\label{EQ:hlph}
\end{equation}
where $k_{+}$ and $k_{-}$ are, respectively, the wavevectors appropriate for
for particle and hole motion in the normal region.  The energy dependence of
these wavevectors is given by
\begin{equation}
k_{\pm }=\sqrt{\mu\pm E}.
\end{equation}
Following the standard optics-type approach, we envision some process (for an
example see Fig.~\ref{FIG:lloyd}) and then, by using Eqs.~(\ref{EQ:elph})
and (\ref{EQ:hlph}), we calculate the total phase acquired during this process.
In the classical limit, the dominant process is the one (or ones) that make
stationary this total phase, and hence determines information such as
relationships between angles of incidence and reflection.  Thus, in effect we are
finding the rules of classical dynamics for Andreev billiards.

At this point we have to introduce the physics of Andreev reflection
\lq\lq by hand,\rlap \rq\rq\ and do so by requiring that after one reflection from the
billiard boundary an electron is converted into a hole (and {\it vice versa\/}).
Thus we are demanding that there is no electron-to-electron (or
hole-to-hole) scattering (due to a single reflection).  Under these
conditions, any scattering process can be analyzed in terms of
the basic electron-to-hole and hole-to-electron processes.
We need only examine one of these because the corresponding phases
are identical and thus, at stationarity the two processes have the
same geometry.  In other words, the stationary path describing an
incoming electron and scattered hole may be reversed (by reversing
the direction of propagation of each particle) to give the
stationary path of an incoming hole and the scattered electron.

\begin{figure}[hbt]
\epsfxsize=10.0cm
\centerline{\epsfbox{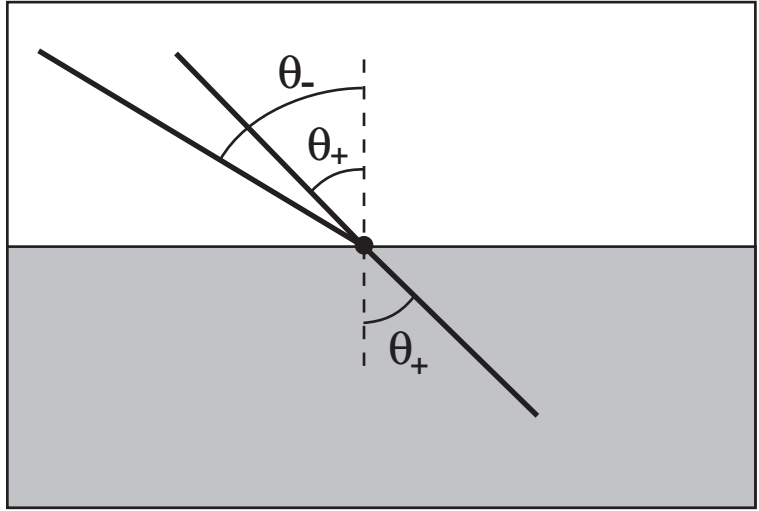}}
\vskip+0.4truecm
\caption{The analogy between Andreev reflection and optical refraction}
\label{FIG:refraction}
\end{figure}%
It might be useful to note the similarity between the phenomena of Andreev reflection and the
the refraction of light.  The common feature is that, in both
settings, before and after scattering the waves have the same frequency
but differing wavevector magnitudes.  In the case of the refraction of
light, the wavevector is changed because the wave enters a medium
with a distinct index of refraction.  In the case of Andreev reflection,
the wavevector is instead changed because the electron wave is converted
into a hole wave, the latter having a distinct dispersion relation.
In fact, by reversing the sign of the phase for a hole wave, as well as
the direction of propagation of the hole wave, one transforms the
Andreev reflection process into the familiar optical refraction process.
Thus, one has an electron-to-hole reflection law that is essentially
identical to  Snell refraction (rather than specular reflection).
Thus, Andreev reflection looks like optical refraction but with the
outgoing direction reversed with respect to the scattering point
(see Fig.~\ref{FIG:refraction}).

To quantify these remarks, consider the problem depicted in
Fig.~\ref{FIG:lloyd}, in which an electron arriving from a fixed point (1)
is reflected and converted at the variable point $x$ into a hole, which
then propagates to another fixed point (2).  The classical path corresponds
to the value of $x$ at which the total phase for the process is stationary
with respect to variations of $x$.  For the process at hand, total
phase is given by
\begin{equation}
k_{+}\sqrt{h_{+}^{2}+x^{2}}-k_{-}\sqrt{h_{-}^{2}+(x-l)^{2}},
\end{equation}
for which the stationarity condition reads
\begin{equation}
 k_{+}\frac{x}{\sqrt{h_{+}^{2}+x^{2}}}
-k_{-}\frac{x-l}{\sqrt{h_{-}^{2}+(x-l)^{2}}}
=0.
\end{equation}
By rewriting this condition in terms of the angles of incidence
and reflection (i.e.~$\theta_+$ and $\theta_-$ shown in Fig.~\ref{FIG:lloyd}) one recovers the
Snell's law form:
\begin{equation}
\label{EQ:snell1}
k_{+}\sin\theta_{+}=k_{-}\sin\theta_{-}\,.
\end{equation}
By using Eq.~(\ref{EQ:snell1}) one can construct the stationary paths for
an Andreev billiard, just as one does in the case of geometrical optics.
When there is more than one reflection, Eq.~(\ref{EQ:snell1}) must be
satisfied at each one.

One feature of Eq.~(\ref{EQ:snell1}) is that it makes evident the fact that
the reflected particle is not, in general, perfectly retro-reflected
(i.e.~$\theta_{+}\ne\theta_{-}$).  However, when $k_{+}$ and $k_{-}$ are
very close to each other, $\theta_{+}$ and $\theta_{-}$ will be, too.
Let us now calculate this small deflection angle $\theta_{-}-\theta_{+}$
in terms of $\mu$ and $E$.  To do this, let us assume that
$\theta_{+}$ and $\theta_{-}$ are close and that $E/\mu\ll 1$,
and expand Eq.~(\ref{EQ:snell1}) to obtain
\begin{equation}
\label{EQ:snell2}
\theta_{-}-\theta_{+}
\approx
(E/\mu)\,\tan\theta_{+}\,.
\end{equation}
As expected, the deflection angle is ${\cal O}(E/\mu)$, unless the
incident direction grazes the boundary.  This qualification divides
the space of incoming trajectories into two classes:
(i)~a large fraction, occupying most of the phase space, in which
$\tan\theta_{+}$ is of order unity, and
(ii)~the rest, in which the deflection angle $\theta_{-}-\theta_{+}$
is not small.  Equation~(\ref{EQ:snell2}), although approximate,
provides a guide for addressing whether or not deflections
(i.e.~imperfectness in retro-reflection) needs to be
taken into account.

\begin{figure}[hbt]
\epsfxsize=10.0cm
\centerline{\epsfbox{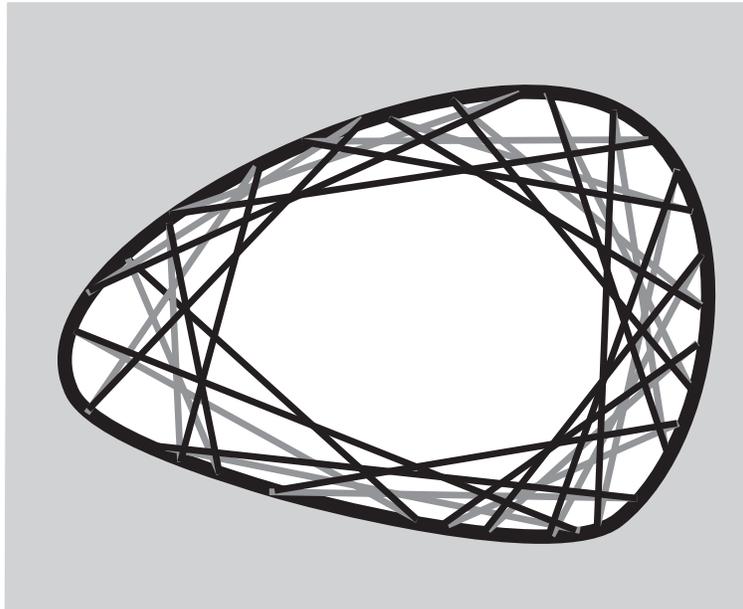}}
\vskip+0.4truecm
\caption{Typical closed classical trajectory in an Andreev billiard.
Black lines depict electron paths; gray lines depict hole paths.
The imperfectness of the retro-reflection is exaggerated.}
\label{FIG:imp_tra}
\end{figure}%
We are now at the point where we can construct classical dynamics in an
Andreev billiard.  For any trajectory, the reflection rule in Eq.~(\ref{EQ:snell1})
has to be satisfied at every reflection point.
This will generate a type of dynamics that differs from that generated
by the usual specular reflection rule (in which the outgoing angle equals the
incoming one).  An example of a closed trajectory in an Andreev billiard
is shown in Fig.~\ref{FIG:imp_tra}.

\subsection{Asymptotics of the fundamental Green functions
and their derivatives}
\label{SEC:GFasymptotics}
In the present section we investigate the asymptotic behavior of
the fundamental Green functions $\gf^{\rm N}$ and $\gf^{\rm S}$ for both
small and large values of their
(position) arguments.  This investigation will allow us to estimate the
relative dominance of various processes, and thus to organize the multiple
scattering expansion for the exact \bdg\ Green function into a form
suitable for establishing its approximate behavior at large $k_{F}L$.

The asymptotic behavior of $\gf^{\rm N,S}$ is related to the
corresponding Helmholtz Green function $g_{\pm}^{\rm N,S}$,
which can be represented through the Fourier integral
\begin{equation}
g_{\pm}^{\rm N,S}({\bf l})=
\int\frac{d^d p}{(2\pi)^d}\,
\frac{{\rm e}^{i{\bf p}\cdot{\bf l}}}{p^2-k^2}\,\,,
\label{EQ:fourier_gf}
\end{equation}
where $k=k_{\pm}^{\rm N,S}$, depending on the Green function in question.
Then the asymptotic behavior of $g_{\pm}^{\rm N,S}$ for large $k l$ can
be obtained from the asymptotic evaluation of this Fourier integral, which
gives
\begin{equation}
\label{EQ:lr_asym_gf}
g_{\pm}^{\rm N,S}(l)
\approx
\pm i\left(\frac{k_{\pm}^{\rm N,S}}{2\pi l}\right)^{\frac{d-1}{2}}
\frac{\exp({\pm ik_{\pm}^{\rm N,S}l\mp i\pi(d-1)/4})}{2k_{\pm}^{\rm N,S}}.
\end{equation}
The derivatives of the Green functions for large $kl$ can be obtained by differentiating this asymptotic expression.

Determining the small $kl$ asymptotics of $g_{\pm}^{\rm N,S}(l)$
is more tricky.  By scaling $p$ with $l$ we get
\begin{equation}
g(l)=
l^{(2-d)}\int\frac{d^d a}{(2\pi)^d}\,
\frac{{\rm e}^{i\bbox{a}\cdot {\bf \hat{l}}}}{a^2-(kl)^2},
\end{equation}
where $\bbox{a}\equiv{\bf p} l$, and $g$ is shorthand for any of
the four $g_{\pm}^{\rm N,S}$.  Thus for small $kl$ we have
\begin{equation}
g({\bf l})\sim
\cases{\dis
  l^{(2-d)},& if $d\neq 2$; \cr
  \ln(kl), & if $d=2$. \cr}
\label{EQ:asym_small}
\end{equation}

\begin{figure}[hbt]
\epsfxsize=10.0cm
\centerline{\epsfbox{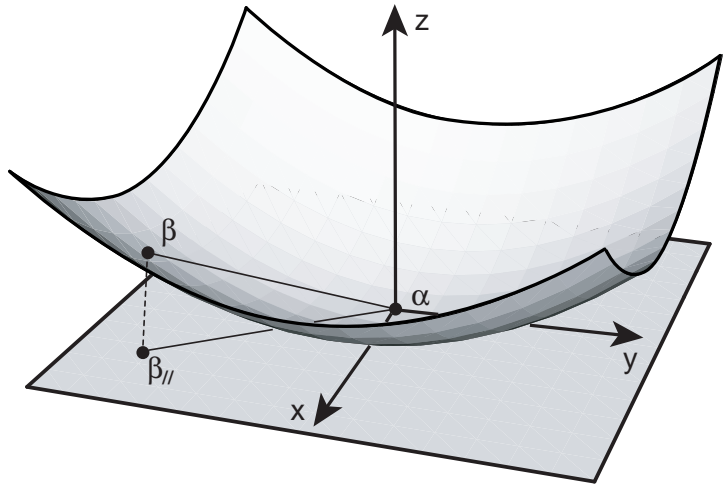}}
\vskip+0.4truecm
\caption{Approximating the surface by its tangent plane}
\label{FIG:app_sur}
\end{figure}%
Having determined the form of $g$ for small $kl$, we now investigate what
can be said about $g(l)$, $\partial g(l)$ and $\partial\delta g(l)$ when ${\bf l}$
is a vector connecting two nearby points on the surface $\surS$.  The
geometry (for the $d=3$ case ) is illustrated in Fig.~\ref{FIG:app_sur}.
In this figure, ${\bf l}=\bbox{\beta}-\bbox{\alpha}$, where $\bbox{\beta}$
and $\bbox{\alpha}$ are points on the surface.  In the following, we
shall work in $d=3$, although results obtained will be applicable for all
$d$.  Without loss of generality, we choose $\bbox{\alpha}$ to be the
origin of our coordinate system, with the $z$ direction coinciding with the
inward normal direction of the surface at $\bbox{\alpha}$.  The remaining
directions are chosen arbitrarily (at least for the time being), the only
constraint being that the coordinate-system be right handed.  In this
coordinate system the surface may be defined locally through an equation
of the form $z=f(x,y)$, where $x$ and $y$ span the tangent plane, and
$\partial_x f\big\vert_{(x,y)=(0,0)}=\partial_y f\big\vert_{(x,y)=(0,0)}=0$. Then a point $\bbox{\beta}$ (near
$\bbox{\alpha}$) on the surface will approximately have the coordinates
\begin{equation}
\bbox{\beta}=
\Big(x,y,
 \frac{\partial^2 f}{\partial x^2}\Big\vert_0 x^2
+\frac{1}{2}\frac{\partial^2 f}{\partial x\partial y}\Big\vert_0 xy
+\frac{\partial^2 f}{\partial y^2}\Big\vert_0 y^2
\Big).
\end{equation}
However, by a suitable rotation within the tangent plane the equation
for $\bbox{\beta}$ may be written as
\begin{equation}
\bbox{\beta}=
\Big(x,y,
\frac{x^2}{2R_1}+\frac{y^2}{2R_2}
\Big),
\end{equation}
where $R_1$ and $R_2$ are the two principal radii of curvature of the
surface at the point $\bbox{\alpha}$.  For reasons that should soon become
clear, we define the vector $\bbox{\beta}_{/\negthinspace/}$ to be
the projection of $\bbox{\beta}$ on to the tangent plane at $\bbox{\alpha}$.
Next, we make two assumptions:
\begin{enumerate}
\item
the radii of curvature are on the order of the linear size of the
billiard, i.e., $R_{1,2}/L={\cal O}(1)$.
\item
the region we of interest around $\bbox{\alpha}$ has a linear size
on the order of $k_{F}^{-1}$.
\end{enumerate}
Under these assumptions, we have that
\begin{equation}
l\equiv
\vert\bbox{\beta}-\bbox{\alpha}\vert=
\vert\bbox{\beta}_{/\negthinspace/}\vert
\left\{1+{\cal O}\left((k_{F}L)^{-2}\right)\right\}.
\end{equation}

Let us now turn to $\partial g$.
In the coordinate system specified above, the normal vector
${\bf n}_{\beta}$ at the surface point $\bbox{\beta}$ is given by
\begin{equation}
{\bf n}_{\beta}=\Big(\frac{x}{R_1},\frac{y}{R_2},1\Big)+{\cal O}\left((k_{F}L)^{-2}\right).
\end{equation}
Then the quantity $\partial l$ is given by
\begin{equation}
\partial l\equiv{\bf n}_{\beta}\cdot\bbox{\nabla}_{\beta}
\vert\bbox{\beta}-\bbox{\alpha}\vert
=\frac{1}{\vert\bbox{\beta}_{/\negthinspace/}\vert}
\left(\frac{x^2}{R_1}+\frac{y^2}{R_2}\right)+
{\cal O}\left((k_{F}L)^{-2}\right)=
{\cal O}\left(\frac{l}{L}  \right).
\end{equation}
Hence, we have that the normal derivative of the Green function is
given by
\begin{equation}
\partial g(l)\approx\frac{\partial g(l)}{\partial l}
\frac{1}{l}
\left(\frac{x^2}{R_1}+\frac{y^2}{R_2}\right).
\end{equation}
By generalizing to arbitrary dimensionality $d$ we obtain
\begin{equation}
\partial g(l)=
{\cal O}\left(L^{-1}l^{(2-d)}\right).
\end{equation}

In order to evaluate $\partial\delta g$, we consider a third point on
$\surS$, which we denote by $\bbox{\gamma}$, focus on the quantity
$\partial\delta l^\prime$
(where $l^\prime\equiv\vert\bbox{\beta}-\bbox{\gamma}\vert$)
and, at the end of our calculation, let
$\bbox{\gamma}$ tend to $\bbox{\alpha}$ .
We shall use primed coordinates $x'$ and $y'$ for $\bbox{\gamma}$.
In this way, we find that
\begin{eqnarray}
\partial\delta l^\prime
&\equiv&
({\bf n}_{\beta}\cdot\bbox{\nabla}_{\beta})
({\bf n}_{\gamma}\cdot\bbox{\nabla}_{\gamma})
\vert\bbox{\beta}-\bbox{\gamma}\vert
\approx\left(
\frac{x}{R_1}\frac{\partial}{\partial x}+
\frac{y}{R_2}\frac{\partial}{\partial y}+
\frac{\partial}{\partial z}\right)
\left(
\frac{x'}{R_1}\frac{\partial}{\partial x'}+
\frac{y'}{R_2}\frac{\partial}{\partial y'}+
\frac{\partial}{\partial z'}\right)
\vert\bbox{\beta}-\bbox{\gamma}\vert
\nonumber \\
&=&\left(
\frac{\partial^2}{\partial z\,\partial z'}
\sqrt{(x-x')^2+(y-y')^2+(z-z')^2}\right)_{
{z=\frac{x^2}{2R_1}+\frac{y^2}{2R_2}},\,
{z'=\frac{x'^2}{2R_1}+\frac{y'^2}{2R_2}}
}
\nonumber \\
&=&
-\frac{1}{l'^3}\left(\frac{x^2-x'^2}{2R_1}+
\frac{y^2-y'^2}{2R_2}\right)^2-\frac{1}{l'}
=-\frac{1}{l'}\left(1+
{\cal O}\left(\frac{l'^2}{L^2}\right) \right).
\end{eqnarray}
We are now in a position to evaluate $\partial\delta g$, for which we find
\begin{equation}
\partial\delta g(l)=
\frac{\partial g}{\partial l}\,\partial\delta l+
\frac{\partial^2 g}{\partial l^2}(\partial l)(\delta l)
=-\frac{1}{l}\frac{\partial g}{\partial l}
\left(1+{\cal O}\left(\frac{l'^2}{L^2}\right)\right).
\end{equation}

In the MSE, for each term that includes
the product $g\,\partial\delta g$ there is a corresponding term
in which $g\,\partial\delta g$ is replaced by $\partial g\,\partial g$.
It is therefore desirable to estimate relative size of these terms
for small values of their arguments.  The asymptotic formulas
given in the pressent section are useful for this comparison, giving
\begin{equation}
\vert\partial g(l)\,\partial g(l)\vert=
\vert g(l)\,\partial\delta g(l)  \vert\times
\cases{ {\cal O}\left(l^2/L^2\right), & for $d\ne2$;  \cr
        {\cal O}\left(l^2/L^2 \ln{kl}\right), & for $d=2$. \cr}
\end{equation}
\subsection{Asymptotic expansion for the quantal amplitude}
\label{SEC:asym_quan_amp}
The MSE of Sec.~\ref{SEC:MuScEx} has provided us with a series expansion for the Green
function (i.e.~the full quantum-mechanical amplitude for the
propagation of a quasiparticle excitation from one point ${\bf x}$
to another ${\bf x}^{\prime}$).  This series expresses the correction
to the free-space propagation of quasiparticles caused by multiple
scattering processes of the quasiparticles from the boundary that
separates the normal and superconducting regions of the billiard.
A generic term features the following possible processes:
(i)~inner reflections (which are marked by the occurrence of an
adjacent pair of normal-state Green functions in the algebraic
expression for the contribution),
(ii)~transmissions (marked by the occurrence of an adjacent pair
of Green functions, one normal and one superconducting),
(iii)~and outer reflections (marked by the occurrence of an
adjacent pair of superconducting-state Green functions).
Throughout the present section, we shall assume that $d=3$.
(The extension of the following discussion to general $d$ is
straightforward.)

The generic contribution to the amplitude involving a total of $n$
reflections and scatterings can be written as
\begin{eqnarray}
\genco({\bf x},{\bf x}^{\prime})
\equiv
\int\amp({\bf x}, {\bbox{\alpha}}_1, {\bbox{\alpha}}_2,
\ldots,  {\bbox{\alpha}}_n, {\bf x}^{\prime})\,
\exp{ik_{\rm F}
\phfunc({\bf x}, {\bbox{\alpha}}_1, {\bbox{\alpha}}_2,
  \ldots, {\bbox{\alpha}}_n,{\bf x}^{\prime})},
\label{EQ:amplitude}
\end{eqnarray}
where the integral is taken over all values of
$\big\{{\bbox{\alpha}}_1,{\bbox{\alpha}}_2,
\ldots,{\bbox{\alpha}}_{n-1},{\bbox{\alpha}}_n\big\}$,
each element ranging over the surface $\surS$.
The {\it modulus function\/}
$\amp({\bf x,}{\bbox{\alpha}}_1,{\bbox{\alpha}}_2,{\bbox{\alpha}}_3,\ldots,
{\bbox{\alpha}}_n,{\bf x}^{\prime})$ is real and, as can be seen from
the iterative solution of Eqs.~(\ref{EQ:decorate}-\ref{EQ:MaGrOu}),
is a sum of products of functions such as
$|{\bbox{\alpha}}_i-{\bbox{\alpha}}_{i+1}|^{-1}$,
first and second normal derivatives of this function,
as well as a polynomial in $k_{\rm F}$ and $k_{\pm}^{\rm N,S}$.
The {\it phase function\/}
$\phfunc({\bf x,}{\bbox{\alpha}}_1,{\bbox{\alpha}}_2,{\bbox{\alpha}}_3,
\ldots,{\bbox{\alpha}}_n,{\bf x}^{\prime})$
is, in general, complex and, as can also be seen from the iterative
solution of Eqs.~(\ref{EQ:decorate}-\ref{EQ:MaGrOu}), is a sum of
terms each of the form
$\big({k_{\pm}^{\rm N,S}}/{k_{\rm F}}\big)\,
|{\bbox{\alpha}}_i-{\bbox{\alpha}}_{i+1}|$.

The approximation scheme that we shall invoke involves the short-wave
asymptotic expansion of this quantal amplitude $\genco$, valid for large
$k_{\rm F}L$.  The method used for the construction of this expansion
is the asymptotic expansion of the multiple integrals appearing in the
terms in the MSE.  (See, e.g., Ref.~\cite{REF:Blei} for a discussion
of the asymptotic expansion of multiple Fourier integrals.)\thinspace\
This method allows one to approximate $\genco({\bf x},{\bf x}^{\prime})$
as a sum over critical points (c.p.) of the domain of integration
(which, in this case, is the $2n$-dimensional manifold
$\proman=\surS\times\surS\times\cdots\times\surS$):
\begin{equation}
\genco({\bf x},{\bf x}^{\prime})
\approx
\sum_{\rm c.p.}\genco_{\rm c.p.}({\bf x},{\bf x}^{\prime}),
\label{EQ:asymp_amp}
\end{equation}
where $\genco_{\rm c.p.}({\bf x},{\bf x}^{\prime})$ depends solely
on the local properties of $\amp$ and $\phfunc$ at the critical point.
As $k_{\rm F}L\rightarrow\infty$, corrections to this formula
vanish faster than any power of $(k_{\rm F}L)^{-1}$~\cite{FT:CPt_DEF}.
In other words, for large $k_{\rm F}L$ contributions from the critical
points {\it dominate\/} the total amplitude.  A critical point
$({\bbox{\alpha}}_1^{\rm c},{\bbox{\alpha}}_2^{\rm c},
{\bbox{\alpha}}_3^{\rm c},\cdots,{\bbox{\alpha}}_n^{\rm c})$ can be
interpreted as a trajectory (although not necessarily a classical one)
in which an excitation travels from point ${\bf x}$ to ${\bf x}^\prime$,
along the way scattering at the points
${\bbox{\alpha}}_1^{\rm c},{\bbox{\alpha}}_2^{\rm c},
\ldots,{\bbox{\alpha}}_n^{\rm c}$, so that
$\genco_{\rm c.p.}({\bf x},{\bf x}^{\prime})$ can be interpreted as the
amplitude corresponding to this trajectory.

Before actually proceeding with the construction of the expansion of the
multiple integrals appearing in the terms in the MSE, we first classify
the critical points of $\proman$: these are
\begin{enumerate}
\item
Points at which the gradient of the phase, i.e.,
$\nabla_{{\bbox{\alpha}}_i}
S(
{\bf x},
{\bbox{\alpha}}_1,
{\bbox{\alpha}}_2,
{\bbox{\alpha}}_3,\cdots,
{\bbox{\alpha}}_n,
{\bf x}^{\prime})$
vanishes for all $i$.
\item
Points at which
$\amp(
{\bf x},
{\bbox{\alpha}}_1,
{\bbox{\alpha}}_2,\cdots,
{\bbox{\alpha}}_{n-1},
{\bbox{\alpha}}_n,
{\bf x}^{\prime})$
has a singularity.
\item
Points at which $\amp$ or $\phfunc$ fail to be infinitely differentiable.
\item
All points on the boundary of the manifold $\proman$.
\item
Points satisfying criteria~(1-4) in a mixed sense, i.e.,
points satisfying criterion~(1) within a submanifold of
points satisfying criterion~(2) within a submanifold of
points satisfying criterion~(3) within a submanifold of
points satisfying criterion~(4).
\end{enumerate}
In the present setting, $\proman$ has no boundaries
and, thus, there are no Type~4 critical points.
As for Type~3 critical points, when $\surS$ is infinitely
differentiable, so are $\amp$ and $\phfunc$ and, hence, there are no
Type~3 critical points either~\cite{FT:surdisc}.
Thus, the only possible types of critical point are (1), (2) and (5).
In present case, $\amp$ consists of products of functions such as
${|{\bbox{\alpha}}_i-{\bbox{\alpha}}_{i+1}|^{-1}}$
and its first and second normal derivatives.
Thus, critical points of types~2 and~5 occur whenever one or more
of the propagation distances
$|{\bbox{\alpha}}_{i}-{\bbox{\alpha}}_{i+1}|$ vanish.
In accordance with the trajectory interpretation of critical points,
in which the sequence
$\{{\bbox{\alpha}}_1,{\bbox{\alpha}}_2,\cdots,{\bbox{\alpha}}_n\}$
defines the trajectory, we call the part of the critical trajectory
having vanishing propagation distance a {\it zero-length path\/}.
Then, all Type~5 critical points can be generated from Type~1 critical
points by the insertion of zero-length paths.
Stated technically, if
$({\bbox{\alpha}}_{1},
  {\bbox{\alpha}}_{2},  \cdots,
  {\bbox{\alpha}}_{i},
  {\bbox{\alpha}}_{i+1},\cdots,
  {\bbox{\alpha}}_{n-1})$
is a critical point of the [$(2n-2)$-dimensional] manifold
$\proman^{\prime}$ then the point
$({\bbox{\alpha}}_{1},
  {\bbox{\alpha}}_{2},\cdots
  {\bbox{\alpha}}_{i},
  {\bbox{\alpha}}_{i},
  {\bbox{\alpha}}_{i+1},\cdots,
  {\bbox{\alpha}}_{n-1})$ will be a critical point of Type~5 in the
[$2n$-dimensional] manifold $\proman$.  Criterion~(1) amounts to the
familiar stationary-phase approximation for the amplitude function,
because the points at which all gradients of $\phfunc$ vanish are the
stationary phase points.

Throughout this Paper we are interested in {\it leading-order\/}
contributions to quantal amplitudes.  Thus, it is useful to determine
whether or not a process contributes to the full amplitude at leading
order.  To do this, we must be able to estimate the order-of-magnitude
of contributions from different types of critical point.  To this end,
let us consider a type~1 critical point (i.e.~a trajectory defined via
the principle of stationary phase), and a Type~5 critical point
constructed from this Type~1 critical point via the insertion of
a zero-length path.  For the sake of simplicity, let us consider
as our Type~1 critical point a very simple amplitude, i.e., one having
just one reflection:
\begin{equation}
\genco({\bf x},{\bf x}^{\prime})=
\int\partial g({\bf x},\bbox{\alpha})\,
g(\bbox{\alpha},{\bf x}^{\prime}),
\end{equation}
where $g$ is a generic Helmholtz Green function.  [For the purposes of
determining the order-of-magnitude of the contribution from various
critical points, whether the Green function is $g^{\rm N}$ or $g^{\rm S}$
is irrelevant.]\thinspace\  By using the asymptotic formulas for $g$
presented in Sec.~\ref{SEC:GFasymptotics}, it is possible to write
asymptotically (up to from numerical factors)
\begin{equation}
\genco({\bf x},{\bf x}^{\prime})\sim
\int\frac{ik}
{\vert{\bf x}-\bbox{\alpha}\vert\,\vert\bbox{\alpha}-{\bf x}^{\prime}\vert}
\exp{\big(ik \vert{\bf x}-\bbox{\alpha}\vert+
ik^\prime\vert\bbox{\alpha}-{\bf x}^{\prime}\vert\big)}.
\end{equation}
Here, $k$ and $k^\prime$ can be $\pm k_{+,-}^{\rm N,S}$.
The position $(\bbox{\alpha}_c)$ of the critical point will depend on
the chosen values of $k$ and $k^\prime$ via the stationarity condition.
However, owing to the facts that
$k$ and $k^\prime$ are both ${\cal O}(k_{\rm F})$ and that
$\vert{\bf x}-\bbox{\alpha}_c\vert$ and
$\vert\bbox{\alpha}_c-{\bf x}^{\prime}\vert$ are both ${\cal O}(L)$,
it is possible to estimate the order-of-magnitude of the contribution
associated with $(\bbox{\alpha}_c)$ to be
\begin{equation}
\genco({\bf x},{\bf x}^{\prime})=
{\cal O}\left({k_{\rm F}}/{L^2}\right)
\exp{\big(ik \vert{\bf x}-\bbox{\alpha}_c\vert+
ik^\prime\vert\bbox{\alpha}_c-{\bf x}^{\prime}\vert\big)}
\int dx\,dy\,\exp \big(ik_{\rm F}L(A x^2+Bxy+Cy^2)\big),
\end{equation}
where we have expanded the phase to second order in deviations from
$(\bbox{\alpha}_c)$.  From dimensional considerations we know that
$A$,$B$ and $C$ are all ${\cal O}(L^{-2})$ and, thus, that
\begin{equation}
\genco({\bf x},{\bf x}^{\prime})=
{\cal O}\left({1}/{L}\right)
\exp{\big(ik \vert{\bf x}-\bbox{\alpha}_c\vert+
ik^\prime\vert\bbox{\alpha}_c-{\bf x}^{\prime}\vert\big)}.
\end{equation}
Now let us consider the insertion of a zero-length path.  The rules
described in Sec.~\ref{SEC:reorganize} for constructing a term in the
MSE allow three possible types of such insertions:
(i)~the insertion of $g$,
(ii)~the insertion of $\partial\delta g$, and
(iii)~the insertion of $\partial g$.
More specifically, we are interested in the amplitudes
\begin{mathletters}
\begin{eqnarray}
\genco_{\rm i}({\bf x},{\bf x}^{\prime})
\equiv\int\partial g({\bf x},\bbox{\alpha})\,
g(\bbox{\alpha},\bbox{\beta})\,
\delta g(\bbox{\beta},{\bbox x}^\prime),
\\
\genco_{\rm ii}({\bf x},{\bf x}^{\prime})
\equiv\int g({\bf x},\bbox{\alpha})\,
\partial\delta g(\bbox{\alpha},\bbox{\beta})\,
g(\bbox{\beta},{\bbox x}^\prime),
\\
\genco_{\rm iii}({\bf x},{\bf x}^{\prime})
\equiv\int\partial g({\bf x},\bbox{\alpha})\,
\partial g(\bbox{\alpha},\bbox{\beta})\,
g(\bbox{\beta},{\bbox x}^\prime),
\end{eqnarray}%
\end{mathletters}%
and their leading-order asymptotic contribution due to the critical point
$(\bbox{\alpha},\bbox{\beta})=(\bbox{\alpha}_c,\bbox{\alpha}_c)$.
Let us start with $\genco_{\rm i}$.  By using a coordinate system centered
at $\bbox{\alpha}_c$ and the short-distance asymptotics of $g$, and
expanding $\phfunc$ to second order around $\bbox{\alpha}_c$, we find that
\begin{eqnarray}
\genco_{\rm i}({\bf x},{\bf x}^{\prime})
&=&
{\cal O}\left(\frac{k_{\rm F}^2}{L^2}\right)
{\rm e}^{iS_c}
\int d\sigma_\alpha\,d\sigma_\beta\,
\frac{{\rm e}^{ik_{F}\big(D(x_\alpha-x_\beta)+E(y_\alpha-y_\beta)\big)}}
{\sqrt{(x_\alpha-x_\beta)^2+(y_\alpha-y_\beta)^2}}
{\rm e}^{ik_{\rm F}L(A x_\alpha^2+Bx_\alpha \, y_\alpha+Cy_\alpha^2+
A' x_\beta^2+B'x_\beta \, y_\beta+C'y_\beta^2)}
\nonumber \\
&=&{\cal O}\left(\frac{k_{\rm F}^2}{L^2}\right)
{\cal O}\left(\frac{1}{k_{\rm F}}\right){\rm e}^{iS_c}
\int dx\,dy\,\exp\big(ik_{\rm F}L(A'' x^2+B''xy+C''y^2)\big)
={\cal O}\left(\frac{1}{L}\right){\rm e}^{iS_c},
\end{eqnarray}
where $A$, $A'$, $A''$, $B$, $B'$, $B''$, $C$, $C'$ and $C''$ are
${\cal O}(1/L^2)$, $D$ and $E$ are ${\cal O}(1)$, and
$S_c=k \vert{\bf x}-\bbox{\alpha}_c\vert
    +k^\prime\vert\bbox{\alpha}_c-{\bf x}^{\prime}\vert$.
Thus, $\genco_{\rm i}$ contributes at the same order as $\genco$.
A similar calculation provides an estimate of the order-of-magnitude
of $\genco_{\rm ii}$, and show it to be of the same order as
$\genco$ and $\genco_{\rm i}$.  Next, let us consider $\genco_{\rm iii}$.
By using the short-distance behavior
\begin{equation}
\partial g(l)\approx {\cal O}\left(\frac{1}{L}\right)
\frac{\partial g(l)}{\partial l} \,l
\propto{\cal O}\left(\frac{1}{L}\right) g(l),
\end{equation}
the order of magnitude of $\genco_{\rm iii}$ can be deduced from the
estimate of $\genco_{\rm i}$:
\begin{equation}
\genco_{\rm iii}=
{\cal O}\left(\frac{1}{k_{\rm F}L^2}\right){\rm e}^{iS_c}.
\end{equation}
Thus, we see that $\genco_{\rm iii}$ contributes to the full amplitude
only at {\it subleading\/} order.
\hfill

In the usual case of a Schr\"odinger billiard with hard walls
(i.e.~Dirichlet boundary conditions), the asymptotic contribution coming
from the critical points of Type~2 and Type~5 have been shown to be
smaller than the stationary-phase (i.e.~Type~1 critical points)
contributions, and by a factor of $k_{\rm F}L$~\cite{REF:BaBoThree}.
The reason for this is that in the hard wall case {\it all\/} the
singularities of $\amp$ are due to $\partial g$'s.  Thus, for
the Schr\"odinger billiard with hard walls, as far as the leading-order
contribution is concerned the Type~2 and Type~5 critical points are
irrelevant. Thus, in such billiards, the
leading asymptotic contribution comes from the stationary-phase points.

In constrast with the case of Schr\"odinger billiards with hard
walls, in Andreev billiards (and Schr\"odinger billiards with soft
walls, i.e., with a finite bounding potential) the Type~2 and Type~5
critical points do {\it not\/} necessarily give only subleading-order
contributions.  More specifically, in Andreev billiards, in addition to
its $\partial g$ singularities, $\amp$ can have additional singularities
due to $g$ and $\partial\delta g$.  As shown above, both of these
singularities contribute at leading order in the $(k_{\rm F}L)^{-1}$
expansion~\cite{FT:dbledrvGF}.  Thus, Type~2 and Type~5 critical points
are relevant for Andreev billiards.

Having determined the significance of Type~5 critical points for
Andreev billiards, we now examine these critical points more
closely.  The first important observation is that the insertion of a
zero-length path does not change the value of the phase function $\phfunc$.
As all Type~5 critical points can be regarded as originating from
Type~1 critical points via the insertion of a suitable number of
zero-length paths, the phase of any Type~5 critical point is
equal to that of the originating Type~1 critical point.
The second important observation is that, because the phase is not
changed by the insertion of zero-length paths, all amplitudes
originating (via insertions) from a given Type~1 critical point carry
a common phase, and therefore add coherently.  Thus, the effect
of the Type~5 critical points is to modify (but not necessarily increase)
the amplitude of the contribution to the originating Type~1 critical point.

In order to make less abstract the issue discussed in the previous
paragraph, consider the example of reflection from an infinite plane
boundary in the short-wave asymptotic limit.  For the case of the
hard Schr\"odinger billiard there is a single critical point, which
is of the stationary-phase type: it is the classical reflection point
(for which the angle of incidence equals the angle of reflection).
For the case of the Andreev billiard there are two possible electron
reflections: electron-to-hole and electron-to-electron.  These two
processes have differing phases and, correspondingly, differing
stationary-phase (i.e.~classical reflection) points.  If one were to
take into account only the stationary-phase (i.e.~Type~1) points then one would find
that the amplitude for electron-to-hole reflection would vanish,
whereas that for electron-to-electron reflection would be of order unity.
However, this finding would be misleading, owing to the fact that the set
of critical points that contribute at leading order (in the the short-wave
asymptotic limit) is much larger.  To see this,
focus on the case of electron-to-electron reflection.
Let us label the classical reflection point by
${\bbox{\alpha}}_{\rm c}$.
Then set of critical points is
$({\bbox{\alpha}}_{\rm c})$,
$({\bbox{\alpha}}_{\rm c},{\bbox{\alpha}}_{\rm c})$,
$({\bbox{\alpha}}_{\rm c},{\bbox{\alpha}}_{\rm c},
  {\bbox{\alpha}}_{\rm c})$,
$({\bbox{\alpha}}_{\rm c},{\bbox{\alpha}}_{\rm c},
  {\bbox{\alpha}}_{\rm c},{\bbox{\alpha}}_{\rm c})$,
etc., i.e., there is the possibility of multiple scatterings from the
boundary, all taking place in the vicinity of the classical reflection
point.  These additional critical points correct the amplitude for this
scattering process, and yield the expected result, namely that the net
electron-to-electron amplitude is very small.  The origin of this
correction, then, is that multiple virtual propagations, inside the
superconductor but near to the classical reflection point, decrease the
amplitude of the electron-to-electron reflection process.
{\it Mutatis mutandis\/}, this mechanism increases the electron-to-hole
reflection amplitude, leading to the familiar physics of Andreev
reflection.

There is a simple physical explanation for this mechanism.
Quasiparticle propagation in the bulk of the normal region is not
affected by the superconducting surround, except via those Feynman
paths that pass nearby the boundary.  In the short-wave asymptotic
limit, this occurs near reflection points.  Thus, quantum mechanically,
there is an effective volume around the boundary in which propagation
is modified due to the amplitude for electron-to-hole conversion
(and {\it vice versa\/}).  Hence, there is an effective volume around the
classical reflection points, and in this volume multiple scatterings
convert electrons arriving from the interior of the normal region
into holes departing for the interior. Classically, the volume
for such processes is zero, i.e., the conversion takes place {\it precisely\/}
{\it at\/} the reflection point.  Thus, zero-length propagation at the
boundary is responsible for the electron-hole interconversion aspect of
Andreev reflection, whereas the requirement of phase-stationarity,
applied to propagation in the interior of the normal region, is
responsible for the retro-reflection aspect of Andreev reflection.
\subsection{Integrating out propagation in the
superconducting region: Effective reflection}
\label{SEC:MRE}
We now actually evaluate the contribution to the short-wave asymptotic
approximation to the Green function that arises from all critical points
involving zero-length propagation.  In doing this, we collect
contributions from Type~5 critical points, and arrive at the expected
result that reflection leads to {\it almost complete\/} electron-hole
interconversion.

We start with the expression~(\ref{EQ:NewIter}) for
${\bbox{\mu}}^{\rm ii}$ which, for the sake
of convenience, we rewrite here along with the explicit form
of ${\bf K}^{\rm oo}$ obtained from definition~(\ref{EQ:ImplDefK}):
\begin{eqnarray}
{\bbox{\mu}}^{\rm ii}&=&
-2\psm{3}\,\gf^{\rm N}
+4\psm{3}\,\gf^{\rm S}\,{\bf K}^{\rm oo}\,\psm{3}\,\delta\gf^{\rm N}
+\left\{
-2\psm{3}\,\partial\gf^{\rm N}
+4\psm{3}\,\gf^{\rm S}\,{\bf K}^{\rm oo}\,
  \psm{3}\,\partial\delta\gf^{\rm N}
\right\}{\bbox{\mu}}^{\rm ii},
\nonumber\\
{\bf K}^{\rm oo}&\equiv&
\left(
{\bf I}+2\psm{3}\cdot\delta\gf^{\rm S}
\right)^{-1}.
\nonumber
\end{eqnarray}
As we have shown in Sec.~\ref{SEC:asym_quan_amp}, in the short-wave
asymptotic limit, critical trajectories that have been obtained from a
Type~1 critical trajectory (i.e.~a pure stationary-phase trajectory)
by the insertion of zero-length propagations of
$\partial\delta\gf^{\rm N}$ and $\gf^{\rm S}$ contribute at the same
order as the original Type~1 critical trajectory.
Thus, such contributions should be summed to all orders.  On the
other hand, critical trajectories obtained from Type~1 critical
trajectories by the insertion of $\partial\gf^{\rm N}$ and
$\delta\gf^{\rm S}$ contribute only at sub-leading orders.  Thus, it
is appropriate to ignore such contributions.

Moreover, we are considering situations in which the range of
$\gf^{\rm S}$ is much smaller than the size of the billiard.  Thus,
all critical points that include finite-range superconducting
propagation are suppressed exponentially (in the size of the billiard),
despite their being formally of leading order.  We shall therefore
neglect them, at least for the time being.  Such contributions constitute
the single-particle tunneling amplitude through the classically-inaccessible S region.
Below, in App.~\ref{APP:nonconvex}, we shall study the consequences of
relaxing the condition that the range of $\gf^{\rm S}$ be much
smaller than the size of the billiard.  We shall then show that
in settings involving convex billiards (i.e.~billiards for which all chords lie
inside the billiard) such contributions cancel each other at
leading asymptotic order.  Thus, for the purposes of a
leading-order calculation we can make the approximation
$\delta\gf^{\rm S}\approx 0$.  It follows that
 \begin{equation}
{\bf K}^{\rm oo}\approx{\bf I}.
\end{equation}
The only remaining appearances of the superconducting Green function
(i.e.~$\gf^{\rm S}$ with {\it no\/} derivatives) in the MSE generate critical
points having zero-length superconducting propagation.  Moreover,
the only Green function that contributes at leading order to both
zero-length and nonzero-length propagation is $\partial\delta\gf^{\rm N}$.
In order to re-sum all possible zero-length propagations it is natural to
separate the operator $\partial\delta\gf^{\rm N}$ into two parts:
one solely generating zero-length propagation; the other solely
generating finite-length propagation~\cite{FT:fnte_lngth_prop}.
A convenient way to do this is to define the following operators:
\begin{mathletters}
\begin{eqnarray}
\left(\partial\delta
\gf^{\rm N}_{\rm z}\,{\bf F}\right)({\bbox{\alpha}})
&\equiv&
\int_{\surS}d\sigma_{\beta}\,
\partial_{\alpha}\,
\partial_{\beta}\,
\gf^{\rm N}({\bbox{\alpha}},{\bbox{\beta}})
\,{\bf F}({\bbox{\beta}})\,
w({\bbox{\alpha}}-{\bbox{\beta}}),
\\
\left(\partial\delta
\gf^{\rm N}_{\rm f}\,{\bf F}\right)({\bbox{\alpha}})
&\equiv&
\int_{\surS}d\sigma_{\beta}\,
\partial_{\alpha}\,
\partial_{\beta}\,
\gf^{\rm N}({\bbox{\alpha}},{\bbox{\beta}})
\,{\bf F}({\bbox{\beta}})\,
\big(1-w({\bbox{\alpha}}-{\bbox{\beta}})\big),
\end{eqnarray}
\end{mathletters}
where $w({\bbox{\alpha}}-{\bbox{\beta}})$ is a smooth function
that equals unity whenever ${\bbox{\alpha}}$ and ${\bbox{\beta}}$
are close to one another and vanishes whenever  ${\bbox{\alpha}}$
and ${\bbox{\beta}}$ are far away from one another.  The effect of
this function is to isolate the critical point at
${\bbox{\beta}}={\bbox{\alpha}}$ from the remaining critical
points, the latter having finite-length propagation involving
$\partial\delta\gf^{\rm N}$.
For the purposes of our asymptotic expansion, the isolating function
$w$ is only a convenience, and its particular form does
not affect the final results (as long as the range of $w$ precludes
its enveloping simultaneously any pairs of critical points).
Then the equation for ${\bbox{\mu}}^{\rm ii}$ becomes
 \begin{equation}
{\bbox{\mu}}^{\rm ii}\approx
-2\psm{3}\,\gf^{\rm N}
+4\psm{3}\,\gf^{\rm S}
\,\psm{3}\,\delta\gf^{\rm N}
+\left\{
 4\psm{3}\,\gf^{\rm S}\,\psm{3}\,
  \partial\delta\gf^{\rm N}_{\rm z}
+\left(
-2\psm{3}\,\partial\gf^{\rm N}
+4\psm{3}\,\gf^{\rm S}\,\psm{3}\,
  \partial\delta\gf^{\rm N}_{\rm f}
\right)
\right\}{\bbox{\mu}}^{\rm ii}.
\label{EQ:NeglIter}
\end{equation}
Having decomposed the kernel of this equation into two pieces (the first
consisting of critical points having zero-length propagation and the
second consisting of critical points having finite-length propagation)
we invert this equation with respect to the former piece, obtaining
 \begin{equation}
{\bbox{\mu}}^{\rm ii}\approx
\left({\bf I}-
  4\psm{3}\,\gf^{\rm S}\,\psm{3}\,
  \partial\delta\gf^{\rm N}_{\rm z}\right)^{-1}
    \left\{
-2\psm{3}\,\gf^{\rm N}
+4\psm{3}\,\gf^{\rm S}\,\psm{3}\,\delta\gf^{\rm N}
+   \left(
-2\psm{3}\,\partial\gf^{\rm N}
+4\psm{3}\,\gf^{\rm S}\,\psm{3}\,
  \partial\delta\gf^{\rm N}_{\rm f}
    \right){\bbox{\mu}}^{\rm ii}
    \right\}.
\label{EQ:InvertIter}
\end{equation}
We now define the renormalized Green function
$\gf^{\rm R}({\bbox{\alpha}},{\bf x}^{\prime})$:
 \begin{equation}
\gf^{\rm R}({\bbox{\alpha}},{\bf x}^{\prime})
\equiv
\left({\bf I}-
  4\psm{3}\,\gf^{\rm S}\,\psm{3}\,
  \partial\delta\gf^{\rm N}_{\rm z}\right)^{-1}
    \left\{
-\psm{3}\,\gf^{\rm N}
+2\psm{3}\,\gf^{\rm S}\,\psm{3}\,\delta\gf^{\rm N}
    \right\}.
\label{EQ:GeffDef}
\end{equation}
In terms of $\gf^{\rm R}$, Eq.~(\ref{EQ:InvertIter}) becomes
 \begin{equation}
{\bbox{\mu}}^{\rm ii}
\approx
 2\gf^{\rm R}
+2\partial\gf^{\rm R}\,
    {\bbox{\mu}}^{\rm ii}.
\label{EQ:EffIter}
\end{equation}
We note that this equation, no critical points containing
zero-length propagation contribute to $\bbox{\mu}^{\rm ii}$ at leading order.  Thus,
the summation of short-range critical orbits is achieved via the
calculation of $\gf^{\rm R}$ in the short-wave asymptotic
approximation.

The main contributions to the integrals implied in Eq.~(\ref{EQ:GeffDef})
come from the neighborhood of ${\bbox{\alpha}}$.  This follows
from:
(i)~the fact that isolating function is short-ranged;
(i)~the fact that $\gf^{\rm S}$ is finite-ranged; and
(iii)~the assumption that the billiard is large enough to exponentially
suppress any finite-range critical points produced by $\gf^{\rm S}$.
Therefore, it is adequate to approximate the boundary surface $\surS$
around ${\bbox{\alpha}}$.  The lowest-order approximation to $\surS$ is
the tangent plane at ${\bbox{\alpha}}$.  The corrections to this
approximation are smaller by a factor of $(k_{\rm F}R)^{-1}$,
where $R$ is the radius of curvature at the point ${\bbox{\alpha}}$.
Throughout this Paper we are assuming that the surface $\surS$
is sufficiently smooth that $R$ is of order $L$, i.e., the radius of
curvature is comparable to the billiard size.  Thus, corrections due
to the curvature of the surface do not contribute at leading order.
Having replaced $\surS$ by a tangent plane, the integral equation for
$\gf^{\rm R}({\bbox{\alpha}},{\bf x}^{\prime})$ becomes solvable, owing
to the resulting translational invariance in all directions parallel to
the tangent plane.  Thus, by introducing the two-dimensional Fourier
transform (2DFT; see App.~\ref{APP:2DFT}) of all Green functions
appearing in the integral equation~(\ref{EQ:GeffDef}), it is straightforward to obtain
the following algebraic result for the 2DFT of the renormalized Green
function:
\begin{equation}
\gf^{\rm R}(p,z^{\prime})
=   \left\{
{\bf I}-4\psm{3}\,\gf^{\rm S}(p)\,\psm{3}\,
\partial\delta\gf^{\rm N}(p)
    \right\}^{-1}
    \left\{
- \psm{3}\,
  \gf^{\rm N}(p,z^{\prime})
+2\psm{3}\,
\gf^{\rm S}(p)\,\psm{3}\,
\delta\gf^{\rm N}(p,z^{\prime})
    \right\},
\label{EQ:FTGC}
\end{equation}
to which there are corrections of order $\big(k_{\rm F}R\big)^{-1}$.
(This Grenn function is exactly the Green function for a planar boundary.)\thinspace\
Here and elsewhere, $p$ denotes the magnitude of the 2D vector ${\bf p}$
conjugate to the position-vector in the tangent plane.

We now embark on the task of inverting the 2DFT
$\gf^{\rm R}(p,z^{\prime})$ in order to obtain the approximate
real-space renormalized Green function
$\gf^{\rm R}({\bbox{\alpha}},{\bf x})$.
Thus, we need to evaluate the integral
\begin{equation}
\gf^{\rm R}({\bbox{\alpha}},{\bf x}^{\prime})
=\int\frac{d^2p}{(2\pi)^2}\,
\gf^{\rm R}(p,z^{\prime})\,
\exp{i{\bf p}\cdot({\bbox{\alpha}}-{\bbox{\beta}})},
\end{equation}
where ${\bbox{\beta}}$ is the component of ${\bf x}$ parallel to the
plane and $z^{\prime}$ is the perpendicular component.  The
quantities
$\gf^{\rm N}(p,z^{\prime})$ and
$\delta \gf^{\rm N}(p,z^{\prime})$,
needed to construct $\gf^{\rm R}(p,z^{\prime})$,
are derived in App.~\ref{APP:2DFT}, where they are found to be
given by
\begin{mathletters}
\begin{eqnarray}
\gf^{\rm N}(p,z^{\prime})
&=&
\pmatrix{
\frac 1{2a_{+}(p)}
{\romE}^{-a_{+}(p)|z^{\prime}|} & 0 \cr
0 & -\frac 1{2a_{-}(p)}
{\romE}^{-a_{-}(p)|z^{\prime}|}
}
=
\pmatrix{
\frac 1{2a_{+}(p)} & 0 \cr
0 & -\frac 1{2a_{-}(p)}
}
\pmatrix{
{\romE}^{-a_{+}(p)|z^{\prime}|} & 0 \cr
0 &
{\romE}^{-a_{-}(p)|z^{\prime}|}
},
\label{EQ:FTGN}
\\
\delta \gf^{\rm N}(p,z^{\prime})
&=&
\pmatrix{
\frac{1}{2}{\romE}^{-a_{+}(p)|z^{\prime}|} & 0 \cr
0 & -\frac 12{\romE}^{-a_{-}(p)|z^{\prime}|}
}
=
\pmatrix{
\frac 12 & 0 \cr
0 & -\frac 12
}
\pmatrix{
{\romE}^{-a_{+}(p)|z^{\prime}|} & 0 \cr
0 & {\romE}^{-a_{-}(p)|z^{\prime}|}
},
\label{EQ:FTGP}
\end{eqnarray}
\end{mathletters}
in which $a_{\pm }(p)\equiv\sqrt{p^2-k_{\pm }^2}$, the square roots
being taken such that their real parts are always positive.
Note that the only $z^{\prime}$ dependence in
$\gf^{\rm R}(p,z^{\prime})$ comes from
$\gf^{\rm N}(p,z^{\prime})$ and
$\delta\gf^{\rm N}(p,z^{\prime})$,
and it is only in these terms that there is exponential dependence
on $p$.  By inserting Eqs.~(\ref{EQ:FTGN}) and (\ref{EQ:FTGP}) into
Eq.~(\ref{EQ:FTGC}) we find that
$\gf^{\rm R}({\bbox{\alpha}},{\bf x}^{\prime})$
is given by
\begin{mathletters}
\begin{equation}
\gf^{\rm R}({\bbox{\alpha}},{\bf x}^{\prime})
=\int\frac{d^2p}{(2\pi)^2}\,{\bf R}(p)
\pmatrix{
\frac{1}{2a_{+}(p)}{\romE}^{-a_{+}(p)|z^{\prime}|} & 0 \cr
0 & -\frac{1}{2a_{-}(p)}{\romE}^{-a_{-}(p)|z^{\prime}|}
}
\exp{i{\bf p}\cdot({\bbox{\alpha}}-{\bbox{\beta}})},
\label{EQ:GFc}
\end{equation}
Here, ${\bf R}(p)$ is a certain algebraic function of $p$,
and is defined by
\begin{eqnarray}
{\bf R}(p)
&\equiv&
\Big\{
{\bf I}
-4\psm{3}\,
\gf^{\rm S}(p)
\,\psm{3}\,\partial\delta
\gf^{\rm N}(p)
\Big\}^{-1}
\,
\left\{
\pmatrix{
1 & 0 \cr
0 & -1
}
+2\psm{3}\,
\gf^{\rm S}(p)
\,
\pmatrix{
a_{+}(p) & 0 \cr
0 & -a_{-}(p)
}
\right\},
\label{EQ:AfromG}
\end{eqnarray}
\end{mathletters}%
The analytic expression for ${\bf R}(p)$ is obtained using the 2DFTs of $\gf^{\rm S}(p)$
and $\partial\delta\gf^{\rm N}(p)$:
\begin{mathletters}
\begin{eqnarray}
\gf^{\rm S}(p)&=&\frac{1}{2}
\left\{
\frac{E}{i\sqrt{\Delta ^{2}-E^{2}}}{\bf I}
-\frac{\Delta}{i\sqrt{\Delta^{2}-E^{2}}}\psm{2}
\right\}
\left(
 \frac{1}{2a_{+}^{{\rm S}}(p)}
-\frac{1}{2a_{-}^{{\rm S}}(p)}
\right)
+\frac{1}{2}\,{\psm{3}}
\left(
 \frac{1}{2a_{+}^{{\rm S}}(p)}
+\frac{1}{2a_{-}^{{\rm S}}(p)}
\right),
\label{EQ:2dftGS}
\\
\partial\delta\gf^{\rm N}(p)&=&-\frac{1}{4}\,{\bf I}\,\big(a_+(p)-a_-(p)\big)
                                -\frac{1}{4}\,\psm{3}\,\big(a_+(p)+a_-(p)\big).
\label{EQ:2dftGN}
\end{eqnarray}%
\end{mathletters}%
We evaluate the integral in Eq.~(\ref{EQ:GFc}) via the method of
stationary phase, which becomes exact in the limit
$k_{\rm F}|\bbox{\alpha}-{\bf x}^{\prime}|\rightarrow\infty$.
From the form of $a_{\pm}(p)$ we see that for values of $p$ for
which $a_{\pm}(p)$ is essentially imaginary (i.e.~for $p<{\rm Re} e\,k_{\pm }$)
there exists the possibility of a stationary-phase point.
Note, however, that $a_{+}$ behaves differently from $a_{-}$, due to the
fact that the imaginary parts of $k_{\pm}^2$ have opposing signs.
For $p<{\rm Re} k_{-}$ (note that $k_{-}$ is always smaller than $k_{+}$)
we have
\begin{equation}
a_{\pm}(p)=\mp i\sqrt{k_{\pm}^2-p^2}.
\end{equation}
The stationary-phase point is defined by the condition
\begin{equation}
\frac{\partial}{\partial{\bf p}}
\left(
{\bf p}\cdot({\bbox{\alpha}}-{\bbox{\beta}})
\pm z^{\prime}\sqrt{k_{\pm}^2-|{\bf p}|^{2}}
\right)={\bf 0},
\end{equation}
from which we see that stationary-phase point
${\bf p}_{\rm c}$ satisfies
\begin{equation}
\frac{({\bbox{\alpha}}-{\bbox{\beta}} )}{z^{\prime}}=
\pm\frac{{\bf p}_{\rm c}}{\sqrt{k_{\pm }^2-|{\bf p}_{\rm c}|^2}},
\label{EQ:SPCGF1}
\end{equation}
and that $|{\bf p}_{\rm c}|^2$ has the value
$k_{\pm }^2\sin ^2\theta_{\alpha x^{\prime}}$,
where $\theta_{{\bbox{\alpha}}x^{\prime}}$ is defined to be the angle
between the normal vector at the surface point ${\bbox{\alpha}}$ and
the vector ${\bf x}^{\prime}-{\bbox{\alpha}}$.  Then the effective
Green function can be asymptotically approximated as
\begin{eqnarray}
\gf^{\rm R}({\bbox{\alpha}},{\bf x}^{\prime})
&\approx&
\pmatrix{R_{++}(k_{+}^2\sin ^2\theta_{\alpha x^{\prime}})&
R_{+-}(k_{-}^2\sin ^2\theta_{\alpha x^{\prime}}) \cr
R_{-+}(k_{+}^2\sin ^2\theta_{\alpha x^{\prime}})&
R_{--}(k_{-}^2\sin ^2\theta_{\alpha x^{\prime}})
}
\int\frac{d^2p}{(2\pi)^2}\,
\pmatrix{
\frac{{\romE}^{-a_{+}(p)|z^{\prime}|}}{2a_{+}(p)} & 0 \cr
0 & -\frac{{\romE}^{-a_{-}(p)|z^{\prime}|}}{2a_{-}(p)}
}
{\romE}^{i{\bf p}\cdot({\bbox{\alpha}}-{\bbox{\beta}})}
\nonumber \\
&=&
\pmatrix{R_{++}&
R_{+-} \cr
R_{-+}&
R_{--}
}
\pmatrix{g_{+}({\bbox{\alpha}}-{\bf x}^{\prime})&0
\cr 0&-g_{-}({\bbox{\alpha}}-{\bf x}^{\prime})}.
\label{EQ:GenGeff}
\end{eqnarray}
The second line is obtained by noting that the integral is, in fact, the 2DFT of $\gf^{\rm N}$
(see App.~\ref{APP:2DFT}).
The amplitudes
$R_{++}$, $R_{+-}$, $R_{-+}$ and $R_{--}$
can now be respectively interpreted as the electron-electron,
electron-hole, hole-electron and hole-hole reflection amplitudes,
and can be obtained from Eq.~(\ref{EQ:AfromG}).
These amplitudes
are, in general, nonvanishing.
However the charge-preserving amplitudes
(i.e.~$R_{++}$ and $R_{--}$) are smaller than
the charge interconverting amplitudes (i.e.~$R_{+-}$ and $R_{-+}$)
by a factor of $\Delta/\mu\cos^2\theta$.
In order to evaluate ${\bf R}$ to leading order in
$\Delta/\mu\cos^2\theta$, we first note that for
$\theta_{\alpha x^{\prime}}$ not near $ \pi/2$
(i.e.~not near grazing) one has the following approximations
for $a(p)$:
\begin{mathletters}
\begin{eqnarray}
a_{\pm}(k_{\rm F}\sin\theta_{\alpha x^{\prime}})
&=&
\mp ik_{\rm F}\cos\theta_{\alpha x^{\prime}}
+{\cal O}\left(E/\mu\right),
\\
a_{\pm}^{\rm S}(k_{\rm F}\sin\theta_{\alpha x^{\prime}})
&=&
\mp ik_{\rm F}\cos\theta_{\alpha x^{\prime}}
+{\cal O}\left({\Delta_0}/\mu\right).
\end{eqnarray}
\end{mathletters}
By applying these approximations to Eqs.~(\ref{EQ:2dftGS}-\ref{EQ:2dftGN})
we find
\begin{mathletters}
\begin{eqnarray}
\gf^{\rm S}(k_{\rm F}\sin\theta_{\alpha x^{\prime}})
&\simeq&
\left(
\frac E{2i\sqrt{\Delta _0^2-E^2}}\,{\bf I}
-\frac{\Delta _0}{2i\sqrt{\Delta_0^2-E^2}}\,
\psm{2}
\right)
\frac 1{-ik_{\rm F}\cos\theta_{{\alpha}x^{\prime}}},
\\
\partial\delta
\gf^{\rm N}(k_{\rm F}\sin\theta_{{\alpha}x^{\prime}})
&\simeq&
\frac 12{\bf I}\,
ik_{\rm F}\cos\theta_{{\alpha}x^{\prime}}\, .
\end{eqnarray}
\end{mathletters}
By using these expressions in Eq.~(\ref{EQ:AfromG}) we obtain
${\bf R}(k_{\rm F}\sin\theta_{{\alpha}x^{\prime}})
\approx{\rm e}^{-i\varphi}\psm{1}$
and, thus, from Eq.~(\ref{EQ:GenGeff})
we obtain
\begin{equation}
\gf^{\rm R}({\bbox{\alpha}},{\bf x}^{\prime})
\approx\exp\left({-i\varphi+i\frac{\pi}{2}}\right)
\pmatrix{
0 & -g^{\rm N}_{-}({\bbox{\alpha}},{\bf x}^{\prime})
\cr
\noalign{\medskip}
g^{\rm N}_{+}({\bbox{\alpha}},{\bf x}^{\prime})& 0
}.
\label{EQ:GenGeff_PCIM}
\end{equation}
The off-diagonal structure represents the total electron-hole
interconversion that occurs for large perpendicular momenta.
However, strictly speaking, electron-hole interconversion is not
perfect, i.e., the amplitudes $R_{++}$ and $R_{--}$ do not vanish.
Moreover, these charge-preserving amplitudes increase, as the
angle of incidence approaches $\pi/2$. Note
the difference between the regimes of validity for the approximate
expressions Eq.~(\ref{EQ:GenGeff}) and Eq.~(\ref{EQ:GenGeff_PCIM}): the former
becomes valid for $\Kf L\gg 1$ whereas the latter becomes valid
for $\Kf L\gg 1$ {\it and\/} $\sudel/\Ef\ll 1$. However, in either case
the MSE can be cast into the following effective Multiple Reflection Expansion:
\begin{eqnarray}
\gf^{\rm ii}({\bf x,x}^{\prime})
&=&
\gf^{\rm N}({\bf x}-{\bf x}^{\prime})
+2\int_{\surS}d\sigma_{\alpha}\,
\partial_{\alpha}
\gf^{\rm N}({\bf x}-{\bbox{\alpha}})\,
\gf^{\rm R}({\bbox{\alpha}},{\bf x}^{\prime})
+4\int_{\surS}d\sigma_{\alpha}\,d\sigma_{\beta}\,
\partial_{\alpha}
\gf^{\rm N}({\bf x}-{\bbox{\alpha}})\,
\partial_{\beta}
\gf^{\rm R}({\bbox{\alpha}},{\bbox{\beta}})\,
\gf^{\rm R}({\bbox{\beta}},{\bf x}^{\prime})
\nonumber\\
&&\qquad\qquad
+8\int_{\surS}d\sigma_{\alpha}\,d\sigma_{\beta}\,d\sigma_{\gamma}\,
\partial_{\alpha}
\gf^{{\rm N}}({\bf x}-{\bbox{\alpha}})\,
\partial_{\beta}
\gf^{\rm R}({\bbox{\alpha}},{\bbox{\beta }})\,
\partial_{\gamma}
\gf^{\rm R}({\bbox{\beta}},{\bbox{\gamma}})\,
\gf^{\rm R}({\bbox{\gamma}},{\bf x}^{\prime})+\cdots\,\,.
\label{EQ:appMRE}
\end{eqnarray}
We stress two points about Eq.~(\ref{EQ:appMRE}):
(i)~it is free of leading-order short-range critical points
[i.e.~the contributions of short-range critical points to
$\gf^{\rm ii}$ are smaller, by at least a factor of
${\cal O}(1/{k_{\rm F}R})$, than the leading-order
contribution, and this is what we were aiming for]; and
(ii)~all propagations inside the superconductor have been integrated
out, leading to the effective {\it reflection\/} expansion for
$\gf^{\rm ii}$.

\section{Density of states oscillations}
\label{SEC:DenOfStates}
In Sec.~\ref{SEC:MRE} we integrated out superconducting propagation by
evaluating the short-range critical points in the MSE and, hence, we
obtained an effective expansion for the Green function, which we have
termed an MRE.  In the present Section we shall focus on the density
of states $\rho(E)$, expressing this quantity in terms of the Green
function which, in turn, we express via the MRE.  In this section we
shall ignore the effects of normal reflection, returning to them only
in Sec.~\ref{SEC:IncorOR}. Thus, by using the results of the previous
section, we have
\begin{eqnarray}
&&
\rho(E)\approx
\frac{1}{\pi}\int d^{d}x\,
{\rm Im}\Bigg\{
g^{\rm N}_+({\bf x},{\bf x}^{\prime})+
g^{\rm N}_-({\bf x},{\bf x}^{\prime})
+4\int_{\surS}\,
\partial
g^{\rm N}_+({\bf x},{\bbox{\alpha}})\,
\partial
g^{\rm R}_-({\bbox{\alpha}},{\bbox{\beta}})\,
g^{\rm R}_+({\bbox{\beta}},{\bf x}')+
\partial g^{\rm N}_-({\bf x},{\bbox{\alpha}})\,
\partial g^{\rm R}_+({\bbox{\alpha}},{\bbox{\beta}})\,
g^{\rm R}_-({\bbox{\beta}},{\bf x}')
\nonumber \\
&&
\qquad\qquad
\cdots +2^{2n}\int_{\surS}\,\left(
\partial g^{\rm N}_+\,\partial g^{\rm R}_-\,\partial g^{\rm R}_+\,\partial g^{\rm R}_-
\cdots\partial g^{\rm R}_-\,g^{\rm R}_+\right)
+2^{2n}\int_{\surS}\,\left(
\partial g^{\rm N}_-\,\partial g^{\rm R}_+\,\partial g^{\rm R}_-\,\partial g^{\rm R}_+
\cdots\partial g^{\rm R}_+\,g^{\rm R}_-\right)
+\cdots
\Bigg\}_{\bf x'=x},
\label{EQ:DOS_MRE}
\end{eqnarray}
where terms with odd numbers of reflections vanish, owing to the
off-diagonal structure of $\gf^{\rm R}$.  First, let us note that
the first two terms on the right hand side of Eq.~(\ref{EQ:DOS_MRE}), which have
zero-length propagation and hence vanishing action, do not introduce
any oscillations into the DOS.  As these terms do not
involve any surface integrals (and, hence, do not involve any surface
effects), they produce the bulk DOS of
a homogeneous N region~\cite{FT:leadingWEYL}:
\begin{equation}
\frac{1}{\pi}\int_{\volV} d^{d}x\,
{\rm Im}\,g^{\rm N}_{\pm}(\bbox{x}-\bbox{x}')\Big\vert_{x'=x}
=\frac{{\cal S}^{d-1}}{2 (2\pi)^d}\, V\, k_{\pm}^{d-2},
\label{EQ:WEYL_DOS}
\end{equation}
where ${\cal S}^{d-1}$ is the $(d-1)$-dimensional surface area of a
$d$-dimensional unit sphere and $V$ is the volume of $\volV$.
We now deal with the (remaining) critical points that contribute at
leading order.  These critical points are closed classical trajectories
consisting of the propagation of quasiparticles through the bulk of the
billiard (i.e.~the N region), connected by reflections from the
billiard boundary (i.e.~the N-S interface).

We shall distinguish between two asymptotic approximation schemes for
$\rho$, both of which are obtained by evaluating the integrals in
Eq.~(\ref{EQ:DOS_MRE}) within the stationary-phase approximation,
valid for for large $k_{\rm F} L$ and small $\Delta/\Ef$.
From a technical point of view, the difference between the two schemes
concerns the stationary-phase points they use, which must be in
accordance with the particular limits assumed for the parameters
$\Kf L$ and $\sudel/\Ef$
(which the approximation becomes exact).

\noindent\textsc{Scheme~A\/}:
The first scheme is, in essence, equivalent to the (by now conventional)
adiabatic approximation to the wave function, first introduced by Andreev~\cite{REF:AFAndreev};
it becomes exact when energy-level spacing
goes to zero, which occurs for the following limit:
\begin{equation}
\Kf L\rightarrow\infty,
\qquad
\sudel/\Ef\rightarrow 0,
\quad
{\rm and}
\quad
L\sudel/\Kf\rightarrow {\rm constant}.
\end{equation}
In this scheme, an excitation undergoes perfect retro-reflection
(i.e.~perfect velocity-reversal) because in this limit the difference
between $k_{+}$ and $k_{-}$ is ignored in the calculation of the
critical trajectories.  The resulting classical dynamics is confined to
the chords of the billiard and, thus, is integrable, regardless of the
shape of the billiard.  However, for finite values of the parameters
(i.e.~large but finite $\Kf L$ and small but finite $\sudel/\Ef$),
Scheme~A produces the {\it locally-energy-averaged\/} DOS, which
becomes numerically accurate only around the DOS
singularities that it correctly captures.  However, it fails to capture
the DOS oscillations arising from the confinement of quasiparticles to the
billiard.  To capture these oscillations is the main motivation of the
following scheme.

\noindent\textsc{Scheme~B\/}:
In this scheme we shall take into account the previously neglected
difference in electron and hole wavevectors.  This leads to
imperfectness in retro-reflection because, upon reflection, the
wavelengths of the incoming and outgoing waves are no longer identical
(as happens with refraction except, of course, that the waves are now on
the same side of the boundary).  Technically speaking, approximation
scheme~B becomes exact when
\begin{equation}
\Kf L\rightarrow\infty,
\quad
{\rm and}
\quad
\sudel/\Ef
\quad
{\rm is\,\,a\,\,small\,\, parameter}.
\end{equation}
Now the classical trajectories are determined by the reflection
rule given by Eq.~(\ref{EQ:snell1}) and, consequently, the dynamics
is no longer {\it a priori\/} integrable;
on the contrary, it is weakly chaotic for most
billiard shapes~\cite{FT:magchaos}.

In order to understand the distinction between asymptotic
schemes~A and B, consider the phase function $S$ for a process
having $2n$ reflections:
\begin{equation}
S(\bbox{x},\bbox{\alpha}_1,\bbox{\alpha}_2,\cdots,\bbox{\alpha}_{2n})=
k_+\ell_{x,\alpha_1}-k_-\ell_{\alpha_1,\alpha_2}
+k_+\ell_{\alpha_2,\alpha_3}-k_-\ell_{\alpha_3,\alpha_4}
\cdots
-k_-\ell_{\alpha_{2n-1},\alpha_{2n}}+k_+\ell_{\alpha_{2n},x}\,\, ,
\end{equation}
where $\ell_{\alpha_i,\alpha_i+1}
\equiv|\bbox{\alpha}_{i}-\bbox{\alpha}_{i+1}|$.
Notice that $S$ can be separated into two parts,
a large one $S_{\rm And}$
and a small one $S_{\rm imp}$
so that $S=S_{\rm And}+S_{\rm imp}$,
where
\begin{mathletters}
\begin{eqnarray}
S_{\rm And}(\bbox{x},\bbox{\alpha}_1,\bbox{\alpha}_2,
\cdots,\bbox{\alpha}_{2n})
&\equiv&
\frac{k_++k_-}{2}\left(
 \ell_{x,\alpha_1}
-\ell_{\alpha_1,\alpha_2}
+\ell_{\alpha_2,\alpha_3}-
 \ell_{\alpha_3,\alpha_4}\cdots
-\ell_{\alpha_{2n-1},\alpha_{2n}}
+\ell_{\alpha_{2n},x}\right),
\\
S_{\rm imp}(\bbox{x},\bbox{\alpha}_1,\bbox{\alpha}_2,
\cdots,\bbox{\alpha}_{2n})
&\equiv&
\frac{k_+-k_-}{2}\left(\ell_{x,\alpha_1}+
\ell_{\alpha_1,\alpha_2} +\ell_{\alpha_2,\alpha_3}+
\ell_{\alpha_3,\alpha_4}\cdots+\ell_{\alpha_{2n-1},
\alpha_{2n}}+\ell_{\alpha_{2n},x}\right).
\end{eqnarray}
\end{mathletters}
For $n$ not too large (i.e.~$n\ll k_{F}L$),
$S_{\rm And}={\cal O}(k_F L)$ and
$S_{\rm imp}={\cal O}(L\Delta/k_{F})$.
Although it is clear that $S_{\rm And}$ is large and must be
included in any stationary phase calculation,
whether or not $S_{\rm imp}$ should be included in a stationary
phase calculation depends precisely on the nature of the limiting scheme.
If one solely uses $S_{\rm And}$ for the calculation of stationary-phase
points then the critical trajectories feature retro-reflection;
this is the content of Scheme~A.  On the other hand, with the inclusion
of $S_{\rm imp}$ in the calculation of stationary-phase points, the
critical trajectories feature small deviations from retro-reflection,
due to the difference in electron and hole wavevectors.
(Note that $S_{\rm imp}$ is proportional to $k_+-k_-$.)\thinspace\
This, in turn, is Scheme~B.  For larger $n$ (and thus higher
resolution contributions to the DOS),
$S_{\rm imp}$ becomes larger, so that Scheme~B should be
used~\cite{FT:mixedAB}.  These higher-resolution contributions
show up in the DOS as mesoscale oscillations due to the
superconducting confinement of quasiparticles.

\subsection{Scheme~A: Andreev approximation}
\label{SEC:SAandreev}
In the present section we shall evaluate Eq.~(\ref{EQ:DOS_MRE}) for
the DOS within the stationary-phase approximation via
asymptotic Scheme~A.  In this scheme, the stationary-phase points
(i.e.~the closed classical trajectories) are obtained by making
stationary the phase $S_{\rm And}$ alone.  The factor
$\exp iS_{\rm imp}$ is considered to be slowly-varying,
and thus is evaluated at the critical points determined  from
$S_{\rm And}$ alone.  In all other factors, the difference
between $k_+$ and $k_-$ can be neglected.  The reflection rule can
be obtained from Eq.~(\ref{EQ:snell1}) by letting
$k_+,k_-\rightarrow\Kf$.  In this limit, velocity vectors are, upon
reflection, exactly reversed.  The classical trajectories obtained
by this reflection rule are tracings of the chords of the billiard.
This allows us to label every classical trajectory by two boundary
points, along with the number of reflections (or tracings).
Therefore, the closed classical trajectories of Scheme~A with $n$
tracings are degenerate (in the sense that their phases
$S_{\rm And}$ are identical) and they belong to a $(2d-2)$-parameter
family.

Let us first explore the underlying physics of Scheme~A.
In conventional billiards, degeneracy of closed trajectories
is usually related to the
symmetry of the billiard (e.g.~rotational symmetry of a circular billiard).
However, in Andreev billiards this degeneracy is due to the underlying
electron-hole symmetry at the Fermi surface, and is broken explicitly at
nonzero energies by the term $S_{\rm imp}$~\cite{FT:ehsym}.  This
(approximate) non-geometric symmetry is the reason for the (approximate)
integrability of Andreev billiards, whatever the shape of the billiard.

The method for evaluating the DOS is as follows.
We fix two reflection points (so that the degeneracy is lifted),
evaluate the rest of the (surface and volume) integrals in the
stationary-phase approximation, and then evaluate the remaining
two surface integrals.  Without loss of generality, we choose the first and
last reflection points as those to be fixed, and label them $\bbox{\alpha}$ and $\bbox{\beta}$,
respectively.  We focus on the contribution to the DOS
from the term having $2n+2$ reflections:
\begin{equation}
\rho_{2n}^{\pm}(E)
\equiv
{\rm Im}\,\frac{2^{2n+2}}{\pi}\,
{\rm e}^{2ni\varphi}
\int \negthinspace\negthinspace
d\sigma_{\alpha}\,
d\sigma_{\beta}\,
d\sigma_{\alpha_1}\cdots
d\sigma_{\alpha_{2n}}\,
\partial g_{\mp}(\bbox{\alpha},\bbox{\alpha}_1)\,
\partial g_{\pm}(\bbox{\alpha}_1,\bbox{\alpha}_2)\cdots
\partial g_{\mp}(\bbox{\alpha}_{2n},\bbox{\beta})\,
\int_\volV \negthinspace\negthinspace d^{d}x\,
g_{\pm}(\bbox{\beta},{\bf x})\,
\partial g_{\pm}({\bf x},\bbox{\alpha}),
\label{EQ:RhoTwoN}
\end{equation}
where the $+/-$ sign represents the contribution from the
electron/hole sector.

\begin{figure}[hbt]
 \epsfxsize=10.0cm
\centerline{\epsfbox{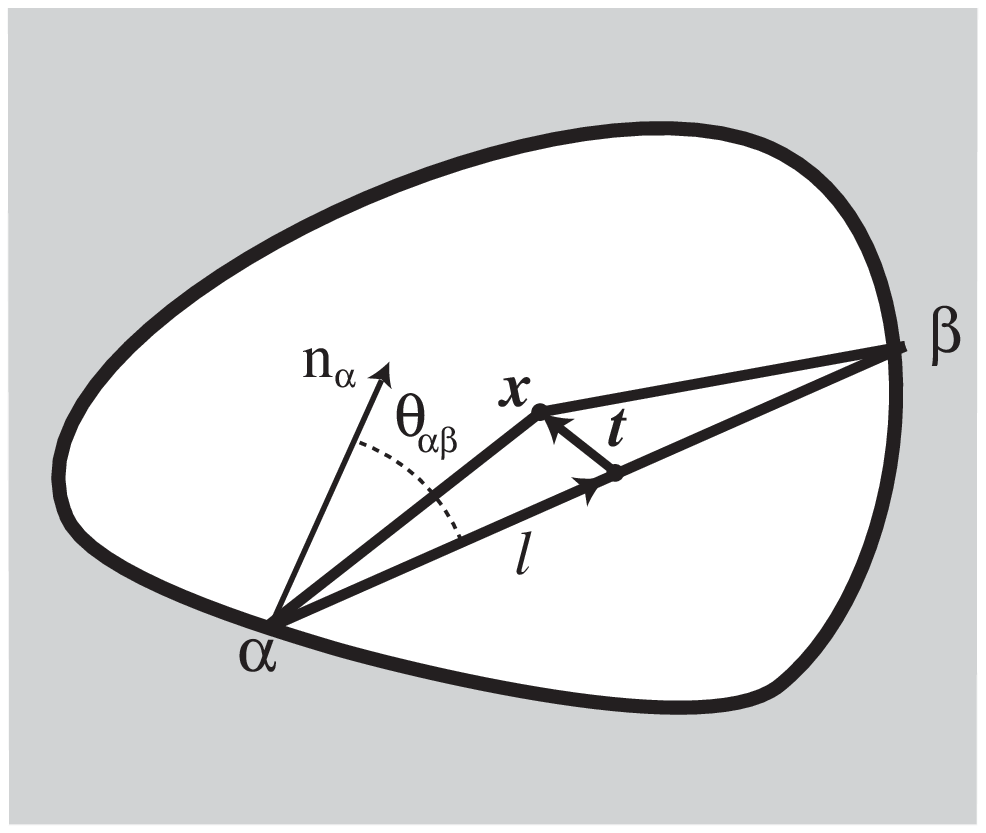}}
\vskip+0.4truecm
\caption{Geometry for $\bbox{x}$ integration in
Eq.~(\ref{EQ:asym_x_int})}.
\label{FIG:geo_x_int}
\end{figure}%
We first evaluate the ${\bf x}$ integral.  The stationary-phase points
of the ${\bf x}$ integration lie on the line joining $\bbox{\alpha}$ to
$\bbox{\beta}$.  Because the phase does not vary as one moves the point
${\bf x}$ along this line, it
is natural to separate the ${\bf x}$ integration into longitudinal
(i.e.~parallel to $\bbox{\alpha}-\bbox{\beta}$) and transverse
(i.e.~perpendicular to $\bbox{\alpha}-\bbox{\beta}$) components;
see Fig.~\ref{FIG:geo_x_int} for the nomenclature and geometry).
We use the asymptotic expression~(\ref{EQ:lr_asym_gf}) for the
homogeneous Green functions at large argument, together with the
asymptotic expression for $\partial g$, i.e.,
\begin{equation}
2\,\partial g_{\pm}(\bbox{\beta},\bbox{\alpha})
\approx
\left(
\frac{k_{\pm}}{2\pi \ell_{\alpha\beta}}\right)^{\negthinspace\frac{d-1}{2}}
\cos\theta_{\alpha\beta}\,\,
\exp\big({\pm ik_{\pm}\ell_{\alpha\beta}\mp i\pi(d-1)/4}\big),
\label{EQ:partial_gf}
\end{equation}
where $\theta_{\alpha\beta}$ is the angle between the ray joining
$\bbox{\alpha}$ to $\bbox{\beta}$ and the normal direction at the
point $\bbox{\alpha}$.  Then, we fix $l$, evaluate the ${\bf t}$
integral in the stationary-phase approximation, and evaluate
the $l$ integral:
\begin{eqnarray}
\label{EQ:asym_x_int}
\int_{\volV}
d^{d}x\,
g_{\pm}(\bbox{\beta},\bbox{x})\,
\partial g_{\pm}(\bbox{x},\bbox{\alpha})
&\approx&
\pm i \frac{k_{\pm}^{d-2}{\rm e}^{\mp i\pi(d-1)/2}}{2^{d+1} \pi^{d-1}}\cos\theta_{\alpha\beta}
\int dl \,
\Big(l(\ell_{\alpha\beta}-l)\Big)^{\frac{1-d}{2}}
\int d^{d-1}{t}\,
{\rm e}^{\pm ik_{\pm}
\left(\sqrt{l^2+|\bbox{t}|^2}+
\sqrt{(\ell_{\alpha\beta}-l)^2
+|\bbox{t}|^2}\right)}
\nonumber\\
&\approx&
\pm i \frac{k_{\pm}^{d-2}{\rm e}^{\mp i\pi(d-1)/2}}{2^{d+1} \pi^{d-1}}\cos\theta_{\alpha\beta}
\int dl \,
\Big(l(\ell_{\alpha\beta}-l)\Big)^{\frac{1-d}{2}}
\int d\bbox{t}\,
{\rm e}^{\pm ik_{\pm}\ell_{\alpha\beta}
\left(1+
\frac{|\bbox{t}|^2}{2l(\ell_{\alpha\beta}-l)}\right)}
\nonumber\\
&=&
\pm i \frac{k_{\pm}^{d-2}{\rm e}^{\mp i\pi(d-1)/2}}{2^{d+1} \pi^{d-1}}\cos\theta_{\alpha\beta}
\int dl \,
\Big(l(\ell_{\alpha\beta}-l)\Big)^{\frac{1-d}{2}}
{\rm e}^{\pm ik_{\pm}\ell_{\alpha\beta}}
\left(\frac{2\pi l(\ell_{\alpha\beta}-l)}{k_{\pm}\ell_{\alpha\beta}}\right)^{\frac{d-1}{2}}
{\rm e}^{\pm i \pi (d-1)/4}
\nonumber \\
&=&
\pm i \left(\frac{\ell_{\alpha\beta}}{4k_{\pm}}\right)
\left(\frac{k_{\pm}}{2\pi\ell_{\alpha\beta}}\right)^{\frac{(d-1)}{2}}
\cos\theta_{\alpha\beta}\,
\exp{\Big(\pm ik_{\pm}\ell_{\alpha\beta}\mp i\pi(d-1)/4\Big)}.
\label{EQ:PDint}
\end{eqnarray}
Next, we use Eqs.~(\ref{EQ:partial_gf}) and (\ref{EQ:PDint}), together
with Eq.~(\ref{EQ:RhoTwoN}), to obtain the following asymptotic
expression for $\rho_{2n}^\pm(E)$:
\begin{mathletters}
\begin{eqnarray}
\label{EQ:dos_2n}
\rho_{2n}^\pm(E)
&\approx&
{\rm Re}\int
d\sigma_{\alpha}\,
d\sigma_\beta\,
(\cos\theta_{\alpha\beta}\,\cos\theta_{\beta\alpha})^{n}\,
I_n(\bbox{\alpha},\bbox{\beta};\Kf)
\left(\frac{\Kf}{2\pi\ell_{\alpha\beta}}\right)^{nd-n-1}
\exp\Big({-i2n\varphi+in(k_{+}-k_{-})\ell_{\alpha\beta}}\Big),
\\
I_n(\bbox{\alpha},\bbox{\beta};\Kf)
&\equiv&
\int_{\rm sp}
d\sigma_{\alpha_1}
\cdots
d\sigma_{\alpha_{2n-2}}\,
{\rm e}^{\pm i\Kf S_{\rm And}},
\nonumber\\
S_{\rm And}&\equiv&\ell_{\beta\alpha} - \ell_{\alpha\alpha_1}
+\ell_{\alpha_1\alpha_2}\cdots-\ell_{\alpha_{2n-2}\beta}\, ,
\end{eqnarray}
\end{mathletters}%
and $\int_{\rm sp}$ indicates that the surface integrals over
$\bbox{\alpha}_1\cdots\bbox{\alpha}_{2n-2}$
should be evaluated in the stationary phase approximation.
In order to do this evaluation, we expand the functions
$\{\ell_{\alpha_i\,\alpha_j}\}$ around
$(\bbox{\alpha}_{2i},\bbox{\alpha}_{2i+1})=
(\bbox{\alpha},\bbox{\beta})$.

\begin{figure}[hbt]
 \epsfxsize=10cm
\centerline{\epsfbox{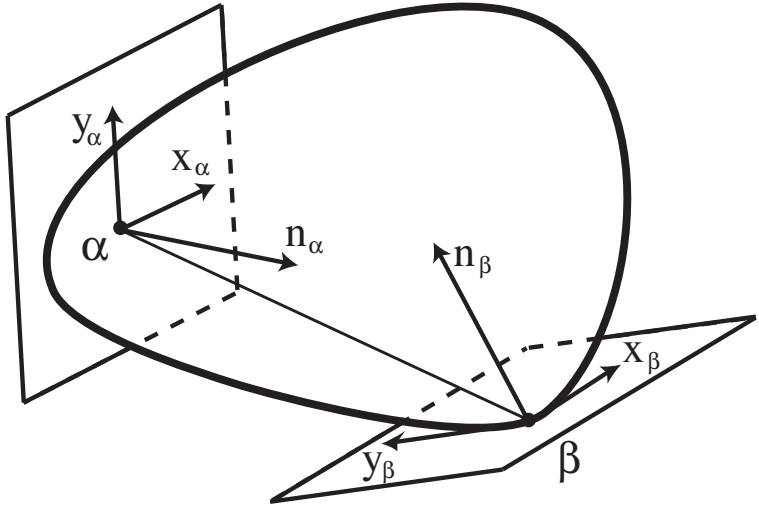}}
\vskip+0.4truecm
\caption{Geometry for fluctuations of the reflection points}
\label{FIG:geo_fluc}
\end{figure}%
We now focus on the evaluation of the stationary phase integrals in the
definition of $I_n(\bbox{\alpha},\bbox{\beta};\Kf)$.  In the remainder
of this section we shall work in three dimensions; after going through
the calculation for this $d=3$ case, the extension to higher dimensions is
straightforward.  The coordinate system for evaluating these stationary
phase integrals is as follows (see Fig.~\ref{FIG:geo_fluc}).  We use a
pair of coordinate systems: one for the set $\{\bbox{\alpha}_{2i}\}$
(which are near $\bbox{\alpha}$); the other for the set
$\{\bbox{\alpha}_{2i+1}\}$ (which are near $\bbox{\beta}$.)  Each of
these coordinate systems is constructed from the normal vector at the
point in question, along with two unit vectors in the tangent plane at this
point.  Thus, the coordinate system doublet comprises two sets:
one formed by the unit vectors
$\{\hat{\bf x}_\alpha,
   \hat{\bf y}_\alpha,
   \hat{\bf n}_\alpha\}$;
the other formed by the unit vectors
$\{\hat{\bf x}_\beta,
   \hat{\bf y}_\beta,
   \hat{\bf n}_\beta\}$, where
$\hat{\bf x}_{\alpha/\beta}$ and
$\hat{\bf y}_{\alpha/\beta}$
respectively lie on the tangent planes at the points
$\bbox{\alpha}/\bbox{\beta}$.
The choices of orientation of the axes within the tangent planes are
somewhat arbitrary; to ease the calculation we shall choose
$\hat{\bf x}_\alpha$ to be parallel to $\hat{\bf x}_\beta$
(and thus parallel to the line of intersection of the two tangent planes),
which fixes $\hat{\bf y}_\alpha$ and $\hat{\bf y}_\beta$.
In this coordinate doublet, we shall label the {\it local\/} coordinates
of point $\bbox{\alpha}_i$ by $(x_i,y_i)$.  The three-dimensional position
vectors are then determined as follows:
(i)~If $i$ is even, use the coordinate system at point $\bbox{\alpha}$;
otherwise use the coordinate system at point $\bbox{\beta}$.
(ii)~Then the three-dimensional position vector in the chosen coordinate
system can be expressed as
$\bbox{\delta}_i\equiv\big(x_i,y_i,z_{\eta_i}(x_i,y_i)\big)$,
where
\begin{equation}
\bbox{\eta}_i\equiv\cases{\bbox{\alpha},& for $i$ even,\cr
                          \bbox{\beta},& for $i$ odd,\cr}
\end{equation}
and $z_{\eta_i}(x,y)$ is the local equation of $\surS$ near the
point $\bbox{\eta}_i$.  We first expand
$\ell_{\alpha_{2i} \alpha_{2i+1}}$ to second order:
\begin{mathletters}
\begin{eqnarray}
\ell_{\alpha_{2i-1} \alpha_{2i}}&\cong& \ell_{\alpha\beta}
+ {\bf n}_{\alpha\beta}\cdot (\bbox{\delta}_{2i-1}-\bbox{\delta}_{2i})
+\frac{\vert\bbox{\delta}_{2i-1}-\bbox{\delta}_{2i}\vert^2}{2\ell_{\alpha\beta}}
-\frac{({\bf n}_{\alpha\beta}\cdot\bbox{\delta}_{2i-1}-{\bf n}_{\alpha\beta}\cdot\bbox{\delta}_{2i})^2}
{2\ell_{\alpha\beta}},
\\
\ell_{\alpha_{2i} \alpha_{2i+1}}&\cong& \ell_{\alpha\beta}
+ {\bf n}_{\alpha\beta}\cdot (\bbox{\delta}_{2i+1}-\bbox{\delta}_{2i})
+\frac{\vert\bbox{\delta}_{2i+1}-\bbox{\delta}_{2i}\vert^2}{2\ell_{\alpha\beta}}
-\frac{({\bf n}_{\alpha\beta}\cdot\bbox{\delta}_{2i+1}-{\bf n}_{\alpha\beta}\cdot\bbox{\delta}_{2i})^2}
{2\ell_{\alpha\beta}},
\end{eqnarray}%
\end{mathletters}%
\noindent
where ${\bf n}_{\alpha\beta}\equiv(\bbox{\beta}
        -\bbox{\alpha})/\ell_{\alpha\beta}$.
We then use this approximation to re-write $S_{\rm And}$:
\begin{equation}
\label{EQ:S_And_exp}
S_{\rm Andreev}\approx-\frac{1}{\ell_{\alpha\beta}}\sum_{i=1}^{n-1}
\Big(
\bbox{\delta}_{2i-1}\cdot\bbox{\delta}_{2i}
+(\bbox{\delta}_{2i-1}\cdot{\bf n}_{\beta\alpha})
 (\bbox{\delta}_{2i}  \cdot{\bf n}_{\alpha\beta})
\Big).
\end{equation}
Note that $z_{\eta_i}(x,y)$ is a quadratic function of $x$ and $y$
and therefore, for the purposes of expanding $S_{\rm And}$ to
second order in $\{x_i,y_i\}$, we can neglect the ${\bf n}_{\eta_i}$
component of $\bbox{\delta}_i$.  Thus, the curvature of $\surS$ does
not feature in $I_n(\bbox{\alpha},\bbox{\beta};\Kf)$, which implies
that the stability of these trajectories does not depend on the
curvature~\cite{FT:ind_cur}. Then, $S_{\rm And}$ can be written as
follows:
\begin{mathletters}
\begin{eqnarray}
S_{\rm Andreev}&\approx&\frac{1}{\ell_{\alpha\beta}}\sum_{i=1}^{n-1}
\pmatrix{x_{2i-1} & y_{2i-1}}\cdot{\bf D}\cdot\pmatrix{x_{2i} \cr y_{2i}},
\\
\noalign{\medskip}
{\bf D}&\equiv&\pmatrix{1-n_x^2     &  -n_x(n_y\cos\phi  +n_z\sin\phi )\cr
                        -n_x n_y    &   \cos\phi-n_y(n_y\cos\phi  +n_z\sin\phi )},
\end{eqnarray}
\end{mathletters}
where $n_x$, $n_y$ and $n_z$ are, respectively, the $x$, $y$ and $z$ coordinates of ${\bf n}_{\alpha\beta})$ in
the coordinate system at point $\bbox{\alpha}$, and $\cos\phi$ is the angle between the $y$ axes of the
coordinate systems at the points $\bbox{\alpha}$ and $\bbox{\beta}$.  Then
\begin{mathletters}
\begin{eqnarray}
&&
I_n(\bbox{\alpha},\bbox{\beta};\Kf)=
\int \prod_{i=1}^{2n-2} dx_i\,dy_i\,
\exp\Big(
(i\Kf/\ell_{\alpha\beta})
{\bf X}^\dagger \cdot \bbox{\cal D}(n)
\cdot{\bf X}
\Big),
\\
\noalign{\medskip}
&&
{\bf X}\equiv\pmatrix{x_1 \cr y_1\cr \vdots\cr x_{2n-2} \cr y_{2n-2}},
\qquad\qquad
\bbox{\cal D}(n)\equiv
\pmatrix{{\bf 0}              & {\bf D}        &{\bf 0}               &\cdots           &{\bf 0}       \cr
        {\bf D}^\dagger &{\bf 0}               &-{\bf D}        &                 &        \cr
        {\bf 0}               &-{\bf D}^\dagger&  \ddots        &                 &        \cr
        \vdots          &                &                &                 &-{\bf D}\cr
        {\bf 0}               &                &                &-{\bf D}^\dagger & {\bf 0}     }_{(4n-4)\times(4n-4)}.
\end{eqnarray}
\end{mathletters}
Thus we arrive at the following expression for $I_n$:
\begin{eqnarray}
\label{EQ:fluc_det}
I_n(\bbox{\alpha},\bbox{\beta};\Kf)
&=&
\left(
2\pi\ell_{\alpha\beta}/\Kf
\right)^{2n-2}
\Big(\det\bbox{\cal D}(n)\Big)^{-1/2}
\exp\left(i\pi\,{\rm sgn}\bbox{\cal D}/4\right)
\nonumber\\
&=&
\left(2\pi\ell_{\alpha\beta}/\Kf\right)^{2n-2}
\left(\cos\theta_{\alpha\beta}\,
      \cos\theta_{\beta\alpha}\right)^{1-n},
\end{eqnarray}
where ${\rm sgn}\bbox{\cal D}$ denotes the signature of the
matrix $\bbox{\cal D}$ (i.e.~the number of positive eigenvalues
minus the number of negative ones).  By inserting this expression
into Eq.~(\ref{EQ:dos_2n}) we obtain the contribution to the oscillatory
part of the DOS associated with the $2n$-reflection term:
\begin{eqnarray}
&&\rho_{2n}^\pm(E)
\approx
{\rm Re}\int
d\sigma_{\alpha}\,
d\sigma_\beta\,
\left(\frac{\Kf\cos\theta_{\alpha\beta}\,
    \cos\theta_{\beta\alpha}}{2\ell_{\alpha\beta}}\right)
\exp\big({-2in\varphi+in(k_{+}-k_{-})\ell_{\alpha\beta}}\big).
\end{eqnarray}
By collecting together the contributions from all numbers of reflections
($n=1,2,\ldots$) we arrive at the following formula for the oscillatory
part of the DOS in Scheme~A:
\begin{eqnarray}
&&\delta\rho_{\gamma}(E)
\approx\int d\sigma_{\alpha}\,d\sigma_{\beta}\,
\frac{\Kf\cos\theta_{\alpha\beta}\,
\cos\theta_{\beta\alpha}}{4\pi^2\ell_{\alpha\beta}}
\,\,{\rm Re}\,
\frac{\exp\left({-i2\varphi+i\frac{E}{\Kf}\ell_{\alpha\beta}
        -\frac{\gamma}{\Kf}\ell_{\alpha\beta}}\right)}
{1-\exp\left({-i2\varphi+i\frac{E}{\Kf}\ell_{\alpha\beta}
        -\frac{\gamma}{\Kf}\ell_{\alpha\beta}}\right)}.
\end{eqnarray}
A similar calculation for the $d$-dimensional case leads to the result
\begin{eqnarray}
\label{EQ:And_dos_osc}
&&\delta\rho_\gamma(E)
\approx
\int d\sigma_{\alpha}\,d\sigma_{\beta}\,
\frac{\cos\theta_{\alpha\beta}\,
      \cos\theta_{\beta\alpha}}{(2\pi)^d}
\left(\frac{\Kf}{\ell_{\alpha\beta}}\right)^{d-2}
{\rm Re}\,\frac{\exp\left({-i2\varphi+i\frac{E}{\Kf}\ell_{\alpha\beta}
        -\frac{\gamma}{\Kf}\ell_{\alpha\beta}}\right)}
{1-\exp\left({-i2\varphi+i\frac{E}{\Kf}\ell_{\alpha\beta}
        -\frac{\gamma}{\Kf}\ell_{\alpha\beta}}\right)}.
\end{eqnarray}
Formula~(\ref{EQ:And_dos_osc}) for the oscillatory part of
the DOS  simplifies further in the $\gamma\rightarrow 0$ limit.
To see this, consider the following familiar identity:
\begin{equation}
{\rm Re}\,
\frac{{\rm e}^{-i2\varphi+i\frac{E}{\Kf}\ell_{\alpha\beta}
        -\frac{\gamma}{\Kf}\ell_{\alpha\beta}}}
{1-{\rm e}^{-i2\varphi+i\frac{E}{\Kf}\ell_{\alpha\beta}
        -\frac{\gamma}{\Kf}\ell_{\alpha\beta}}}
=\frac{1}{2}\sum_{n=-\infty}^{\infty}
{\rm e}^{-i2n\varphi+in\frac{E}{\Kf}\ell_{\alpha\beta}}
        -\frac{1}{2}
=\pi\sum_{m=-\infty}^{\infty}
        \delta\left(\frac{E}{\Kf}\ell_{\alpha\beta}-2\varphi-2m\pi\right)
        -\frac{1}{2}.
\end{equation}
Applying this identity to Eq.~(\ref{EQ:And_dos_osc}), we obtain
\begin{equation}
\label{EQ:dos_shrp}
\delta\rho_\gamma(E)
\approx
\sum_{m=-\infty}^{\infty}
\int d\sigma_{\alpha}\,d\sigma_{\beta}\,
\frac{\cos\theta_{\alpha\beta}\,\cos\theta_{\beta\alpha}}{2(2\pi)^{d-1}}
\left(\frac{\Kf}{\ell_{\alpha\beta}}\right)^{d-2}
\delta\left(\frac{E}{\Kf}\ell_{\alpha\beta}-2\varphi-2m\pi\right)
-\int d\sigma_{\alpha}\,d\sigma_{\beta}\,
\frac{\cos\theta_{\alpha\beta}\,\cos\theta_{\beta\alpha}}{2(2\pi)^d}
\left(\frac{\Kf}{\ell_{\alpha\beta}}\right)^{d-2}.
\end{equation}

\begin{figure}[hbt]
\epsfxsize=10.0truecm
\centerline{\epsfbox{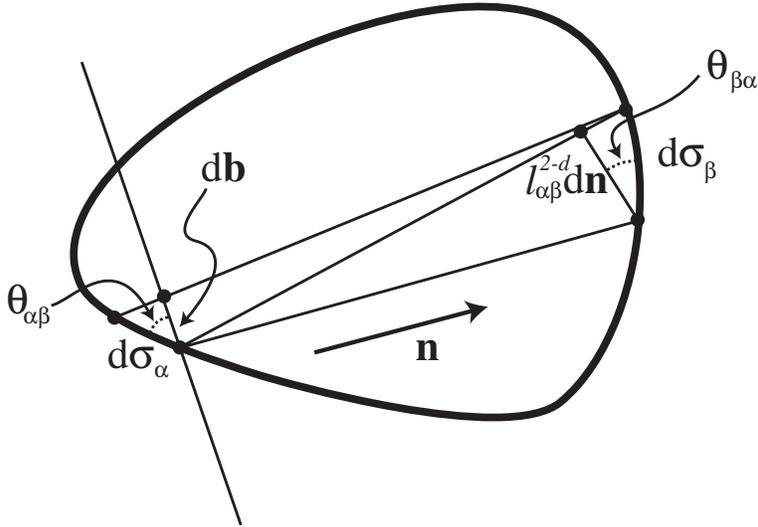}}
\vskip+0.4truecm
\caption{Jacobian of the transformation from boundary points to scattering parameters}
\label{FIG:sur_elmnt}
\end{figure}%
The last term in Eq.~(\ref{EQ:dos_shrp}) is a constant term, and it
exactly cancels the leading-order Weyl (i.e.~bulk) term.  In order
to see this, consider the coordinate transformation from
$(\bbox{\alpha},\bbox{\beta})$ to $({\bf b},{\bf n})$, where
${\bf n}$ is the direction of the chord and ${\bf b}$ is the
position-vector specifying the intersection of the chord with the
plane perpendicular to ${\bf n}$ (i.e.~the impact parameter).  The
transformation of the surface elements are as follows
(see Fig.~\ref{FIG:sur_elmnt}):
\begin{eqnarray}
d\sigma_{\alpha}\,d\sigma_{\beta}\,
\cos\theta_{\alpha\beta}\,
\cos\theta_{\beta\alpha}\,
\ell_{\alpha\beta}^{1-d}
=d{\bf n}\,d{\bf b}.
\end{eqnarray}
By using the $({\bf b},{\bf n})$ coordinate system, the last term
in Eq.~(\ref{EQ:dos_shrp}) can be cast into the following form:
\begin{equation}
 -\frac{1}{2(2\pi)^d} \Kf^{d-2}
  \int d{\bf n}\,d{\bf b}\,\ell_{\alpha\beta}
=-\frac{1}{2(2\pi)^d}\,\Kf^{d-2}\,V\int d{\bf n}
=-\frac{1}{2(2\pi)^d}\,\Kf^{d-2}\,V\,{\cal S}^{d-1}.
\label{EQ:canWeyl}
\end{equation}
A direct comparison with the leading-order Weyl term, given in
Eq.~(\ref{EQ:WEYL_DOS}), shows that the term in Eq.~(\ref{EQ:canWeyl})
identically cancels with the leading-order Weyl term.  Thus, the
DOS in Scheme~A can be written as
\begin{equation}
\rho(E)
\approx
\sum_{m=-\infty}^{\infty}
\frac{\Kf^{d-2}\ell_{\alpha\beta}}{2(2\pi)^{d-1}}\,
\int d{\bf n}\,d{\bf b}\,
\delta\left(\frac{E}{\Kf}\ell_{\alpha\beta}-2\varphi-2\pi m\right),
\end{equation}
which is precisely the result found via the Andreev approximation,
as stated in Sec.~\ref{SEC:Andreev}.
\hfill

\subsection{Scheme~B: Mesoscale oscillations beyond the
resolution of the Andreev approximation}
\label{SEC:SchemeB}
The main motivation of Scheme~B is to capture the mesoscale
oscillations in the DOS that are caused by confinement
of quasiparticles by the superconducting surround.  The reason that Scheme~A (and thus the
Andreev approximation) is not capable of capturing such
oscillations is that in Scheme~A the transverse degrees of
freedom are not quantized (i.e.~quasiparticle motion on the
chords is quantal, but there are chords arbitrarily close to one
another, indicating that the transverse degrees of freedom are
treated classically).  On the other hand, in Scheme~B we take
into account:
(i)~the imperfectness in retro-reflection (arising from the
previously-neglected difference between the wave vectors of
incident and reflected electrons and holes); and
(ii)~the imperfectness in charge-interconversion, the amplitude
for which is ${\cal O}(\sudel/\Ef)$.
{\it A priori\/}, we know that mesoscale oscillations in the DOS
must originate from one or more of the processes not yet taken into account,
these being (i) and (ii), above, as well as the subleading quantal
corrections (that are ignored in both schemes).  In fact, as we shall
see, it is the imperfectness in retro-reflection that is primarily
responsible for the mesoscale oscillations in the DOS.  (Note that
the imperfectness in retro-reflection occurs {\it transverse\/} to the
incoming direction.)\thinspace\  The imperfectness in
charge-interconversion does modify these DOS oscillations.  However,
by itself (such as in a model with perfect retro-reflection but
imperfect charge-interconversion) it is not capable of producing
them.  In order to clarify this issue, and thus to assess the
significance of these processes, we shall define a useful intermediate
model: the Perfectly Charge-Interconverting Model (PCIM).
This PCIM has the feature of being fully quantum-mechanical; however,
in it, any single reflection from the boundary is certain to have the
effect of converting electrons to holes (and {\it vice versa\/}).  As it
contains all quantal effects, the comparison of its predictions with
those from the semiclassical approach enable us to assess the
importance of quantal effects beyond the semiclassical limit.  Moreover,
from this comparison it is possible to draw conclusions regarding
whether it is imperfectness in retro-reflection that is capable of
capturing the mesoscale oscillations in the DOS.
\subsubsection{Perfectly Charge-Interconverting Model}
We are now at a position to define a {\it Perfectly Charge-Interconverting
Model\/} (PCIM).  We start with the expansion for $\gf$ in terms of
$\gf^{\rm R}$ in Eq.~(\ref{EQ:appMRE}).  However, we shall replace
$\gf^{\rm R}$ by its leading-order form, i.e.,
$-i{\rm e}^{-i\varphi}\psm{1}\gf^{\rm N}$.
Then the resulting model is defined as an integral equation for $\gf$,
residing inside the billiard:
\begin{eqnarray}
\gf({\bf x},{\bf x}')=
  \gf^{\rm N}({\bf x},{\bf x}')
-2i{\rm e}^{-i\varphi}
\int_{\surS}\surel_{\alpha}\,
\partial\gf^{\rm N}({\bf x},\bbox{\alpha})\,
\psm{1}\,\gf(\bbox{\alpha},{\bf x}').
\end{eqnarray}
The off-diagonal matrix $\psm{1}$ ensures that, upon each reflection
from the boundary, electrons are totally converted into holes (and {\it vice versa\/}).
Moreover, this model is fully quantum-mechanical, in the sense
that it retains wave-propagation effects (as implied by the presence of
surface integrals.)  Let us now focus on the DOS of the PCIM, treated at
the semiclassical level.

Due to the imperfectness in retro-reflection in Scheme~B, the
corresponding classical dynamics is no longer {\it a priori\/} integrable;
on the contrary, it is weakly chaotic for generic
shapes~\cite{REF:Kosztin}.  However, the closed periodic orbits do
fall into two quite distinct classes:
one consists of multiple tracings of each stationary chord
(i.e.~chord of stationary length, which we refer to as a {\SC});
the other of much longer trajectories that \lq\lq creep\rq\rq\
around the billiard boundary (see Fig.~\ref{FIG:imp_tra}).
Correspondingly, the DOS is the sum of:
(i)~an average term, which depends on the volume of the
billiard (i.e.~the leading Weyl term);
(ii)~a finer-resolution term, having a universal lineshape
that depends solely on the length and endpoint-curvatures of
the {\SC}s; and
(iii)~highest resolution terms, which depend on the classical
dynamics of the billiard in question.

\subsubsection{Stationary chords}
Here, we focus on the contribution to the DOS coming from stationary
chords.  As we shall derive below, it is possible to give a closed-form
expression for the contribution of a \SC\/ to the DOS for billiards of generic
shape.  It is, however, necessary to distinguish between isolated
{\SC}s and degenerate {\SC}s.  First, let us consider an isolated {\SC}
in a 2D billiard, with endpoint curvatures $R_1$ and $R_2$.  Then the
\SC\/ contribution to the DOS from $2n$ reflections is given by
\begin{equation}
\rho_{\SCm,\pm}^n=\frac{\ell_\SCm}{2\pi k_\pm}{\rm Re}
\left(\frac{k_+}{2\pi\ell_\SCm}\right)^{n/2}
\left(\frac{k_-}{2\pi\ell_\SCm}\right)^{n/2}
\exp\Big({in(k_+-k_-)\ell_\SCm-i2n\varphi}\Big)\, I_n.
\end{equation}
Here, $I_n$ is the Gaussian integral resulting from the expansion of the action
to second order:
\begin{equation}
I_n\equiv\int\prod_{i=1}^{2n} dx_i\,\exp{
\sum_{i=1}^{n}\,i\left(\frac{k_+}{\ell_\SCm}x_{2i-1}\,x_{2i}
        -\frac{k_-}{\ell_\SCm}x_{2i}\,x_{2i+1}+\frac{k_+-k_-}{2}
\left(\frac{1}{\ell_\SCm}-\frac{1}{R_1}\right)x_{2i-1}^2
        +\left(\frac{1}{\ell_\SCm}-\frac{1}{R_2}\right)x_{2i}^2
\right)},
\end{equation}
where $x_{2n+1}\equiv x_1$.  In order to evaluate this integral, we define a matrix $M$ such that:
\begin{equation}
\sum_{i,j} x_i\,M_{ij}\,x_j\equiv
\sum_{i=1}^{n}\,i
\left(\frac{k_+}{\ell_\SCm}x_{2i-1}\,x_{2i}-\frac{k_-}{\ell_\SCm}x_{2i}\,x_{2i+1}
+\frac{k_+-k_-}{2}\left(\frac{1}{\ell_\SCm}-\frac{1}{R_1}\right)x_{2i-1}^2
+\left(\frac{1}{\ell_\SCm}-\frac{1}{R_2}\right)x_{2i}^2
\right)\,.
\end{equation}
(Note that, by definition, an isolated stationary chord is one for which
none of the eigenvalues of $M$ is zero; whenever a zero eigenvalue occurs, the
stationary chord is said to be degenerate.)\thinspace\ Then, for an isolated \SC\ , $I_n$ can
be expressed in terms of the determinant and signature of $M$:
\begin{equation}
I_n=\frac{\pi^n}{\sqrt{\vert\det M\vert}}\,
\exp\left({i\frac{\pi}{4}{\rm sgn} M}\right).
\end{equation}
The eigenvalues of $M$ can be obtained from the eigenvalue
equation $\sum_j M_{ij}\,x_j=\lambda x_i$.
The symmetry of $M$ under the transformation
$x_i\rightarrow x_{i+2}$ restricts the form of the eigenvectors
to be
\[
x_{2j}  =A_m \,\romE^{i2m\pi\left(\frac{2j}{2n}\right)},\qquad
x_{2j+1}=B_m \,\romE^{i2m\pi\left(\frac{2j+1}{2n}\right)},
\]
where $m=1,2,\cdots,n$.
Then the eigenvalue equation reduces to
\begin{equation}
\Bigg(\matrix{
 \frac{k_+-k_-}{2}\left(\frac{1}{\ell_\SCm}-\frac{1}{R_2}\right) &
 \frac{k_+}{\ell_\SCm}\romE^{-i\pi\frac{m}{n}}
-\frac{k_-}{\ell_\SCm}\romE^{i\pi\frac{m}{n}} \cr
-\frac{k_-}{\ell_\SCm} \romE^{-i\pi\frac{m}{n}}
+\frac{k_+}{\ell_\SCm} \romE^{i\pi\frac{m}{n}} &
 \frac{k_+-k_-}{2}\left(\frac{1}{\ell_\SCm}-\frac{1}{R_1}\right)
}\Bigg)
\Bigg(\matrix{A_m\cr B_m}\Bigg)=
\lambda_m
\Bigg(\matrix{A_m\cr B_m}\Bigg).
\label{EQ:eigenfluc}
\end{equation}
The determinant of $M$ is readily obtained as
\begin{eqnarray}
\det M
&=&
\prod_{m=1}^{n}
\det
\pmatrix{
 \frac{k_+-k_-}{2}\left(\frac{1}{\ell_\SCm}-\frac{1}{R_2}\right) &
 \frac{k_+}{\ell_\SCm}\romE^{-i\pi\frac{m}{n}}
-\frac{k_-}{\ell_\SCm}\romE^{i\pi\frac{m}{n}} \cr
-\frac{k_-}{\ell_\SCm} \romE^{-i\pi\frac{m}{n}}
+\frac{k_+}{\ell_\SCm} \romE^{i\pi\frac{m}{n}} &
 \frac{k_+-k_-}{2}\left(\frac{1}{\ell_\SCm}-\frac{1}{R_1}\right)
}
\nonumber \\ &=&\prod_{m=1}^{n}
\left[
\left(\frac{\ell_\SCm^2-\ell_\SCm R_1 -\ell_\SCm R_2}{R_1 R_2}\right)
\left(\frac{k_+-k_-}{\ell_\SCm}\right)^2
        -4\frac{k_+k_-}{\ell_\SCm^2}\sin^2\left(\pi\frac{m}{n}\right)
\right].
\end{eqnarray}
For $n\ll\sqrt{k_+k_-}/(k_+-k_-)$, this expression can be simplified to read
\begin{eqnarray}
\det M
&\approx&
\left(\frac{\ell_\SCm^2-\ell_\SCm R_1 -\ell_\SCm R_2}{R_1 R_2}\right)
\left(\frac{k_+-k_-}{\ell_\SCm}\right)^2
\left(\frac{4k_+k_-}{\ell_\SCm^2}\right)^{n-1}
\left(\prod_{m=1}^{n-1}\sin^2\left(\pi\frac{m}{n}\right)\right)^2
\nonumber \\
&=&(-1)^{n-1}
\left(\frac{\ell_\SCm^2-\ell_\SCm R_1 -\ell_\SCm R_2}{R_1 R_2}\right)
\left(\frac{k_+-k_-}{\ell_\SCm}\right)^2
\left(\frac{k_+k_-}{\ell_\SCm^2}\right)^{n-1} n^2\,.
\end{eqnarray}
The signature of $M$ is obtained from the eigenvalue problem~(\ref{EQ:eigenfluc})
via the investigation of the signs of the
eigenvalues $\lambda_m^\pm$ [or, equivalently, from the trace and
the determinant of the matrix in Eq.~(\ref{EQ:eigenfluc})].  If the
determinant is negative then $\lambda_m^+$ and $\lambda_m^-$ are of
differing signs and, hence, the associated pair cancel in the
signature of $M$.  If the determinant is positive then there is a
pair of negative or of positive eigenvalues, the sign of which is
determined by the trace.  These considerations allow us to express
${\rm sgn}\,M$ as follows:
\begin{equation}
{\rm sgn} M=\sum_{m=1}^{n}
\left(
1+{\rm sgn}\left[
\left(\frac{\ell_\SCm^2-\ell_\SCm R_1 -\ell_\SCm R_2}{R_1 R_2}\right)
\left(\frac{k_+-k_-}{\ell_\SCm}\right)^2
        -4\frac{k_+k_-}{\ell_\SCm^2}\sin^2\left(\pi\frac{m}{n}\right)
\right]
\right){\rm sgn}
\left(
\frac{2}{\ell_\SCm}-\frac{1}{R_1}-\frac{1}{R_2}
\right).
\end{equation}
Again there is simplification for $n\ll\sqrt{k_+k_-}/(k_+-k_-)$,
because in this regime only $m=n$ term contributes to the sum above
and, hence, the expression for ${\rm sgn} M$ becomes
\begin{equation}
{\rm sgn}\,M\approx
\left(
1+{\rm sgn}
\left(\frac{\ell_\SCm- R_1 - R_2}{R_1 R_2}\right)
\right){\rm sgn}
\left(
\frac{2}{\ell_\SCm}-\frac{1}{R_1}-\frac{1}{R_2}
\right).
\end{equation}
In particular, when $R_1,R_2>0$ we have
\begin{equation}
{\rm sgn}\,M\approx
{\rm sgn}
\left(R_1 + R_2-\ell_\SCm\right) -1.
\end{equation}
Putting all the pieces together, the expression for the
contribution to the DOS originating from isolated stationary
chords is as follows:
\begin{mathletters}
\begin{eqnarray}
\rho_{\SCm,\pm}^n
&\approx&
\frac{\ell_\SCm}{2\pi k_\pm}{\rm Re}
\left(\frac{k_+}{\ell_\SCm}\right)^{n/2}
\left(\frac{k_-}{\ell_\SCm}\right)^{n/2}
\frac{1}{\sqrt{\vert\det M\vert}}\,
\exp\Big({in(k_+-k_-)\ell_\SCm-i2n\varphi+i\frac{\pi}{4}{\rm sig} M}\Big)
\\
&\approx&
{\rm Re}
\frac{1}{n}\,
\frac{1}{k_\pm}
\sqrt{\frac{k_+k_-\ell_\SCm R_1 R_2}{4\pi^2(k_+-k_-)^2\vert\ell_\SCm-R_1 -R_2\vert}}\,
\exp\Big({in(k_+-k_-)\ell_\SCm-i2n\varphi+i\frac{\pi}{4}
\big({\rm sgn}\left(R_1 + R_2-\ell_\SCm\right) -1\big)}\Big),
\end{eqnarray}%
\end{mathletters}%
where the second line is valid for
$n\ll\sqrt{k_+k_-}/(k_+-k_-)$.
We can now sum this expression to all orders in $n$ to obtain the general
form for the contribution to the DOS originating from an isolated stationary
chord:
\begin{equation}
\rho_{\SCm} \approx {\rm Re}\,\sqrt{\frac{(k_{+}+k_{-})^2
\ell_{\SCm}R_{1}R_{2}}{4\pi^2k_{+}k_{-}
(k_{+}-k_{-})^2|\ell_{\SCm}-R_{1}-R_{2}|}}
\,\romE^{i\frac{\pi}{4}({\rm sgn} \left(R_1 +
R_2-\ell_\SCm\right) -1)}
\ln(1-\romE^{in(k_+-k_-)\ell_\SCm-i2n\varphi}).
\label{EQ:stat_dos1}
\end{equation}
\begin{figure}[hbt]
\epsfxsize=5.0truecm
\centerline{\epsfbox{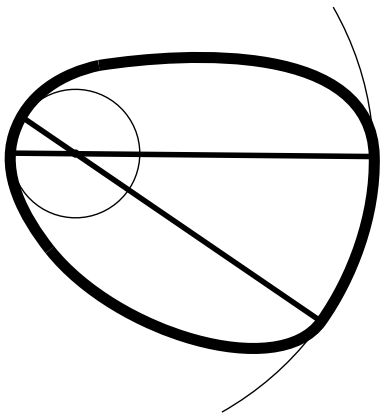}}
\vskip+0.4truecm
\caption{Example of a billiard that is not circular but has
degenerate stationary chords.  Opposite segments of the boundary are
partially coincident with a pair of concentric circles, as shown;
when this happens, $\ell_\SCm=R_1+R_2$.}
\label{FIG:deg_stat}
\end{figure}%
The DOS expression in Eq.~(\ref{EQ:stat_dos1}) is valid when its prefactor
is free of singularities.  Such singularities would indicate that the fluctuation
determinant has a vanishing eigenvalue.  This would mean that the stationary chord
is no longer isolated; instead there is a direction, corresponding to the
vanishing eigenvalue, along which fluctuations do not change the phase.

Let us now focus on these prefactor singularities.  The first kind occurs
when $k_+=k_-$ or, equivalently, when $E=0$.  The reason that this
singularity arises is that when $k_+=k_-$ retro-reflection is perfect and,
thus, all chords become stationary.  As, in Scheme~B, we are implicitly
assuming that $k_+$ and $k_-$ are distinct, this kind of singularity is an
artifact of the approximation scheme, and thus is unphysical.  The second
kind of singularity occurs when $\ell_{\SCm}=R_{1}+R_{2}$.  This is a
geometrical singularity, in the sense that whenever the shape of the billiard
is such that chords of stationary lengths are not isolated, such a singularity
arises.  For example, if the billiard is circular, the chord of stationary
length is the diameter of the circle, and its length does not change when it
is rotated.  In this case $R_1=R_2=\ell_{\SCm}/2$.  Another example is shown
in Fig.~\ref{FIG:deg_stat}; see the caption for details.  Such singularities
can be handled in Scheme~B as follows: we fix one (or more, if necessary)
reflection points, so that the \SC\ is no longer degenerate; we then integrate
the remaining surface integrals in the stationary phase approximations; and,
after that, we evaluate the remaining surface integral.  In this way, we
obtain the general form for the stationary-chord contribution to the DOS:
\begin{equation}
\rho_{\SCm}(E)
\approx
{\rm Re}
\sum_{\ell_{\SCm}}
Z_{\SCm}\,\,{\rm e}^{i\lambda\pi/4}\,{\rm Li}_{d-1-\frac{w}{2}}
(1-{\rm e}^{i(k_{+}-k_{-})\ell_{\SCm}-2i\varphi}).
\label{EQ:stat_dos2}
\end{equation}%
Here,
${\rm Li}_{n}(z)\equiv\sum_{j=1}^{\infty}z^j/j^n$
is the polylogarithm function,
$w$ is the dimensionality of the degeneracy of the \SC\
(e.g.~$w=1$ for a circle),
$Z_{\SCm}$ is a slowly varying real function of energy,
which determines the size of the DOS oscillations, and
$\lambda$ is a measure of the stability of the {\SC}s,
which determines whether the
\lq\lq tail\rq\rq\ goes to higher or lower energies.
For example, for a degenerate \SC\ and $d=2$ we have
\begin{eqnarray}
Z_{\SCm}&=&V_{\SCm}\,\sqrt{\frac{(k_{+}+k_{-})^2
\ell_{\SCm}R}{8\pi^3 k_{+}k_{-}(k_{+}-k_{-})|\ell_{\SCm}-R|}},
\nonumber \\
\lambda&=&{\rm sgn}(R-\ell_{\SCm}), \nonumber
\end{eqnarray}
where $V_{\SCm}$ is the volume of the
degenerate \SC\ and $R\equiv(R_1+R_2)/2$.
\subsubsection{Creeping orbits}
In this section, we shall obtain the finer oscillatory structure that
lies beyond the stationary-chord contribution.  The method for obtaining
this structure consists of:
(i)~finding the classical periodic orbits;
(ii)~evaluating the surface integrals in the stationary phase
approximation (i.e.~expanding the action to second order around each
classical periodic orbit, thus reducing the surface integrals to Gaussian
integrals, and then evaluating the resulting Gaussian integrals); and
(iii)~summing over all periodic-orbit contributions.
For the purposes of illustrating this method, we now focus on the case of a circular Andreev billiard and obtain
an expression for this finer oscillatory structure in DOS.  In addition
to its illustrative purposes, having obtained this expression for the
finer DOS oscillations for circular Andreev billiard will become useful
when we discuss an additional approximation, valid for shapes with slowly
varying curvatures, as we shall do later in this Paper.
\par
\noindent{\it (i) Classical Dynamics\/}:
The classical dynamics of Andreev billiards is defined through the
Scheme~B reflection rule: $k_{+}\sin\theta_+=k_- \sin\theta_-$.
Here, $\theta_{+/-}$ is the angle of incidence (or
reflection) for the electron/hole.  In the case of circular Andreev
billiards, the classical periodic orbits can be specified (up to rigid
rotations of the full trajectory) by the number of reflections ($2n$)
and by the number of times the orbit \lq\lq winds\rq\rq\ around the
billiard ($j$).  The fact that classical periodic orbits of circular
Andreev billiards can be specified by $(n,j)$ is a consequence of the
integrability of the classical dynamics.

\begin{figure}[hbt]
\epsfxsize=10.0cm
\centerline{\epsfbox{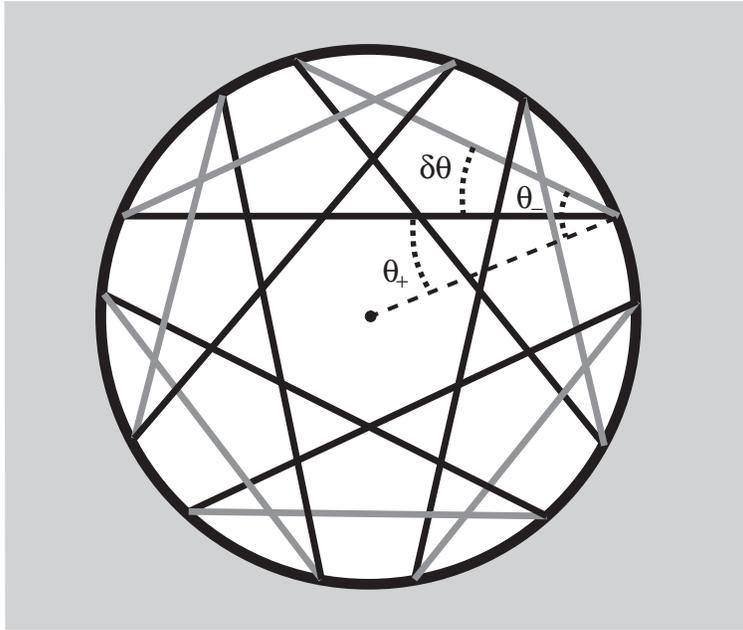}}
\vskip+0.2cm
\caption{A closed periodic orbit with imperfect retro-reflection;
the degree of imperfectness is exaggerated for the purpose of
illustration.}
\label{FIG:circle_cd}
\end{figure}%
Our goal is to express dynamical quantities, such as the action
$S_{\rm cl}(n,m,\Ef,E,R)$, angles of incidence and reflection,
and the length of propagation between two successive reflections
(which we shall denote by $\ell_{\pm}$) in terms of the parameters
$(n,m,\Ef,E,R)$.  Here, $R$ is the radius of the circular billiard.
First note that, for a given orbit, the angle at which an electron is
incident or reflected is the same at all reflections; the same fact
holds for holes (see Fig.~\ref{FIG:circle_cd}).  Next, note that each
reflection rotates the direction of the momentum by
$\delta\theta\equiv\theta_{-} - \theta_{+}$.
Thus, in order for the orbit to close after $2n$ reflections the
momentum direction must be rotated by $2\pi\,j$.  In other words,
$\delta\theta=\pi\,j/n$.
However, $\theta_{\pm}$, must also satisfy the classical
reflection rule
$k_{+}\sin{\theta_+}=k_{-}\sin{\theta_-}$.
By using these conditions we find (after a little algebra) that for a
closed orbit specified by $n$ and $j$ we have
\begin{equation}
\cos\theta_\pm=
\pm\frac{k_{\pm}-k_{\mp}\cos(\pi j/n)}{\Delta k},
\quad
\ell_\pm=2 R\cos\theta_\pm,
\end{equation}
where $\Delta k\equiv
\left(\Ef-2\sqrt{\Ef^2-E^2}\,
\cos(\pi j/n)\right)^{1/2}$.
Observe that, owing to the fact that
$\cos\theta_+>\cos\theta_->0$,
the possible values of $n$ and $j$ are restricted:
\begin{equation}
\cos\left(\frac{\pi j}{n}\right)>\frac{k_-}{k_+}.
\label{EQ:restrict}
\end{equation}
Next we evaluate the action:
\begin{eqnarray}
S_{\rm cl}(n,m,\Ef,E,R)&\equiv& n k_+ \ell_+-nk_- \ell_-
\nonumber\\
&=& 2n (k_{+}R\cos\theta_+-k_{-}R\cos\theta_-)
\nonumber\\
&=& 2nR\left(\Ef-2\sqrt{\Ef^2-E^2}\cos(\pi j/n)\right)^{1/2}
\nonumber \\
&=& 2n\,\Delta k\,R\,.
\end{eqnarray}
\begin{figure}[hbt]
\epsfxsize=10.0cm
\centerline{\epsfbox{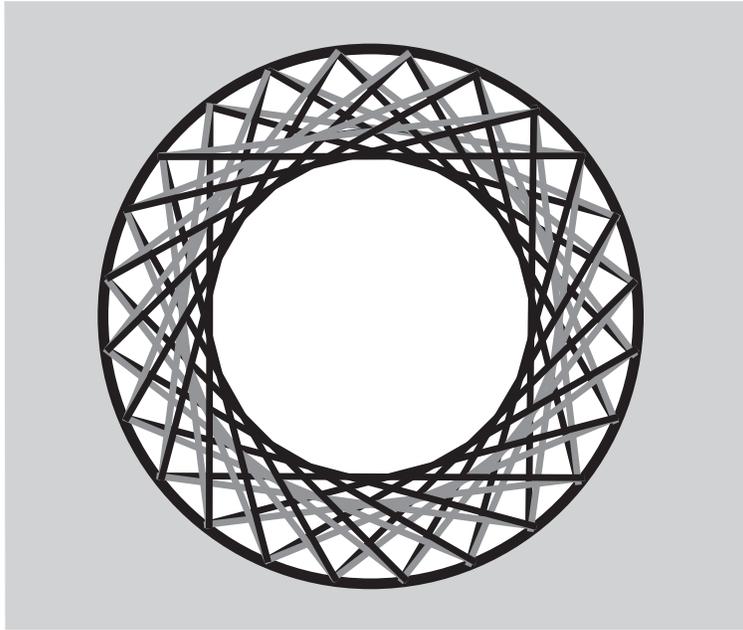}}
\vskip+0.2cm
\caption{A closed periodic orbit with imperfect retro-reflection;
the degree of imperfectness is exaggerated for the purpose of
illustration, but less so than in Fig.~\ref{FIG:circle_cd}.}
\label{FIG:circle_cd_real}
\end{figure}%

\par\noindent{\it (ii)~Evaluating surface integrals\/}:
Having obtained the critical points of the surface integrals in
Eq.~(\ref{EQ:DOS_MRE}), we now proceed to expand the action around these
points.  In order to do this, we expand each
$g_{\pm}(\bbox{\alpha},\bbox{\beta})$ in Eq.~(\ref{EQ:DOS_MRE}) around the
critical values of its arguments
(i.e.~$(\bbox{\alpha}_{\rm c},\bbox{\beta}_{\rm c})$).
This can be accomplished by writing
\begin{equation}
\bbox{\alpha}=\bbox{\alpha}_{\rm c}+\bbox{\delta}_{\bbox{\alpha}},\qquad
\bbox{\beta} =\bbox{\beta}_{\rm c} +\bbox{\delta}_{\bbox{\beta}},
\end{equation}
and expressing
$\bbox{\delta}_{\bbox{\alpha}/\bbox{\beta}}$
in the coordinate system
at $\bbox{\alpha}/\bbox{\beta}$
with coordinate axes specified by the normal
(i.e.~${\bf n}_{\bbox{\alpha}/\bbox{\beta}}$)
and tangent (i.e.~${\bf t}_{\bbox{\alpha}/\bbox{\beta}}$)
directions of $\surS$ at
$\bbox{\alpha}/\bbox{\beta}$.
In this coordinate system, the coordinates of the boundary point
$\bbox{\delta}_{\bbox{\alpha}/\bbox{\beta}}$ can, up to quadratic
order, be parametrized as
\begin{equation}
\bbox{\delta}_{\bbox{\alpha}/\bbox{\beta}}\approx s_{\bbox{\alpha}/\bbox{\beta}}\,{\bf t}_{\bbox{\alpha}/\bbox{\beta}}+
\frac{s_{\bbox{\alpha}/\bbox{\beta}}^2}{2 R_{\bbox{\alpha}/\bbox{\beta}}}\,{\bf n}_{\bbox{\alpha}/\bbox{\beta}},
\end{equation}
where $R_{\bbox{\alpha}/\bbox{\beta}}$
are the radii of curvature at ${\bbox{\alpha}/\bbox{\beta}}$.
The surface elements can also be expressed using the parameter $s_{\bbox{\alpha}/\bbox{\beta}}$:
\begin{equation}
\surel_{\bbox{\alpha}/\bbox{\beta}}=
ds_{\bbox{\alpha}/\bbox{\beta}}+{\cal O}(1/R).
\end{equation}
Then $\partial g_\pm$ can be approximated as
\begin{equation}
2\partial g_\pm(s_{\bbox{\alpha}},s_{\bbox{\beta}})\approx\mp
\sqrt{ \frac{\Kf \cos^2\theta_{\bbox{\beta}}}{2\pi\ell_{\bbox{\alpha}_{\rm c}\,\bbox{\beta}_{\rm c}}} }
~\exp\Big({\pm i k_\pm \ell_{\bbox{\alpha}_{\rm c}\bbox{\beta}_{\rm c}} \mp i \pi/4
\pm i k_\pm \Phi(s_{\bbox{\alpha}},s_{\bbox{\beta}})}\Big),
\end{equation}
where the phase function $\Phi$ is given by
\begin{eqnarray}
\Phi(s_{\bbox{\alpha}},s_{\bbox{\beta}})
\equiv
 s_{\bbox{\beta }}\sin\theta_{\bbox{\beta }}
-s_{\bbox{\alpha}}\sin\theta_{\bbox{\alpha}}
-\frac{s^2_{\bbox{\beta }} \cos\theta_{\bbox{\beta }}  }{ 2R_{\bbox{\beta }} }
-\frac{s^2_{\bbox{\alpha}} \cos\theta_{\bbox{\alpha}}  }{ 2R_{\bbox{\alpha}} }
+\frac{1}{2\ell_{\bbox{\alpha}_{\rm c}\,\bbox{\beta}_{\rm c}}}
(s_{\bbox{\beta}}\cos\theta_{\bbox{\beta}}
        +s_{\bbox{\alpha}}\cos\theta_{\bbox{\alpha}})^2.
\end{eqnarray}
The terms linear in $s_{\bbox{\alpha}/\bbox{\beta}}$ cancel with
the linear terms of the next and previous Green function, due to the
stationarity feature of the critical point.  For the circle we have
$\theta_{\bbox{\alpha}}=\theta_{\bbox{\beta}}=\theta_\pm$,
$R_{\bbox{\alpha}}=R_{\bbox{\beta}}=R$, and
$\ell_{\bbox{\alpha}_{\rm c}\bbox{\beta}_{\rm c}}=2R\cos\theta_\pm$.
This allows us to write
\begin{equation}
\Phi_\pm(s_{\bbox{\alpha}},s_{\bbox{\beta}})=-\frac{\cos\theta_{\pm}}{4 R}(s_{\bbox{\alpha}}-s_{\bbox{\beta}})^2,
\label{EQ:circlephi}
\end{equation}
and thus
$2\partial g_\pm(s_{\bbox{\alpha}},s_{\bbox{\beta}})=2\partial g_\pm(s_{\bbox{\alpha}}-s_{\bbox{\beta}})$.
We are now in a position to evaluate the DOS oscillations due to a
periodic orbit with $2n$ reflections that winds $j$ times around the
circular billiard:
\begin{eqnarray}
\rho_{\pm}^{n,j}
&=&
\frac{\ell_\pm}{2\pi k_\pm}{\rm Re}\,\,
\exp\Big({iS_{\rm cl}(n,j,\Ef,E,R)-i2n\varphi}\Big)\, I_n\,,
\\
I_n
&\equiv&
\int\prod_{i=1}^{2n} ds_i\,
2\partial g_+(s_{1})\,
2\partial g_-(s_{1}-s_2)\cdots
2\partial g_+(s_{2n-1}).
\end{eqnarray}
Although it is possible to evaluate this Gaussian integral by the usual
formula that relates it to the determinant and signature of the quadratic
form in the phase, the (rotationally invariant) form of $\Phi$ in Eq.~(\ref{EQ:circlephi}) allows the evaluation of the integral above in
an easier way, viz., via Fourier decomposition.  By defining the Fourier
transform
\begin{eqnarray}
2\partial g_\pm(p)\equiv\int_{-\infty}^\infty ds \, \romE^{ips}\, 2\partial g_\pm(s)
= i \exp\left(
{\pm ip^2 R/{k_\pm \cos\theta_\pm} }
\right)
\end{eqnarray}
we are able to write $I_{n,j}$ as
\begin{eqnarray}
I_{n,j}&=&\int_{-\infty}^\infty \frac{dp}{2\pi} \,
\big(-2\partial g_+(p)\,2\partial g_-(p)\big)^n \nonumber \\
&=&\int_{-\infty}^\infty \frac{dp}{2\pi} \,(-i)^n
\exp\Big({ inp^2R \left((k_+ \cos\theta_+)^{-1}-(k_- \cos\theta_-)^{-1}\right) }\Big)
\nonumber\\
&=&(-i)^n \sqrt{\frac{k_+\cos\theta_+k_-\cos\theta_-}
{4\pi n R (k_+ \cos\theta_+-k_- \cos\theta_-) } }
\exp{-i\pi/4}.
\end{eqnarray}
Then the contribution to the DOS oscillations originating from a creeping
orbit with $2n$ reflections and winding number $j$ can be written as
\begin{eqnarray}
\rho^{n,j}&=&
{\rm Re}\,
\left(
\frac{\cos\theta_+}{k_+}+\frac{\cos\theta_-}{k_-}
\right)
\sqrt{\frac{R^3 k_+\cos\theta_+\,k_- \cos\theta_-}
        {\pi n (k_+ \cos\theta_+-k_- \cos\theta_-) } }
(-i)^n
\exp\Big({iS_{\rm cl}(n,j,\Ef,E,R)-2in\varphi-i\pi/4}\Big)
\nonumber \\
&=&
{\rm Re}\,
\sqrt{ \frac{R^3\big(k_+-k_-\cos(\pi j/n)\big)(k_+\cos(\pi j/n)-k_-)}
{\pi n\,(\Delta k)^5\, k_+k_-} }\,
E\cos(\pi j/n)\,
\exp\Big({2in\,\Delta k\, R-2in\varphi-i\pi/4}\Big)\,.
\label{EQ:CrPO}
\end{eqnarray}
\par\noindent{\it (iii)~Summing over all creeping orbits\/}:
In order to obtain the creeping-orbit contribution to the DOS oscillations,
we must sum the contributions specified by Eq.~(\ref{EQ:CrPO}) for all $n$
and $j$, obeying the restriction given in Eq.~(\ref{EQ:restrict}).  Although
this summation may appear to be problematic, we remind the reader that $E$
has a nonzero imaginary part $\Gamma$, viz., the smoothing width, which
suppresses exponentially (in $n$) the contributions coming from high $n$
values.  Thus, for a given $\Gamma$ it is possible to perform the sum up to
a value of $n$ such that any contribution from higher $n$ would be invisible
on the scale set by the truncated sum.

\begin{figure}[hbt]
\epsfxsize=10.0cm
\centerline{\epsfbox{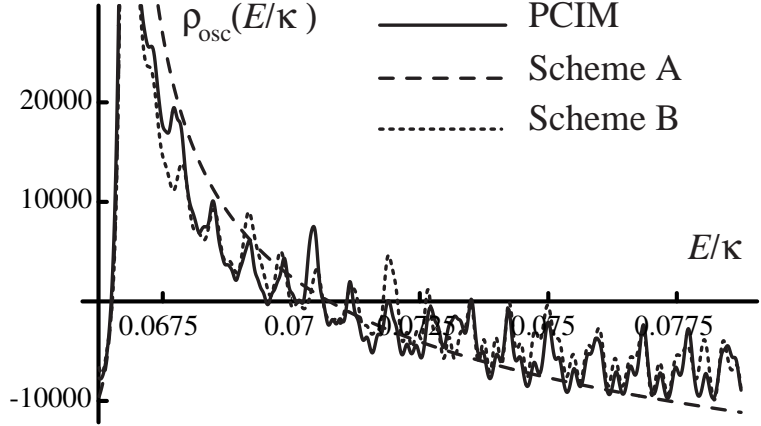}}
\vskip+0.4cm
\caption{Density of states oscillations for a circular Andreev billiard:
$\Kf R=150$; $\sudel/\Ef=0.08$; smoothing width $\Gamma/\Ef=1.1\times
10^{-4}$.}
\label{FIG:CircleDOS_model}
\end{figure}%
In Fig.~\ref{FIG:CircleDOS_model} we plot three versions of the
oscillatory part of the DOS:
the Scheme~A result (dashed line), which includes all chords
but is dominated by the stationary ones;
the Scheme~B result (dotted line),
which includes creeping orbits and stationary chords; and
the exact PCIM result (full line), obtained by the numerical
solution of the PCIM.
Observe that, although the local average behavior of the exact
DOS is essentially that captured by Scheme~A, in order to capture the
mesoscale oscillations beyond this average behavior one must use
Scheme~B.

\par\noindent{\it Poisson summation and the
semiclassical quantization condition:\/}
In the case of a circular \ab\ the summation over $n$ and $j$ can be
performed approximately to all orders.  The procedure for doing this
involves using the Poisson summation formula and evaluating the
integrals in the stationary phase approximation. By using this procedure
it is possible to obtain the energies at which the DOS has simple poles,
viz.,~it is possible to obtain a semiclassical quantization condition.
Our starting point is the expression Eq.~(\ref{EQ:CrPO}) for $\rho^{n,j}$.
In terms of this, the DOS oscillations can be written in the form
\begin{equation}
\rho
\approx
{\rm Re}\,\sum_{n,j}
\frac{a(\pi j/n)}{\sqrt{n}}
\exp\Big({i2 n \Delta k(\pi j/n) R-2in\varphi-i\pi/4}\Big),
\end{equation}
where the amplitude $a$ is defined via
\[
a(x)
\equiv
\sqrt{ \frac{R^3\big(k_+-k_-\cos(x)\big)\big(k_+\cos(x)-k_-\big)}
        {\pi \,\Delta k(x)^5\,\, k_+\,k_-} }
E \cos(x).
\]
By using the Poisson summation formula we first cast the expression for
the DOS in the form
\[
\rho
\approx
{\rm Re}\,\sum_{n,m}
\int dj\,
\frac{a(\pi j/n)}{\sqrt{n}}
\exp\Big({2in \Delta k(\pi j/n) R-2in\varphi-2i\pi m j-i\pi/4}\Big).
\]
Here, $m$ has an interpretation as the angular momentum quantum
number.  We then evaluate the $j$ integral in the stationary phase
approximation. The stationary phase points satisfy
\begin{eqnarray}
\cos\left(\frac{\pi j_c}{n}\right)&=&
\left(\lambda \pm
\sqrt{(1-\lambda)^2-\frac{E^2}{\Ef^2} }\,\right)
\left(1-\frac{E^2}{\Ef^2}\right)^{-1/2}
\nonumber \\
&\equiv& f(m,R,\Ef,E),
\end{eqnarray}
where $\lambda\equiv m^2/(\Kf R)^2$, and only real and positive values of
$\cos(\pi j_c/n)$ are allowed.  The important point to observe here is
that $\cos(\pi j_c/n)$ is independent of $n$, which implies that
$\Delta k$ is independent of $n$, too.  This allows us to write the
expression for $\rho$ as
\begin{eqnarray}
\rho
\approx
\sum_m
\big(\pi\Kf R (f(m)-\lambda)\big)^{-1/2}
a\left(\cos^{-1}f(m)\right)
{\rm Re}\sum_n
\romE^{in\left(2\Delta k(m)\,R-2\varphi-2m\cos^{-1}f(m)\right)}.
\end{eqnarray}
We are now in a position to perform the sum over $n$:
\begin{eqnarray}
\rho\approx
\sum_m
\big(\pi\Kf R (f(m)-\lambda)\big)^{-1/2}
a\left(\cos^{-1}f(m)\right)\,
{\rm Re}\,
\frac{\romE^{i\left(2\Delta k(m)\,R-2\varphi-2m\cos^{-1}f(m)\right)}}
{1-\romE^{i\left(2\Delta k(m)\,R-2\varphi-2m\cos^{-1}f(m)\right)}}.
\end{eqnarray}
Thus, the semiclassical approximation to the eigenenergies
for a given angular momentum quantum number $m$ (i.e.~the
DOS peaks) are given implicitly as the roots of
\begin{equation}
\exp{i\left(2\Delta k(m)\,R-2\varphi-2m\cos^{-1}f(m)\right)}=1.
\end{equation}
In Table~\ref{TAB:eigenvalue} we compare the eigenenergies of the PCIM
calculated via this semiclassical scheme and exactly.  From this Table
we see that, as expected, the semiclassical results agree with the
exact results upto contributions of relative order $1/\Kf R$.
\begin{table}
\caption{Comparison of the eigenenergies of the PCIM
for a circular \ab\ computed semiclassically and
computed exactly (numerically) for
$\Kf R=150$; $\sudel/\Ef=0.08$.
\label{TAB:eigenvalue}}
\begin{tabular}{lcccr}
$m$&$n$&$E_{m,n}/\Ef$\tablenote{PCIM}&$E_{m,n}/\Ef$\tablenote{Semiclassical}&$\Delta E/\Ef$\\
\tableline
0  &4&0.066694&0.066695& 0.000001\\
1  &4&0.066698&0.066696& 0.000002\\
2  &4&0.066699&0.066700&-0.000001\\
3  &4&0.066708&0.066707& 0.000001\\
4  &4&0.066716&0.066716& 0.000000\\
70 &4&0.073978&0.073967& 0.000011\\
71 &4&0.074204&0.074206&-0.000002\\
72 &4&0.074444&0.074450&-0.000006\\
73 &4&0.074711&0.074700& 0.000011\\
74 &4&0.074951&0.074954&-0.000003\\
110&3&0.067310&0.067285& 0.000025\\
111&3&0.067842&0.067880&-0.000038\\
112&3&0.068466&0.068495&-0.000029\\
113&3&0.069169&0.069132& 0.000037\\
114&3&0.069840&0.069791& 0.000049\\
\end{tabular}
\end{table}

\subsubsection{Incorporating ordinary reflection}
\label{SEC:IncorOR}
Thus far in our semiclassical treatment we have ignored all amplitudes
involving ordinary reflection.  For non-grazing incidence
[i.e.~for $\theta-(\pi/2)\sim 1$] the amplitude for ordinary reflection
is very small (in fact, of order $\sudel/\Ef\cos^2\theta$).
However, orbits that contribute dominantly to the oscillatory
structure of the DOS obey $\vert\theta-(\pi/2)\vert\ll 1$, and therefore
ordinary reflection amplitudes are not negligible and must be
incorporated.  This can be done by returning to Eq.~(\ref{EQ:appMRE})
and re-evaluating the trace formula using the full expression for
$\gf^{\rm R}$ (i.e.~not just the leading, off-diagonal, term).
Then, as a result of treating both charge-interconverting and
charge-preserving reflections, the classical limit changes drastically:
the initial conditions of specifying position and momentum (and charge,
in the case of {\ab}s) no longer determines the full orbit; on the
contrary, each reflection splits the incoming ray into two rays: one
(albeit imperfectly) retro-reflecting; the other, ordinarily
reflecting.  Thus, the classical dynamics is no longer deterministic
(i.e.~specifying the position and momentum no longer determines the
full orbit) and, instead, the initial conditions specify a
{\it superposition\/} of orbits.  (A similar situation emerges in the
context of Schr\"odinger billiards when the billiard has sharp jumps
in the single-particle potential~\cite{REF:AGraysplit}.)\thinspace\
Generically, the number of closed orbits increases exponentially as a
function of number of reflections (as opposed to linearly, as is the
case when ordinary reflection is neglected), but the amplitudes for
these orbits are suppressed exponentially (owing to the fact that the
amplitudes for ordinary and Andreev reflections are smaller than
unity), allowing the number of reflections to take care of any
divergences arising from this ray-splitting feature.

The exponential increase in the number of closed orbits, as a function
of the number of reflections, makes evaluation of the periodic-orbit
sum difficult.  For sufficiently smooth shapes there is a way to
circumvent this difficulty, which involves resorting to a different
approximation scheme, as we shall shortly show.  The main motivation
for this approximation scheme is as follows.
\par\noindent
1.~The closer a primitive creeping orbit is to the boundary (i.e.~the
closer $\vert\theta-(\pi/2)\vert$ is to zero), the shorter it its total
length.  As the periodic-orbit contributions are suppressed
exponentially with their lengths, the creeping orbits that are closer
to the boundary contribute more strongly to the DOS oscillations.
\par\noindent
2.~The closer a creeping orbit is to the boundary, the bigger the
ordinary reflection amplitude ($\sim\sudel/\Ef\cos^2\theta$.)
Thus, the orbits that involve ordinary reflection contribute
more strongly if they are close to the boundary.
\par\noindent
3.~For the orbits close to the boundary (which, by the previous
considerations, dominate), consecutive reflections take place very
near to each other, and thus \lq\lq see\rq\rq\ only the local
curvature of the boundary.
\par\noindent
These considerations motivate us to perform an \lq\lq adiabatic\rq\rq\
approximation to the expansion in Eq.~(\ref{EQ:appMRE}), in which we
assume that the curvature of the boundary varies slowly, relative to
the rate at which creeping orbits sample the boundary.

Our starting point is the integral equation~(\ref{EQ:EffIter}) that generates the MRE:
\begin{eqnarray}
\gf^{\rm ii}({\bf x},{\bf x}^{\prime})
&\equiv&
\gf^{\rm N}({\bf x},{\bf x}^{\prime})+
\int_{\surS}d\sigma_{\alpha}\,
\partial_{\alpha}\,\gf^{\rm N}({\bf x},{\bbox{\alpha}})\cdot
        {\bbox{\mu}}^{\rm ii}({\bbox{\alpha}},{\bf x}^{\prime}),
\nonumber \\
{\bbox{\mu}}^{\rm ii}
&\approx&
 2\gf^{\rm R}
+2\partial\gf^{\rm R}
    {\bbox{\mu}}^{\rm ii},
\end{eqnarray}
where $\gf^{\rm R}$ is given in Eq.~(\ref{EQ:GFc}).
These equations can be cast to the following form:
\begin{eqnarray}
\gf^{\rm ii}
\equiv
\gf^{\rm N}+2
\partial\gf^{\rm N}\,
\left(1-2\partial\gf^{\rm R}
\right)^{-1}
\,\gf^{\rm R},
\end{eqnarray}
which shows that the approximate poles of $\gf^{\rm ii}$ are
given by the poles of
${\bf K}\equiv\left(1-2\partial\gf^{\rm R}\right)^{-1}$.
Thus, in order to obtain the energy eigenvalues it is sufficient to
consider the following integral equation defined on the surface $\surS$:
\[
{\bf K}(\bbox{\alpha},\bbox{\beta})=
{\bf I}\,\delta(\bbox{\alpha},\bbox{\beta})+
2\int_\surS \!\!\surel_\gamma\,
\partial\gf^{\rm R}(\bbox{\alpha},\bbox{\gamma})\,
{\bf K}(\bbox{\gamma},\bbox{\beta}).
\]
This equation will have a regular solution (i.e.~one without poles) if and
only if none of the eigenvalues of the operator $2\partial\gf^{\rm R}$ is
equal to unity; conversely, poles of ${\bf K}$ occurs at energies for
which at least one of the eigenvalues of $2\partial\gf^{\rm R}$ is equal to
unity.  Consequentially, we now focus on the following eigenvalue problem defined on
the boundary of the billiard:
\begin{equation}
2\int_\surS\surel_\beta\,
\partial\gf^{\rm R}(\bbox{\alpha},\bbox{\beta})\,
{\bf u}(\bbox{\beta})=\lambda\,{\bf u}(\bbox{\alpha}).
\label{EQ:eigenboun}
\end{equation}
We shall work in the coordinate system in which the boundary is
parametrized by its arc length $s$. (Recall that we are considering 2D billiards.)
Thus, the equation for the
boundary is given by the vector function $\bbox{\alpha}(s)$,
\begin{equation}
{\bf t}(s)\equiv{d\bbox{\alpha}}/{ds}
\end{equation}
is the tangent vector,
\begin{equation}
{\bf n}(s)\equiv -R(s)\,{d^2\bbox{\alpha}}/{ds^2}
\end{equation}
is the unit normal vector, and
\begin{equation}
R(s)\equiv\Big\vert{d^2\bbox{\alpha}}/{ds^2}\Big\vert^{-1}
\end{equation}
is the curvature at the point $\bbox{\alpha}(s)$.
Furthermore, we shall transform to the coordinates
\begin{equation}
t\equiv s-s'\qquad {\rm and}
\qquad
S\equiv\frac{s+s'}{2},
\end{equation}
and define
\begin{equation}
\bar{\gf}^{\rm R}(t,S)
\equiv
\gf^{\rm R}(s,s').
\end{equation}
The virtue of this coordinate system is that for a constant-curvature
boundary (viz.~a circle) $\gf^{\rm R}(s,s')=\bar{\gf}^{\rm R}(s-s')$.
Thus, if the curvature is slowly varying then
$\bar{\gf}^{\rm R}(t,S)$ is a slowly varying function of $S$.
In this coordinate system, then, the eigenvalue equation
Eq.~(\ref{EQ:eigenboun}) becomes
\begin{mathletters}
\begin{eqnarray}
2\int_0^L dt\,
\partial\bar{\gf}^{\rm R}(t,s-t/2)\,{\bf u}(s-t)
&=&
\lambda\,{\bf u}(s),
\label{EQ:surEVeq}
\\
{\bf u}(0)
&=&
{\bf u}(L),
\end{eqnarray}%
\end{mathletters}%
where $L$ is the length of the boundary.  From periodic-orbit theory we
already know that the dominant mesoscale oscillations in the DOS arise
from the part of phase space in which the component of the excitation
momentum lying tangent to the boundary is ${\cal O}(\Kf)$.  Thus, in
order to capture the dominant mesoscale oscillations, it is sufficient
to solve the eigenvalue equation~(\ref{EQ:surEVeq}) for the sector of eigenfunctions
varying on the length scale ${\cal O}(\Kf)$.  In this sector of
rapidly-varying eigenfunctions, only the small-$t$ behavior of the
kernel $\partial\bar{\gf}^{\rm R}(t,S)$ is relevant.  In the following,
we shall obtain an expression for $\partial\bar{\gf}^{\rm R}(t,S)$
valid for small $t$.

Our starting point is the general expression for $\gf$ in polar coordinates,
\begin{equation}
\gf(s,s')=
\sum_m
{\bf R}\cdot
\pmatrix{\frac{i}{4}J_m(k_+r_<)H_m(k_+r_>)
& 0 \cr 0 & -\frac{i}{4}J_m(k_-r_<)H_m(k_-r_>)}
\romE^{im\Theta},
\label{EQ:AMexp}
\end{equation}
where ${\bf R}\big(m/R(s)\big)$ is defined in Eq.~(\ref{EQ:AfromG}).
For $s$ near $s'$ (i.e.~for small $t$), one can choose the
\lq\lq polar\rq\rq\ coordinate system, in which a circle of radius
$R\big((s+s')/2\big)$ coincides locally with the surface at point
$(s+s')/2$.  Then the expansion~(\ref{EQ:AMexp}) can be written as
\begin{equation}
\bar{\gf}(t,S)
\approx
\sum_m
{\bf R}\cdot
\pmatrix{\frac{i}{4}J_m(k_+R(S))H_m(k_+R(S)) & 0
\cr 0 & -\frac{i}{4}J_m(k_-R(S))H_m(k_-R(S))
}\romE^{imt/R(S)}.
\end{equation}
The normal derivative $\partial\gf$ is given by
\begin{equation}
\partial\bar{\gf}(t,S)
\approx
\sum_m
{\bf R}\cdot\frac{i}{4}
\pmatrix{\frac{d}{dR}\big(J_m(k_+R)H_m(k_+R)\big)\Big\vert_{R=R(S)} & 0
\cr 0 & -\frac{d}{dR}\big(J_m(k_-R)H_m(k_-R)\big)\Big\vert_{R=R(S)}}
\romE^{imt/R(S)}.
\label{EQ:bounGF}
\end{equation}
Having obtained the approximate form of $\partial\bar{\gf}$,
valid for small $t$, we shall now seek the eigenfunctions.  We shall
assume that the rapidly-varying eigenfunctions have the following form:
\begin{equation}
{\bf u}(s)=\bar{\bf u}(s)\,\exp{ims/R(s)},
\end{equation}
{\it viz}., a slowly-varying envelope and a rapidly varying part.
Then the eigenvalue equation becomes
\begin{mathletters}
\begin{eqnarray}
\lambda\,\bar{\bf u}(s)\,
\romE^{im(s)/R(s)}
&=&
2\int_0^{L} dt\,
\partial\bar{\gf}^{\rm R}(t,s-t/2)\,
\bar{\bf u}(s-t)\,
\romE^{im(s-t)/R(s-t)},
\\
\lambda\,\bar{\bf u}(s)
&\approx&
2\int_0^{L} dt\,
\partial\bar{\gf}^{\rm R}(t,s)\,
\romE^{-imt/R(s)}\,
\bar{\bf u}(s),
\\
\lambda(m,s)\,
\bar{\bf u}(s)
&=&
2\int_0^{L} dt\,
\partial\bar{\gf}^{\rm R}(t,s)\,
\romE^{-imt/R(s)}\,\bar{\bf u}(s),
\\
\lambda(m,s)\,\bar{\bf u}(s)
&=&
\sum_{m'}2\partial\bar{\gf}^{\rm R}(m',s)
\int_0^{2\pi R(s)} dt\,
\romE^{-i(m-m')t/R(s)}
-2\partial\bar{\gf}^{\rm R}(m',s)\int_0^{2\pi R(s)-L} dt\,
\romE^{-i(m-m')t/R(s)}\,\bar{\bf u}(s),
\\
\lambda(m,s)\,\bar{\bf u}(s)
&=&
\sum_{m'}2\partial\bar{\gf}^{\rm R}(m',s)
 2\pi R(s)\delta_{m,m'}
-2\partial\bar{\gf}^{\rm R}(m',s)
\int_0^{2\pi R(s)-L} dt\,\romE^{-i(m-m')t/R(s)}
\,\bar{\bf u}(s),
\\
&\approx&
\sum_{m'}2\partial\bar{\gf}^{\rm R}(m',s)
\left(2\pi R(s)\delta_{m,m'}-(2\pi R(s)-L)\right)
\,\bar{\bf u}(s),
\\
\lambda(m,s)\,
\bar{\bf u}(s)
&\approx&
2\partial\bar{\gf}^{\rm R}(m,s)\,
2\pi R(s)\,
\bar{\bf u}(s)
+{\cal O}
\left(\left\vert{dR}/{ds}\right\vert\,L\right).
\end{eqnarray}%
\end{mathletters}%
By ignoring the term
${\cal O}\left(\left\vert{dR}/{ds}\right\vert\,L\right)$
we have reduced the eigenvalue equation to
\begin{equation}
i\pi R(s)\,{\bf R}\big(m/R(s)\big)
\cdot
\pmatrix{\frac{d}{dR}\big(J_m(k_+R)H_m(k_+R)\big)\Big\vert_{R=R(s)} & 0
\cr 0 & -\frac{d}{dR}\big(J_m(k_-R)H_m(k_-R)\big)\Big\vert_{R=R(s)}}
\bar{\bf u}(s)=
\lambda(m,s)\,\bar{\bf u}(s),
\label{EQ:FinEP}
\end{equation}
The important point to note about Eq.~(\ref{EQ:FinEP}) is that it
is simply a $2\times 2$ matrix eigenvalue equation and, thus, its
solution is straightforward.  After obtaining the eigenvalues
$\lambda(m,s)$, the DOS is given by the following formula:
\begin{equation}
\rho(E)=\int_0^L \frac{ds}{2\pi R(s)} \sum_m \delta\big(1-\lambda(m,s;E)\big)
\left\vert\frac{\partial\lambda(m,s;E)}{\partial E}\right\vert
\end{equation}

\begin{figure}[hbt]
\epsfxsize=10.0cm
\centerline{\epsfbox{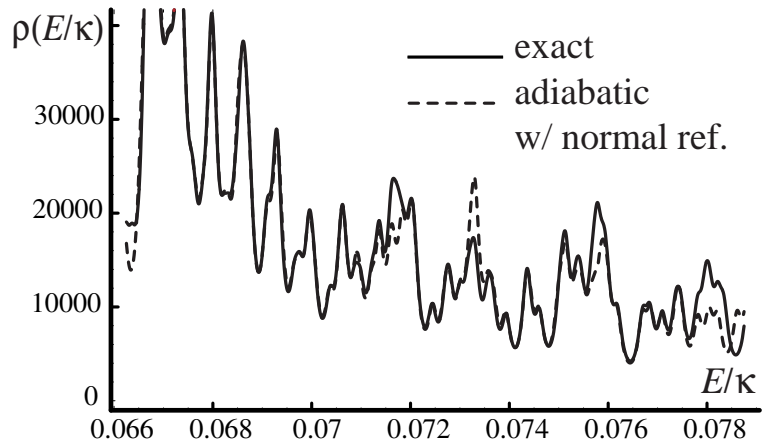}}
\vskip+0.4truecm
\caption{Density of states oscillations for a circular Andreev Billiard:
$\Kf R=150$; $\sudel/\Ef=0.08$; smoothing width $\Gamma/\Ef=1.1\times
10^{-4}$.}
\label{FIG:normal_ref}
\end{figure}%
In Fig.~\ref{FIG:normal_ref} the DOS arising from the solution
of the eigenvalue equation Eq.~(\ref{EQ:FinEP}) is compared to the
the one arising from the exact solution of the full \bdg\
eigenproblem for the case of a circular Andreev billiard, which we have
computed numerically.

In this section we have shown that mesoscale oscillations in DOS
essentially arise from imperfectness in retro-reflection.  However,
in order to correctly account for these oscillations it is necessary
to account for the effects of imperfect charge-interconversion as
well.  The latter effects can be incorporated via an extension of the
trace formula, in which a generic closed periodic orbit has both
charge-interconverting and charge-preserving reflections.  However,
for billiard shapes having slowly-varying curvatures, it is possible
to obtain an adiabatic approximation to the DOS that bypasses the
periodic-orbit summation (which has a number of terms that increases
exponentially with the number of reflections) and, hence, reduces
the task to the solution of $2\times 2$ matrix eigenvalue equation.
The strength of this method is that it relates the DOS to the closed
form of ${\bf R}\big(m/R(s)\big)$ and, thus, it is readily extendible
to cases in which the N-S boundary is not clean (i.e.~reflection
amplitudes are modified).  As seen above, in order to calculate the
DOS one simply needs the reflection amplitudes at the billiard boundary.
\section{Concluding remarks; perspectives}
\label{SEC:Summ}
%
In the present Paper we have explored semiclassical approaches to the
oscillatory part of the density of states of Andreev billiards.
We have done this
by deriving two semiclassical trace formulas, each corresponding to
one of two limiting schemes of the physical parameters specifying the
billiard.
The first of these trace formulas (viz.~the Scheme~A trace formula,
discussed in Sec.~\ref{SEC:SAandreev}) is essentially equivalent to
the conventional quasiclassical approximation scheme first introduced
by Andreev.
The physical ingredients of this scheme are perfect charge-interconversion
and perfect retro-reflection.  It captures the coarsest oscillations in
the DOS.
The second trace formula (viz.~the Scheme~B trace formula, discussed in
Sec.~\ref{SEC:SchemeB}) not only captures the coarsest oscillation, but
also goes beyond this resolution by capturing mesoscale oscillations.
At the semiclassical level, mesoscale oscillations arise from orbits
featuring imperfect retro-reflection.  Although such oscillations are
sensitive to charge-preserving reflection amplitudes (in addition to
charge-interconverting reflection amplitudes) from the N-S interface,
they are present even if there is no charge-preserving reflection.

The methods developed in the present Paper readily apply to settings
such as the superconducting proximity effect or the Josephson effects.
Cases in which the phase of the pair-potential is relevant can be
addressed by the appropriate modification of the renormalized Green
function for complex $\sudel$.  In particular, by using the methods
described in Sec.~\ref{SEC:MRE}, it is possible to show that the
leading-order behavior of $\gf^{\rm R}$ is modified
[cf.~Eq.~(\ref{EQ:GenGeff_PCIM})], becoming
\begin{equation}
\gf^{\rm R}({\bbox{\alpha}},{\bf x}^{\prime})
\approx\exp\left({-i\varphi+i\phi}\right)
\pmatrix{
0 & -g^{\rm N}_{-}({\bbox{\alpha}},{\bf x}^{\prime})
\cr
\noalign{\medskip}
g^{\rm N}_{+}({\bbox{\alpha}},{\bf x}^{\prime})& 0
},
\label{EQ:GenGeff_PCIM_phase}
\end{equation}
where the phase $\varphi\equiv\cos^{-1}(E/\vert\sudel\vert)$ and
the phase $\phi$ is the phase of the pair-potential.  Thus, as one
might have anticipated, the phase of the pair-potential simply adds
to the phase acquired by the reflection.  By using this modified
form for $\gf^{\rm R}$ one can, e.g., account for the zero-energy
states observed in $\pi$-junctions (i.e.~Josephson junctions through
which the pair-potential undergoes a single sign-change).

The Multiple Scattering Expansion that we have developed applies to
systems consisting of piecewise homogeneous N or S regions.  It is
possible to generalize this expansion to handle smoothly-varying
$\sudel({\bf x})$.  This can be achieved via an
\lq\lq energy-slicing\rq\rq\ construction (rather than the time-slicing
kind familiar from, say, the derivation of the Feynman path integral).
In this way, one arrives at a functional version of the Multiple
Scattering Expansion.  To get a feeling why, let us divide the
pair-potential range $(0,\sudel_0)$
[where $\sudel_0\equiv{\rm max}\,\sudel({\bf x})$]
into a large number $N$ of equally-spaced subintervals.
Let us also break the ${\bf x}$ space into subregions, in each of which
the pair-potential has values lying in solely one of the $N$ energy
subintervals.  For $N$ large, $\sudel({\bf x})$ can be taken to be constant
in each subregion.  Then we can apply the Multiple Scattering Expansion
formalism for this intermediate system to obtain its
Bogoliubov-de~Gennes Green function.
Then, by taking the limit $N\rightarrow\infty$,
one recovers the full Bogoliubov-de~Gennes
Green function associated with $\sudel({\bf x})$.
Other spatial inhomogeneities, such as in the single-particle potential,
can also be handled this way.  Semiclassical approximations, as discussed
here, can be applied to this expansion.  For systems that have distinct
regions of N and S, but in which these regions are modified from the ideal
piecewise-homogeneous state because they feature a smooth variation of pair-potential, can still be regarded as {\ab}s.  In fact, as the amplitude
for charge-preserving reflection is expected to diminish for more slowly
varying pair-potentials, we expect the Perfectly Charge-Interconverting Model (PCIM) to be more accurate for such systems.  On the other hand, billiards constructed by fabricating a normal region inside a superconductor should
feature more charge-preserving reflection, owing to interface effects.
Such cases can be modelled by suitably modifying the reflection matrix
${\bf R}$ in the definition of the renormalized Green function $\gf^{\rm R}$.

Apart from problems related to superconductivity, the methods presented here
are also applicable to more conventional topics in quantum chaos.  One such application is to the so-called Ray-splitting billiards.  These billiards
consist of regions of piecewise homogeneous (single-particle) potential,
and the (sharp) boundaries between these regions serve as ray-splitting
boundaries.  By changing the homogeneous N and S Green functions to the
homogeneous Helmholtz Green functions appropriate for a given constant
potential, the formulation in the present Paper is readily
extended~\cite{REF:AGraysplit}.
Another potential application is to multidimensional tunneling, studied in
path-integral language in Ref.~\cite{REF:MD_tunneling}.  The Green-function
language adopted in the present Paper is especially well-suited to the
study of tunneling, owing to the energy (rather than time)
dependence of the Green function.

One experimentally relevant application of this work is to antidot {\ab}s.
Such billiards (in particular their magnetotransport properties) were
recently studied experimentally by Eroms et al.~\cite{REF:antidot}.
In this realization of Andreev billiards, it is the S region that is embedded
in the N region, rather than the converse.  (An experimental virtue of this
geometry is that it is well suited to the study of the effects of weak
magnetic fields, as the magnetic flux need not be quantized.)\thinspace\
The methods presented in th present Paper are readily applicable to
this antidot geometry.  In particular, in App.~\ref{APP:nonconvex}, we
describe the new features that emerge when the N region is nonconvex,
as it is in antidot billiards.   Moreover, the incorporation of magnetic
fields--at least in the weak field case, so that it is simply excluded
from the S region--is handled by modifying the Green functions ($\gf^{\rm N}$
and $\gf^{\rm S}$) to include the magnetic field.  As the electromagnetic
vector potential further increase the difference between the action (or
accumulated phase) of electrons and holes, it will increase the degree of
imperfectness in the retro-reflection.

The present work, and in particular approximation Scheme~B, provides insight
into the general question of when electron dynamics should be handled
separately from the hole dynamics in inhomogeneous superconductors.  In doing
this, it also provides a semiclassical framework for studying the effects of
electron/hole symmetry breaking beyond Andreev approximation.  From the point
of view of quantum chaos, such electron/hole differences lead to the novel
dynamics featuring in the present work.  Phenomena associated with this should
be accessible via experiment and, indeed, the extremely recent Regenbsurg
experiments are paving the way to a thorough experimental exploration of
the novel quantal dynamics of electrons and holes in Andreev billiards.
\acknowledgments
We gratefully acknowledge useful discussions with
Eric Akkermanns,
Dmitrii Maslov, and
Michael Stone.
This material is based upon work supported by the
U.S.~Department of Energy, Division of Materials
Sciences under Award No.~DEFG02-96ER45439,
through the Frederick Seitz Materials Research
Laboratory at the University of Illinois at Urbana-Champaign,
and by the U.S.~National Science Foundation through Award
No.~NSF-DMR-99-75187.
\appendix
\section{Application of boundary integral technique to the
Helmholtz and \bdg\ wave equations}
\label{APP:BasicIngredients}
The underlying strategy employed in this Paper is the boundary
integral technique~\cite{REF:OnBIE}, the origin of which was Fredholm's
analysis of the existence of soluions of the interior Laplace problem
subject to Dirichlet boundary conditions.  In this scheme, Fredholm
transformed the task of solving the Laplace partial differential equation
(subject to Dirichlet boundary conditions) to one of solving a certain
integral equation residing on the boundary.  This prompted Fredholm
to develop the theory of what are now known as Fredholm integral equations
and, in particular, to prove the existence of a solution of the corresponding
Laplace problem.

In the present context of spectral geometry, the virtue of this boundary integral
technique is that it allows one to harness the piecewise homogeneity of the system
(and the corresponding simplicity of the fundamental Green functions in the
locally homogeneous regions) and, thereby, to study the physical implications
of the boundary in as direct and a natural manner as possible.

The aim of this appendix is to provide a guide to the boundary integral technique,
beginning with the simplest setting and working towards the setting of
the \bdg\ eigenproblem.  When discussing the simplest settings we shall
borrow heavily from Refs.~\cite{REF:Kellog,REF:GueLee}.  The elaborations
that we shall be needing for the \bdg\ setting arise from (i)~the multicomponent
nature of the eigenproblem, and (ii)~the presence of matching rather than
boundary conditions.  We mention that Jackson~\cite{REF:JDJackson} gives a highly
readable discussion of the physics of the potential discontinuities that are a
pivotal feature of boundary integral techniques.
\subsection{Review of elementary ingredients}
In this section we shall discuss the origin of the discontinuities
in the three-dimensional potential and the fields generated by surface charge (which we call
single-layer) and dipole (which we call double-layer) densities,
as well as present derivations of explicit formulas quantifying such discontinuities.
Such formulas will become useful in the next section, where we discuss
parametrizations of wave functions in terms of these single and double layers.

Before turning to the derivation, we consider a simple example which contains
the essential features: a planar charge layer with constant charge density
$\nu$. Without loss of generality, let us assume that the charge layer lies in the
$xy$ plane, so that the normal direction ${\bf n}$is $\hat{\bf z}$. Then the potential
$\varphi$ and the field ${\bf E}$ are given by the following surface integrals:
\begin{mathletters}
\begin{eqnarray}
4\pi\varphi({\bf x})&=&\phantom{-}\nu\int \surel_{\alpha}\,\frac{1}{\vert{\bf x}-\bbox{\alpha}\vert},
\\
4\pi{\bf E}({\bf x})&=&-\nu\int \surel_{\alpha}\,\bbox{\nabla}_x
                    \left(\frac{1}{\vert{\bf x}-\bbox{\alpha}\vert}\right)
=\nu \int \surel_{\alpha}\,\frac{{\bf x}-\bbox{\alpha}}{\vert{\bf x}-\bbox{\alpha}\vert^3},
\end{eqnarray}%
\end{mathletters}%
where ${\bf x}\equiv(x,y,z)$, $\bbox{\alpha}\equiv(x',y',0)$, and $\surel_{\alpha}\equiv dx'\,dy'$.
Owing to translational invariance in the $xy$ plane, $\varphi({\bf x})=\varphi(z)$ and
${\bf E}({\bf x})=\hat{\bf z}E(z)$. Without loss of generality, let us choose $x=y=0$, and focus
on $E(z)$. By introducing the polar coordinates $(x',y')=(r\cos\theta,r\sin\theta)$, the integral
for $E(z)$ becomes
\begin{equation}
E(z)=\nu \int r\,dr\,d\theta\,\frac{z}{4\pi(z^2+r^2)^{3/2}}=
\frac{\nu}{2} \int_0^\infty rdr\,\frac{z}{(z^2+r^2)^{3/2}}=\frac{\nu}{2}\,{\rm sgn}(z)
\label{EQ:plane_disc}
\end{equation}
For $z>0$, $E(z)$ is independent of $z$ and equal to $\nu/2$;
for $z<0$, $E(z)$ is independent of $z$, however it is equal to $-\nu/2$
and for $z=0$, $E(z)$ vanishes. The mathematical origin of this discontinuity is
the noncommutativity of the limit $z\rightarrow 0$ and the surface integral in
Eq.~(\ref{EQ:plane_disc}). Thus the value of $\varphi$ is continuous as it approaches the
surface but its normal derivative (in this case partial derivative with respect to $z$)
is not. This discontinuity can be summarized by the equation
\begin{eqnarray}
&&\lim_{z\rightarrow 0^+}\varphi=\varphi\vert_{z=0},\qquad \lim_{z\rightarrow 0^-}\varphi=\varphi\vert_{z=0},
\\
&&\lim_{z\rightarrow 0^+}\frac{\partial \varphi}{\partial z}
=-\frac{\nu}{2}+\frac{\partial \varphi}{\partial z}\Big\vert_{z=0}, \qquad
\lim_{z\rightarrow 0^-}\frac{\partial \varphi}{\partial z}
=\frac{\nu}{2}+\frac{\partial \varphi}{\partial z}\Big\vert_{z=0}.
\end{eqnarray}

Now consider a generic charge layer $\nu(\bbox{\alpha})$ on the surface $\surS$, and focus on the
potential generated by via the Helmholtz (instead of Coulomb) Green function
\begin{equation}
\varphi({\bf x})=\int_{\surS}\surel_{\alpha} \,g^{\rm H}({\bf x},\bbox{\alpha})\,\nu(\bbox{\alpha}).
\end{equation}
As in the case of the simple example of homogeneous planar charge layer, the potential generated by this generic
charge layer is continuous,
\begin{equation}
\lim\limits_{{\bf x}\in\notV\rightarrow{\bbox{\beta}}\in\surS}\varphi({\bf x})
=\int_{\surS}\surel_{\alpha} \,g^{\rm H}(\bbox{\beta},\bbox{\alpha})\,\nu(\bbox{\alpha}),
\end{equation}
but its normal derivative is not. In order to see this, consider the normal
derivative ${\bf n}_\beta\cdot\bbox{\nabla}_x\,\varphi({\bf x})$ as ${\bf x}$ in $\notV$  tends to a generic
point on $\surS$, which we denote by $\bbox{\beta}$, along the interior normal and divide the
domain of the surface integration into two parts: (i)~a small region
${\cal D}_\delta\equiv{\cal C}_\delta\bigcap\surS$, where ${\cal C}_\delta$ is a
sphere of radius $\delta$ around $\bbox{\beta}$ and (ii)~remaining domain
$\overline{\cal D}_\delta\equiv\surS-{\cal D}_\delta$, then
\begin{equation}
\lim\limits_{{\bf x}\in\notV\rightarrow{\bbox{\beta}}\in\surS}{\bf n}_\beta\cdot\bbox{\nabla}_x\,\varphi({\bf x})=
\lim_{\epsilon\rightarrow 0}\int_{{\cal D}_\delta}\surel_{\alpha}
    \,\frac{\partial}{\partial \epsilon}\,g^{\rm H}(\bbox{\beta}+\epsilon\,{\bf n}_\beta,\bbox{\alpha})\,\nu(\bbox{\alpha})
+\lim_{\epsilon\rightarrow 0}\int_{\overline{\cal D}_\delta}\surel_{\alpha}
    \,\frac{\partial}{\partial \epsilon}\,g^{\rm H}(\bbox{\beta}+\epsilon\,{\bf n}_\beta,\bbox{\alpha})\,\nu(\bbox{\alpha})
\label{EQ:seper}
\end{equation}
where $\epsilon$ is the perpendicular distance between ${\bf x}$ and $\bbox{\beta}$. The virtue of this separation
is that the singularity of $g^{\rm H}(\bbox{\beta},\bbox{\alpha})$ at $\bbox{\beta}=\bbox{\alpha}$ is now contained
in the first term on the right hand side of the Eq.~(\ref{EQ:seper}). As the integrand of second term on the right
hand side of this equation is free of singularities, the limit can be taken inside the integral sign. Moreover
within ${\cal D}_\delta$ and for $\delta$ and $\epsilon$ very small,
\begin{equation}
\frac{\partial}{\partial \epsilon}\,g^{\rm H}(\bbox{\beta}+\epsilon\,{\bf n}_\beta,\bbox{\alpha})
\approx \frac{\epsilon}{4\pi(\rho^2+\epsilon^2)^{3/2}}\,\,,
\qquad
\rho\equiv\vert\bbox{\alpha}-\bbox{\beta}\vert^2.
\end{equation}
Then the integral over the domain ${\cal D}_\delta$ can be evaluated for small $\delta$ as follows
\begin{equation}
\lim_{\epsilon\rightarrow 0}\int_{{\cal D}_\delta}\surel_{\alpha}
    \,\frac{\partial}{\partial \epsilon}\,g^{\rm H}(\bbox{\beta}+\epsilon\,{\bf n}_\beta,\bbox{\alpha})\,\nu(\bbox{\alpha})
=\lim_{\epsilon\rightarrow 0}\nu(\bbox{\beta})
\int_0^{2\pi} d\theta\int_0^\delta \rho\,d\rho \frac{\epsilon}{4\pi(\rho^2+\epsilon^2)^{3/2}}
=\frac{1}{2}\nu(\bbox{\beta}).
\label{EQ:sing_part}
\end{equation}
Now that we have evaluated the singular part of the surface integral we take the limit $\delta$ tends to zero:
\begin{eqnarray}
\lim\limits_{{\bf x}\in\notV\rightarrow{\bbox{\beta}}\in\surS}{\bf n}_\beta\cdot\bbox{\nabla}_x\,\varphi({\bf x})
&=&\frac{1}{2}\nu(\bbox{\beta})+\lim_{\delta\rightarrow 0}
\int_{\overline{\cal D}_\delta}\surel_{\alpha}\,
{\bf n}_{\beta}\cdot\bbox{\nabla}_{\beta}\,g^{\rm H}(\bbox{\beta},\bbox{\alpha})\,\nu(\bbox{\alpha})
\\
&=&\frac{1}{2}\nu(\bbox{\beta})+
\int_{\surS}\surel_{\alpha}\,
{\bf n}_{\beta}\cdot\bbox{\nabla}_{\beta}\,g^{\rm H}(\bbox{\beta},\bbox{\alpha})\,\nu(\bbox{\alpha}).
\end{eqnarray}
We note that this discontinuity originates again from the noncommutativity of limit ${\bf x}\rightarrow{\bbox{\beta}}$
and the surface integral. Moreover the calculation of the amount of this discontinuity involves only the form
of $g^{\rm H}$ for small values of distance between its arguments and this form is simply the Coulomb Green function.
Thus, in essence, this discontinuity is the same as the discontinuity in the simple case of a planar,
electrostatic-charge layer discussed at the beginning of this section.

Let us now consider the potential generated by a generic dipole density $\mu(\bbox{\alpha})$ on $\surS$, viz.
\begin{equation}
\varphi({\bf x})=\int_{\surS}\surel_{\alpha} \,{\bf n}_\alpha\cdot\bbox{\nabla}_\alpha\,g^{\rm H}({\bf x},\bbox{\alpha})\,\mu(\bbox{\alpha}).
\end{equation}
Notice the similarity of this form to that of the normal derivative of the potential generated by the charge layer,
only difference being the normal derivative acting on the second rather than the first argument. However, as the
$\gf^{\rm H}$ is a function of the difference between its arguments, one expects a similar discontinuity on the
{\it potential} across the surface. Let us now consider the case in which ${\bf x}$ approaches to a surface
point denoted by $\bbox{\beta}$ from $\notV$. Indeed,
\begin{eqnarray}
\lim\limits_{{\bf x}\in\notV\rightarrow{\bbox{\beta}}\in\surS}\varphi({\bf x})&=&
\lim\limits_{{\bf x}\in\notV\rightarrow{\bbox{\beta}}\in\surS}
\int_{\surS}\surel_{\alpha} \,{\bf n}_\alpha\cdot\bbox{\nabla}_\alpha\,
g^{\rm H}({\bf x},\bbox{\alpha})\,\mu(\bbox{\alpha})
\\
&=&\lim\limits_{{\bf x}\in\notV\rightarrow{\bbox{\beta}}\in\surS}
\int_{{\cal D}_\delta}\surel_{\alpha} \,{\bf n}_\alpha\cdot\bbox{\nabla}_\alpha\,
g^{\rm H}({\bf x},\bbox{\alpha})\,\mu(\bbox{\alpha})
+\lim\limits_{{\bf x}\in\notV\rightarrow{\bbox{\beta}}\in\surS}
\int_{\overline{\cal D}_\delta}\surel_{\alpha} \,{\bf n}_\alpha\cdot\bbox{\nabla}_\alpha\,
g^{\rm H}({\bf x},\bbox{\alpha})\,\mu(\bbox{\alpha})
\\
&=&-\lim\limits_{{\bf x}\in\notV\rightarrow{\bbox{\beta}}\in\surS}
\int_{{\cal D}_\delta}\surel_{\alpha} \,{\bf n}_\beta\cdot\bbox{\nabla}_x\,
g^{\rm H}({\bf x},\bbox{\alpha})\,\mu(\bbox{\alpha})
+\lim\limits_{{\bf x}\in\notV\rightarrow{\bbox{\beta}}\in\surS}
\int_{\overline{\cal D}_\delta}\surel_{\alpha} \,{\bf n}_\alpha\cdot\bbox{\nabla}_\alpha\,
g^{\rm H}({\bf x},\bbox{\alpha})\,\mu(\bbox{\alpha}),
\label{EQ:sim_disc}
\end{eqnarray}
where in order to get to the third line we have used
$\bbox{\nabla}_x g^{\rm H}({\bf x},{\bf x}')=-\bbox{\nabla}_{x'} g^{\rm H}({\bf x},{\bf x}')$,
and that in ${\cal D}_\delta$, as $\delta$ goes to zero, ${\bf n}_\alpha\rightarrow {\bf n}_\beta$.
Notice that the first term in the right hand side of Eq.~(\ref{EQ:sim_disc}) is equal to the first
term in Eq.~(\ref{EQ:sing_part}). Thus in the limit $\delta$ goes to zero, we have:
\begin{eqnarray}
\lim\limits_{{\bf x}\in\notV\rightarrow{\bbox{\beta}}\in\surS}\varphi({\bf x})
&=&-\frac{1}{2}\mu(\bbox{\beta})
+\int_{\surS}\surel_{\alpha} \,{\bf n}_\alpha\cdot\bbox{\nabla}_\alpha\,
g^{\rm H}(\bbox{\beta},\bbox{\alpha})\,\mu(\bbox{\alpha}).
\end{eqnarray}
The discontinuity in $\varphi$ as ${\bf x}$ approaches to $\bbox{\beta}$ from $\volV$
(rather than $\notV$), is obtained similarly, the result is:
\begin{eqnarray}
\lim\limits_{{\bf x}\in\volV\rightarrow{\bbox{\beta}}\in\surS}\varphi({\bf x})
&=&\frac{1}{2}\mu(\bbox{\beta})
+\int_{\surS}\surel_{\alpha} \,{\bf n}_\alpha\cdot\bbox{\nabla}_\alpha\,
g^{\rm H}(\bbox{\beta},\bbox{\alpha})\,\mu(\bbox{\alpha}).
\end{eqnarray}
\subsection{Single- and double-layer parametrizations of wave functions;
Jump conditions}
\label{APP:DISC}
The first ingredient needed for the construction of a MSE is the
parametrization of a wave function in terms of single or double layers.
As this aspect of classical potential theory might be unfamiliar to
some readers, we first illustrate it in the simpler setting of the Helmholtz
wave equation, before turning to the \bdg\ wave equation.
\subsubsection{Parametrizations for one-component
Helmholtz wave functions}
\label{APP:OCHwf}
Consider the Helmholtz wave equation,
\begin{equation}
(\nabla^2+E)\,\varphi({\bf x})=0,
\label{EQ:Helmholtz}
\end{equation}
for the wave function $\varphi({\bf x})$ in the region $\notV$
(which we define to be the region outside some region $\volV$
bounded by the closed surface $\surS$). Then, from potential
theory~\cite{REF:Kellog,REF:GueLee} it is known that the solutions
$\varphi({\bf x})$ can be parametrized in terms of a function
${\bbox{\nu}}({\bbox{\alpha}})$ defined only on $\surS$, via the integral
\begin{equation}
\varphi({\bf x})=
\int_{\surS}d\sigma_{\alpha}\,
g^{\rm H}({\bf x}-{\bbox{\alpha}})\,
\nu({\bbox{\alpha}}).
\label{EQ:SPB}
\end{equation}
Here, ${\bf x}$ and ${\bf x}^{\prime}$ are positions lying in
$\notV$, Greek letters, such as ${\bbox{\alpha}}$, represent vectors on
the boundary $\surS$ (as they do throughout this Paper),
and $g^{\rm H}({\bf x}-{\bf x}^{\prime})$ is the
fundamental Green function for the Helmholtz wave equation, which
satisfies
\begin{equation}
(\nabla ^2+E)\,
g^{\rm H}({\bf x}-{\bf x}^{\prime})
=\delta^{(3)}({\bf x}-{\bf x}^{\prime}).
\end{equation}
One can interpret the parametrization by saying that the wave function
$\varphi({\bf x})$ is the Helmholtz wave function due to a
{\it single layer of charge\/} of surface density $\nu({\bbox{\alpha}})$.

Where does this parametrization come from?  First, note that the wave
function~(\ref{EQ:SPB}) does indeed satisfy
Eq.~(\ref{EQ:Helmholtz}).  To see this, observe that ${\bf x}$ and
${\bbox{\alpha}}$ are never coincident (${\bbox{\alpha}}$ lying on
$\surS$ but ${\bf x}$ lying in $\notV$) so that
$g^{\rm H}({\bf x}-{\bbox{\alpha}})$ solves Eq.~(\ref{EQ:Helmholtz})
for any ${\bbox{\alpha}}$, and the parametrization~\ref{EQ:SPB} is
simply a superposition of such solutions. Second, recall that
by Green's theorem one has
\begin{equation}
\varphi({\bf x})=
\int_{\surS}
d\sigma_{\alpha}\,\partial_{\alpha}\,
g^{\rm A}\!({\bf x},{\bbox{\alpha}})\,
\varphi({\bbox{\alpha}})
-\int_{\surS}
d\sigma_{\alpha}\,
g^{\rm A}\!({\bf x},{\bbox{\alpha}})\,
\partial_{\alpha}\varphi({\bbox{\alpha}}),
\label{EQ:GreenThm}
\end{equation}
where $g^{\rm A}({\bf x},{\bbox{\alpha}})$ is {\it any\/} Helmhotz Green
function (i.e.~not necessarily the fundamental one). In the most common
setting, one then chooses the Green function that satisfies the
homogeneous version of the boundary condition on $\varphi$, thus
eliminating all absent boundary information and arriving at an
expression for $\varphi({\bf x})$ in terms of information known about
$\varphi$ on $\surS$.  Here, instead, one takes a different tack.  One
selects for $g^{\rm A}$ the {\it fundamental\/} Green function $g^{\rm H}$,
jettisons the first contribution to the right hand side of
Eq.~(\ref{EQ:GreenThm}), and accommodates for this by replacing the
boundary information on $\partial\varphi$ by the (as-yet unknown) single-layer
$\nu({\bbox{\alpha}})$.

Representations of wave functions by surface integrals are available in
other settings, too.  We have considered wave functions
satisfying the Helmholtz wave equation {\it outside\/} the region $\volV$
(i.e.~the so-called {\it exterior\/} problem).  One can also consider
wave functions satisfying the Helmholtz wave equation {\it inside\/}
the region $\volV$ (i.e.~the so-called {\it interior\/} problem).

Furthermore, one can parametrize wave functions satisfying the Helmholtz
wave equation in other ways.  For example, consider wave functions
inside the region $\volV$, which one can parametrize as
\begin{equation}
\varphi({\bf x})=
\int_{\surS}
d\sigma_{\alpha}\,
\partial_{\alpha}\,
g^{\rm H}\!({\bf x}-{\bbox{\alpha}})\,
\mu({\bbox{\alpha}}).
\label{EQ:SPA}
\end{equation}
In this case $\varphi({\bf x})$ is the Helmholtz wave function due to a
{\it layer of dipoles\/} on $\surS$ of local strength
$\mu({\bbox{\alpha}})$ and local orientation normal to $\surS$
at each point $\bbox{\alpha}$.  (We denote such surface normal vectors
as ${\bf n}_{\alpha}$, and adopt the convention that they point towards
the {\it interior\/} of $\volV$.)\thinspace\ One can, of course, regard this
dipole layer as consisting of two single layers, vanishingly close to
one another and locally carrying opposite charges, in the limit that
the charges become large and the layer separation becomes
correspondingly small.  Such layers are referred to as a {\it double
layers\/}. The normal component of the gradient acting on the Green
function accounts for the fact that this parametrization features
opposing, vanishingly close, layers.  As with the single-layer
parametrization~(\ref{EQ:SPB}), that the parametrization~(\ref{EQ:SPA})
satisfies Eq.~(\ref{EQ:Helmholtz}) follows because ${\bf x}$ and
${\bbox{\alpha}}$ are never coincident, so that
$g^{\rm H}({\bf x}-{\bbox{\alpha}})$ solves Eq.~(\ref{EQ:Helmholtz}) for
any ${\bbox{\alpha}}$.  Motivation for the parametrization~(\ref{EQ:SPA})
also follows from consideration of Green's theorem,
Eq.~(\ref{EQ:GreenThm}), but with the {\it second\/} contribution on the
right hand side being jettisoned, rather than the {\it first\/}.

There, of course, remains the issue that whether all solutions can be
expressed in terms of these parametrizations. It turns out that if one uses
the double-layer parametrization for interior wave functions, and
the single-layer parametrization for exterior wave functions
then {\it any\/} solution can be thus parametrized.  The converse problem
(i.e.~parametrizing interior wave functions using double layers and exterior
wave functions using single layers) is also possible, provided certain
supplementary conditions are satisfied; see, e.g.,
Refs.~\cite{REF:Kellog,REF:GueLee}

We note that the strategy that we are adopting can be implemented in
more general settings.  For example, one might consider the case of
disconnected superconducting regions connected by normal regions, and
thus address the issue of Josephson tunneling between them.  One might
also consider disconnected normal regions connected by superconducting
regions, and thus address the issue of single-particle tunneling between
them.

The utility of these parametrizations is that they can be used to transform
partial differential equations in $\volV$ or $\notV$ into Fredholm integral equations that reside on
$\surS$, and as we shall see in the following section, such integral equations
prove useful in some cases, especially if the integral equation
is of the second type, i.e., if an iterative solution is possible.

We now turn to the second ingredient needed for the construction of a
MSE, viz., {\it jump conditions\/}. It is an important result of
potential theory that the parametrization of wave functions in
terms of single and double layers, such as those given in
Eqs.~(\ref{EQ:SPB},\ref{EQ:SPA}), leads to
representations of wave functions that behave in a singular fashion for
field points ${\bf x}$ on the surface $\surS$.  It is precisely this
singular behavior, and the attendant jump conditions, that are
responsible for the utility of these parametrizations and, as we shall
see shortly, lead to the formulation of integral equations for the layer
strengths, single or double, known as {\it boundary integral equations\/}.
These equations incorporate what ever boundary conditions one wishes to
impose on the wave function.  By solving boundary integral equations one
arrives at layer strengths that parametrize the wave functions.

There are three virtues to this boundary integral equations formulation.
First, the boundary integral equations reside solely on the boundary
$\surS$ (i.e.~on a manifold of dimension one fewer than the original
wave equation). From the computational standpoint it is
economical to formulate a problem in terms of functions that reside on
lower-dimensional manifolds (and hence depend on fewer
variables). Second, from the theoretical standpoint, existence theorems
have been established for broad classes of integral equations,
often encompassing the boundary integral equations that emerge
from specific examples.  (In fact, it was the goal of establishing
exitence theorems for solutions to the Laplace equation in various
settings---interior or exterior, Dirichlet or Neumann boundary
conditions---that inspired Fredholm to develop the boundary
intergal equation approach to potential theory, and subsequently to
develop the theory of what we now know as Fredholm integral
equations.)\thinspace\ And third, from the physical standpoint,
boundary integral equations and their iterative solution allow one
to organize the computation of wave functions in terms of the
multiple scattering of waves from interfaces that separate spatially homogeneous
regions, along with free propagation between those scattering
events.  Thus, one is in a position to focus on the boundary scattering
events, and thereby to focus on the geometry of the boundary and the
implications of its shape for the physical problem at hand.
In essence, we are invoking the piecewise homogeneity of the
system to \lq\lq integrate up\rq\rq\ our description of it,
leaving us with the need to consider one fewer independent variable.
As a result of this \lq\lq integrating up\rlap,\rq\rq\ we depart from a
purely local description, in terms of differential equations, and arrive
at a nonlocal formulation in terms of integral equations.
\subsubsection{Jump conditions for
one-component Helmholtz wave functions}
\label{APP:JCOCHwf}
Consider the single-layer parametrization for Helmholtz wave functions
in the region $\notV$, given by Eq.~(\ref{EQ:SPB}).  Now, it is known
from potential theory that this parametrization is singular as ${\bf x}$
goes to any value ${\bbox{\beta}}$ on the boundary (see
App.~\ref{APP:DISC}).  Specifically, $\varphi({\bf x})$ is continuous
whereas ${\bf n}_{\beta}\cdot\nabla_{x}\,\varphi({\bf x})$ is not,
but the discontinuity of the latter has a known and useful form:
\begin{mathletters}
\begin{eqnarray}
\lim\limits_{{\bf x}\in\volV\rightarrow{\bbox{\beta}}\in\surS}
\int_{\surS}d\sigma_{\alpha}\,
g^{\rm H}({\bf x}-{\bbox{\alpha}})\,\nu({\bbox{\alpha}})
&=&
\int_{\surS}d\sigma_{\alpha}\,
g^{\rm H}({\bbox{\beta}}-{\bbox{\alpha}})\,\nu({\bbox{\alpha}});
\label{EQ:disc_sl_nd}
\\
\lim\limits_{{\bf x}\in\volV\rightarrow{\bbox{\beta}}\in\surS}
{\bf n}_{\beta}\cdot\nabla_{x}
\int_{\surS}
d\sigma_{\alpha}
g^{\rm H}({\bf x}-{\bbox{\alpha}})\,\nu({\bbox{\alpha}})
&=&
\frac{1}{2}\nu({\bbox{\beta}})
+\int_{\surS}d\sigma_{\alpha}\,
{\bf n}_{\beta}\cdot\nabla_{\beta}\,
g^{\rm H}({\bbox{\beta}}-{\bbox{\alpha}})\,\nu({\bbox{\alpha}}).
\label{EQ:disc_sl_yd}
\end{eqnarray}
\end{mathletters}%

Now consider the double-layer parametrization of the Helmholtz wave
functions on the region $\volV$, given by Eq.~(\ref{EQ:SPA}).  It is
known from potential theory that this parametrization is also singular
as ${\bf x}$ goes to any value of ${\bbox{\beta}}$ on the boundary (see
App.~\ref{APP:DISC}).  However, in this case $\varphi({\bf x})$ itself
is discontinuous, and ${\bf n}_{\beta}\cdot\nabla_{x}\varphi({\bf x})$ is
even more singular, the discontinuity of $\varphi({\bf x})$ being given by
\begin{eqnarray}
\lim\limits_{{\bf x}\in\volV\rightarrow{\bbox{\beta}}\in\surS}
\int_{\surS}d\sigma_{\alpha}\,\partial_{\alpha}\,
g^{\rm H}({\bf x}-{\bbox{\alpha}})\,\mu({\bbox{\alpha}})
=\frac{1}{2}\mu({\bbox{\beta}})
+\int_{\surS}d\sigma_{\alpha}\,
g^{\rm H}({\bbox{\beta}}-{\bbox{\alpha}})\,\mu({\bbox{\alpha}}).
\label{EQ:disc_dl_nd}
\end{eqnarray}
Although ${\bf n}_{\beta}\cdot\nabla_{x}\varphi({\bf x})$ diverges on
$\surS$, it is yet a further result from potential theory that the
values of the limits of this quantitity, as ${\bf x}$ approaches any
point ${\bbox{\beta}}$ on $\surS$ either from $\volV$ or from $\notV$,
are equal to one another.  It will, therefore, prove to be convenient
to {\it redefine\/} the quantity
${\bf n}_{\beta}\cdot\nabla_{x}\varphi({\bf x})
\vert_{{\bf x}=\bbox{\beta}}$ as its limiting value:
\begin{eqnarray}
\int_{\surS}d\sigma_{\alpha}\,
\partial_{\beta}^{\ourlim}\,\partial_{\alpha}\,
g^{\rm H}({\bbox{\beta}}-{\bbox{\alpha}})\,\nu({\bbox{\alpha}})
\equiv
\lim\limits_{{\bf x}\in\volV\,\,{\rm or}\,\,\notV
    \rightarrow{\bbox{\beta}}\in\surS}
{\bf n}_{\beta}\cdot\nabla_{x}
\int_{\surS}d\sigma_{\alpha}\,
\partial_{\alpha}\,
g^{\rm H}({\bbox{\beta}}-{\bbox{\alpha}})\,\nu({\bbox{\alpha}}).
\end{eqnarray}
Via this definition the normal derivative
${\bf n}_{\beta}\cdot\nabla_{x}\varphi({\bf x})$ is continuous
across $\surS$.
\subsubsection{Parametrizations for two-component \bdg\ wave functions}
We now turn to the issue of parametrizing solutions of the \bdg\
equation~(\ref{EQ:TIBdG}) in terms of single and double layers,
bearing in mind the two-component nature of the wave functions.
For the sake of concreteness, as well as experimental relevance, we
focus on the setting of an Andreev billiard, so that space is
partitioned into two regions by the surface $\surS$, with
$\Delta({\bf x})=0$ in the region $\volV$ inside $\surS$ (i.e.~the
normal-metal region) and $\Delta({\bf x})=\Delta_{0}$ in the region
$\notV$ outside $\surS$ (i.e., the superconducting region). However, the
are no obstacles of principle in applying the present techniques in
other settings, provided they comprise regions of space in which
$\Delta({\bf x})$ is constant, separated by surfaces on which
$\Delta({\bf x})$ varies discontinuously [i.e.~$\Delta({\bf x})$
must be piecewise constant].  The present techniques may also be
applied in settings in which other physical parameters vary in a
piecewise continuous fashion.

First, consider the normal-state interior of an Andreev billiard
[i.e.~a region $\volV$ surrounded by a surface $\surS$ in which
$\Delta({\bf x})=0$] in which the \bdg\ wave functions satisfy
\begin{equation}
\pmatrix{
\nabla^2-\mu+E&0
\cr\noalign{\medskip}
0& -\nabla ^2+\mu +E}
\pmatrix{u({\bf x})
\cr\noalign{\medskip}
         v({\bf x})}
=\pmatrix{0
\cr\noalign{\medskip}
          0}.
\label{EQ:bdgNormal}
\end{equation}
Solutions of this equation can be parametrized in terms
of the two-component double layer
${\bbox{\mu}}({\bbox{\alpha}})$
in the following way:
\begin{mathletters}
\begin{equation}
{\bbox{\Phi}}^{\rm N}({\bf x})
=\int_{\surS}
d\sigma_{\alpha}\,
\partial_{\alpha}\,
\gf^{\rm N}({\bf x}-{\bbox{\alpha}})
\cdot{\bbox{\mu}}({\bbox{\alpha}}).
\label{EQ:TPB}
\end{equation}
Here, $\gf^{\rm N}({\bf x}-{\bf x}^{\prime})$ is the fundamental Green
function for the \bdg\ wave equation in the absence of a pair potential,
and is given explicitly by Eq.~(\ref{EQ:GFhnm}).
Such solutions could also be parametrized in terms
of the two-component single layer ${\bbox{\nu}}({\bbox{\alpha}})$, as
\begin{equation}
{\bbox{\Phi}}^{\rm N}({\bf x})
=\int_{\surS}
d\sigma_{\alpha}\,
\gf^{\rm N}({\bf x}-{\bbox{\alpha}})
\cdot{\bbox{\nu}}({\bbox{\alpha}}),
\label{EQ:TPA}
\end{equation}
\end{mathletters}
but it will prove convenient to adopt the former parametrization,
Eq.~(\ref{EQ:TPB}), rather than the latter, Eq.~(\ref{EQ:TPA}).

The two-component layers,
${\bbox{\mu}}({\bbox{\alpha}})$ and
${\bbox{\nu}}({\bbox{\alpha}})$,
reflect the two-component (i.e.~electron and hole) nature of the wave
functions.  If one were solely concerned with the case of a normal
region, this two-component description would be redundant: as no matrix
elements of the Hamiltonian would connect upper and lower components
there would be no need to adopt a language that embraces wave functions
that describe coherent superpositions of electron and hole states.
However, in an Andreev billiard the normal region is surrounded by a
superconductor, the pair potential of which provides precisely the
matrix element connecting electron and hole wave functions. Therefore it
is necessary to adopt this two-component language for the normal region.

In the superconducting exterior of the Andreev billiard
(i.e.~the region $\notV$ outside the surface $\surS$ in which
$\Delta({\bf x})=\Delta_0$) the \bdg\ wave functions satisfy
\begin{equation}
\pmatrix{
\nabla ^2-\mu+E&-i\Delta_0
\cr\noalign{\medskip}
i\Delta_0      &-\nabla ^2+\mu +E}
\pmatrix{u({\bf x})
\cr\noalign{\medskip}
         v({\bf x})}
=\pmatrix{0
\cr\noalign{\medskip}
          0}.
\label{EQ:bdgSuper}
\end{equation}
Solutions of this equation can be parametrized in terms of
of the two-component single layer ${\bbox{\nu}}({\bbox{\alpha}})$
in the following way:
\begin{mathletters}
\begin{equation}
\bbox{\Psi}^{\rm S}({\bf x})
\equiv
\pmatrix{u({\bf x})\cr
         v({\bf x})}
=\int_{\surS}
d\sigma_{\alpha}\,
\gf^{\rm S}({\bf x}-{\bbox{\alpha}})
\cdot{\bbox{\nu}}({\bbox{\alpha}}),
\end{equation}
where $\gf^{\rm S}({\bf x}-{\bf x}^{\prime})$ is the fundamental Green
function for the \bdg\ wave equation in the presence of a homogeneous
pair potential $\Delta_0$, and is given explicitly by
Eq.~(\ref{EQ:GFhsc}). Two-component layers are mandatory here, inasmuch
as each component of the wave function is determined by both components
of a layer, owing to the presence of the pair-potential.
Such solutions could also be parametrized in terms
of the two-component double layer ${\bbox{\mu}}({\bbox{\alpha}})$, as
\begin{equation}
\bbox{\Psi}^{\rm S}({\bf x})
\equiv
\pmatrix{u({\bf x})\cr
v({\bf x})}
=\int_{\surS}
d\sigma_{\alpha}\,
\partial_{\alpha}\,
\gf^{\rm S}({\bf x}-{\bbox{\alpha}})
\cdot{\bbox{\mu}}({\bbox{\alpha}}),
\end{equation}
\end{mathletters}%
except that under certain circumstances the parametrization must be
augmented by an additional term; see, e.g., Ref.~\cite{REF:GueLee}.
However, it will prove adequate for us to stick with the former
parametrization.
\subsubsection{Jump conditions for two-component
\bdg\ wave functions}
We now echo for the case of \bdg\ wave functions the discussion,
given in App.~\ref{APP:JCOCHwf}, of the behavior of single- and
double-layer parametrizations of Helmholtz wave functions in the
vicinity of the surface $\surS$.
Consider the double-layer parametrization of the \bdg\ wave
functions on the region $\volV$ given by Eq.~(\ref{EQ:TPB}).
This parametrization is singular, as ${\bf x}$ goes to any point
${\bbox{\beta}}$ on the boundary (see
App.~\ref{APP:DISC}), inasmuch as
${\bbox{\Phi}}^{\rm N}({\bf x})$ is discontinuous, and
${\bf n}_{\beta}\cdot\nabla_{x}{\bbox{\Phi}}^{\rm N}({\bf x})$
is even more singular.  From the form of
$\gf^{\rm N}({\bf x}-{\bf x}^{\prime})$,
Eq.~(\ref{EQ:FGFhnr}), it can be shown that
the discontinuity of ${\bbox{\Phi}}^{\rm N}({\bf x})$
is given by
\begin{mathletters}
\begin{eqnarray}
\lim\limits_{{\bf x}\in\volV\rightarrow{\bbox{\beta}}\in\surS}
\int_{\surS}d\sigma_{\alpha}\,\partial_{\alpha}\,
\gf^{\rm N}({\bf x}-{\bbox{\alpha}})
\cdot{\bbox{\mu}}({\bbox{\alpha}})
=\frac{1}{2}\psm{3}\cdot{\bbox{\mu}}({\bbox{\beta}})
+\int_{\surS}d\sigma_{\alpha}\,\partial_{\alpha}\,
\gf^{\rm N}({\bbox{\beta}}-{\bbox{\alpha}})
\cdot{\bbox{\mu}}({\bbox{\alpha}}).
\label{EQ:JCnsWF}
\end{eqnarray}
Although
${\bf n}_{\beta}\cdot\nabla_{x}{\bbox{\Phi}}^{\rm N}({\bf x})$
diverges on $\surS$, it can also be shown that that the values of
the limits of this quantitity, as ${\bf x}$ approaches any
point ${\bbox{\beta}}$ on $\surS$ either from $\volV$ or from $\notV$,
are equal to one another.  It will, therefore, prove to be convenient
to {\it redefine\/} the quantity
${\bf n}_{\beta}\cdot\nabla_{x}{\bbox{\Phi}}^{\rm N}({\bf x})
\vert_{{\bf x}=\bbox{\beta}}$
as its limiting value:
\begin{eqnarray}
\int_{\surS}d\sigma_{\alpha}\,
\partial_{\beta}^{\ourlim}\,\partial_{\alpha}\,
\gf^{\rm N}({\bbox{\beta}}-{\bbox{\alpha}})
\cdot{\bbox{\mu}}({\bbox{\alpha}})
\equiv
\lim\limits_{{\bf x}\in\volV\,\,{\rm or}\,\,\notV
\rightarrow{\bbox{\beta}}\in\surS}
{\bf n}_{\beta}\cdot\nabla_{x}
\int_{\surS}d\sigma_{\alpha}\,\partial_{\alpha}\,
\gf^{\rm N}({\bf x}-{\bbox{\alpha}})
\cdot{\bbox{\mu}}({\bbox{\alpha}}).
\label{EQ:JCnsGR}
\end{eqnarray}
\end{mathletters}

Now consider the single-layer parametrization of the \bdg\ wave
functions in the region $\notV$ given by Eq.~(\ref{EQ:TPA}).  This
parametrization is singular, as ${\bf x}\in\notV$ goes to any value
of ${\bbox{\beta}}\in\surS$ (see App.~\ref{APP:DISC}).  Specifically,
${\bbox{\Phi}}^{\rm S}({\bf x})$ is continuous whereas
${\bf n}_{\beta}\cdot\nabla_{x}{\bbox{\Phi}}^{\rm S}({\bf x})$
is not, but the discontinuity has a known and useful form.
Indeed, from the form of $\gf^{\rm S}({\bf x}-{\bf x}^{\prime})$,
Eq.~(\ref{EQ:GFhsc}), it can be shown that
\begin{mathletters}
\begin{eqnarray}
\lim\limits_{{\bf x}\in\notV\rightarrow{\bbox{\beta}}\in\surS}
\int_{\surS}d\sigma_{\alpha}\,
\gf^{\rm S}({\bf x}-{\bbox{\alpha}})
\cdot{\bbox{\nu}}({\bbox{\alpha}})
&=&
\int_{\surS}d\sigma_{\alpha}\,
\gf^{\rm S}({\bbox{\beta}}-{\bbox{\alpha}})
\cdot{\bbox{\nu}}({\bbox{\alpha}});
\label{EQ:CCscWF}
\\
\lim\limits_{{\bf x}\in\notV\rightarrow{\bbox{\beta}}\in\surS}
{\bf n}_{\beta}\cdot\nabla_{x}
\int_{\surS}
d\sigma_{\alpha}
\gf^{\rm S}({\bf x}-{\bbox{\alpha}})
\cdot{\bbox{\nu}}({\bbox{\alpha}})
&=&
\frac{1}{2}\psm{3}\cdot{\bbox{\nu}}({\bbox{\beta}})
+\int_{\surS}d\sigma_{\alpha}\,
{\bf n}_{\beta}\cdot\nabla_{\beta}\,
\gf^{\rm S}({\bbox{\beta}}-{\bbox{\alpha}})
\cdot{\bbox{\nu}}({\bbox{\alpha}}).
\label{EQ:JCscGR}
\end{eqnarray}
\end{mathletters}%
\noindent The important point here is that all discontinuities of these
parametrizations are generated solely by components of $\gf^{\rm N,S}$
proportional to $\psm{3}$. Other components of $\gf^{\rm N,S}$ are
proportional to the scalar Green function composition
$(g_{+}^{\rm N,S}-g_{-}^{\rm N,S})$, and whatever discontinuity might
be generated by the $+$ term is cancelled by a corresponding one
generated by the $-$ term.  Therefore the discontinuities generated
by the parametrizations have the forms given in Eqs.~(\ref{EQ:JCnsWF})
and (\ref{EQ:JCscGR}).
\section{($\lowercase{d}-1$)-dimensional Fourier transforms}
\label{APP:2DFT}
In this appendix we introduce the $d-1$ dimensional Fourier transform for
functions whose arguments are $d$ dimensional vectors, which we use extensively
throughout the text. Mainly, our functions of interest will be the Green
functions $g^{\rm N,S}_{\pm}$  and the functions related to them, such as $\partial g^{\rm N,S}_{\pm}$.

The $d-1$ dimensional Fourier transform (and its inverse) of a function
$f({\bf x})$ of a $d$ dimensional vector ${\bf x}$, are defined by
\begin{mathletters}
\begin{eqnarray}
\tilde{f}(\bbox{\kappa},z)&=&\int_{\cal P} d{{\bf x}_{/\negthinspace/}}\,
{\rm e}^{i \bbox{\kappa}\cdot{{\bf x}_{/\negthinspace/}}} \,f({\bf x}),
\\
f({\bf x})&=&(2\pi)^{-d+1}\int d\bbox{\kappa}\,
{\rm e}^{-i \bbox{\kappa}\cdot
{{\bf x}_{/\negthinspace/}}}\,\tilde{f}(\bbox{\kappa},z),
\end{eqnarray}%
\end{mathletters}%
where $(\bbox{x_{/\negthinspace/}},z)\equiv\bbox{x}$, and
$\bbox{x_{/\negthinspace/}}$ and $\bbox{\kappa}$ are vectors on the
$d-1$ dimensional hyperplane ${\cal P}$ perpendicular to the $z$ axis.
We now evaluate the
$(d-1)$-dimensional Fourier transforms of the functions that are used throughout
this Paper. We begin with the Helmholtz Green function $g(\bbox{x};k^2)$:
\begin{eqnarray}
\tilde{g}(\bbox{\kappa},z;k^2)
&=&
\int_{\cal P}d{{\bf x}_{/\negthinspace/}}\,
{\rm e}^{i \bbox{\kappa}\cdot{{\bf x}_{/\negthinspace/}}}\,g({\bf x};k^2)
=\int_{\cal P}d{{\bf x}_{/\negthinspace/}}\,
{\rm e}^{i \bbox{\kappa}\cdot{{\bf x}_{/\negthinspace/}}}\,
\int\frac{d{\bf p}}{(2\pi)^d} \frac{{\rm e}^{i{\bf p}\cdot{\bf x}}}{p^2-k^2}
= \int\frac{d{\bf p}}{(2\pi)^d}\int_{\cal P}d{{\bf x}_{/\negthinspace/}}\,
{\rm e}^{i (\bbox{\kappa}+{{\bf p}_{/\negthinspace/}})
\cdot{{\bf x}_{/\negthinspace/}}}\,
\frac{{\rm e}^{ip_{z}z}}{p^2-k^2}
\nonumber \\
&=&\int\frac{d{\bf p}}{(2\pi)^d}
(2\pi)^{(d-1)}\delta(\bbox{\kappa}+{{\bf p}_{/\negthinspace/}})
\frac{{\rm e}^{ip_{z}z}}{p^2-k^2}
=\int_{-\infty}^{\infty}\frac{dp_z}{2\pi}\,
\frac{{\rm e}^{ip_{z}z}}{p_{z}^{2}+\kappa^2-k^2}
=\frac{{\rm e}^{-a(\kappa)|z|}}{2 a(\kappa)},
\end{eqnarray}
where $a(\kappa)\equiv\sqrt{\kappa^2-k^2}$ such that ${\rm Re}(a(\kappa))>0$.
The evaluation of the $(d-1)$-dimensional Fourier transform of $\partial\delta g(\bbox{x};k^2)$
is very similar, except that now ${\bf x}=({{\bf x}_{/\negthinspace/}},0)$
(i.e.~both arguments of the Green function lie on ${\cal P}$ and the normal
direction is $z$):
\begin{eqnarray}
\partial\delta\tilde{g}(\bbox{\kappa};k^2)
&=&
\int_{\cal P}d{{\bf x}_{/\negthinspace/}}\,
{\rm e}^{i \bbox{\kappa}\cdot{{\bf x}_{/\negthinspace/}}}\,
\partial\delta g\left(({{\bf x}_{/\negthinspace/}},0);k^2\right)
\nonumber \\
&=&\int_{\cal P}d{{\bf x}_{/\negthinspace/}}\,
{\rm e}^{i \bbox{\kappa}\cdot{{\bf x}_{/\negthinspace/}}}\,
\left(-\frac{d^2}{dz^2}
\int\frac{d{\bf p}}{(2\pi)^d}
\frac{{\rm e}^{i{\bf p}_{/\negthinspace/}
\cdot{\bf x}_{/\negthinspace/}+ip_z z}}{p^2-k^2}
\right)\Bigg\vert_{z=0}
\nonumber \\
&=&\int_{-\infty}^{\infty}\frac{dp_z}{2\pi}\,\frac{p_z^2\,
{\rm e}^{ip_{z}z}}{p_{z}^{2}+a(\kappa)^2}
=-\frac{a(\kappa)}{2}.
\end{eqnarray}
\section{Propagation outside the N region
and nonconvex shapes: cancellations}
\label{APP:nonconvex}
In the present section we study cancellations between terms in the MSE,
which occur, in the large $\Kf L$ limit, when one (or more) of the
homogeneous region(s) constituting the billiard are nonconvex.  In the
sample billiard shown in Fig.~\ref{FIG:and_bill} the S region is nonconvex,
whereas the N region is convex.  Another example in which such cancellations
arise is provided by antidot-billiard geometries, in which S regions are
embedded in an N region.  For such cases, the N region is certainly nonconvex,
and so may be the S region.  In the former case (i.e.~convex N; nonconvex S),
these cancellations eliminate terms that include
finite--outside-propagation, validating the claim made in
Sec.~\ref{SEC:MRE} that periodic orbits that include such paths do
not contribute at leading order to the semiclassical DOS oscillations.

\begin{figure}[hbt]
\epsfxsize=10.0cm
\centerline{\epsfbox{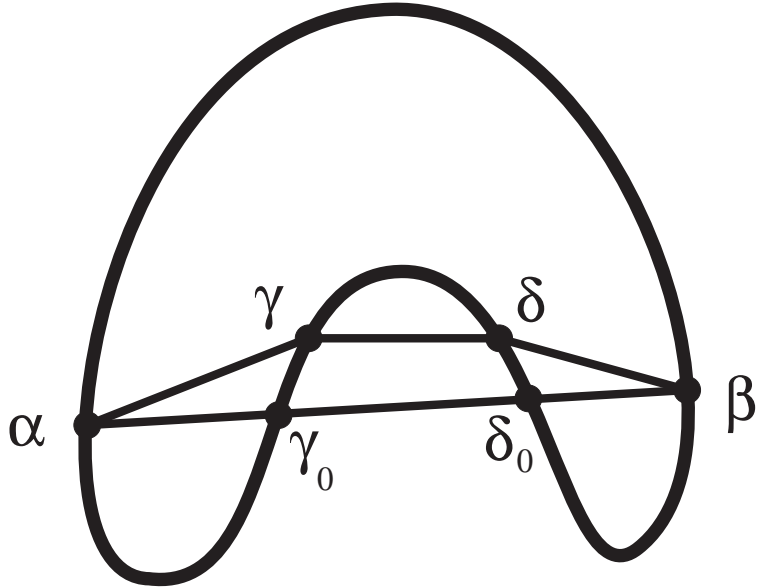}}
\vskip+0.4cm
\caption{Example of a nonconvex inside and outside regions. Also shown
is a propagation directly from $\bbox{\alpha}$ to $\bbox{\beta}$ and two-reflection
correction.}
\label{FIG:nonconvex}
\end{figure}%
In order to bring to the fore the physics underlying these cancellations,
consider a much simpler case: the MRE for the Helmholtz Green function
satisfying homogeneous Dirichlet boundary conditions~\cite{FT:cancel_helmholtz}.
By applying the methods described here, one can express this Green function in
terms of the following MRE:
\begin{equation}
g^{\rm D}({\bf x},{\bf x}')=g^{\rm H}({\bf x},{\bf x}')
-2\int_\surS \partial g^{\rm H}({\bf x},\bbox{\alpha})\,
g^{\rm H}(\bbox{\alpha},{\bf x}')
+4\int_\surS \partial g^{\rm H}({\bf x},\bbox{\beta})\,
\partial g^{\rm H}(\bbox{\beta},\bbox{\alpha})\,
g^{\rm H}(\bbox{\alpha},{\bf x}')+\cdots\,,
\label{EQ:Helmholtz_MRE}
\end{equation}
where $g^{\rm H}$ is the homogeneous Helmholtz Green function.
If the domain $\volV$ over which $g^{\rm D}$ is defined is nonconvex
then one may be concerned by the presence of amplitudes involving
propagations between pairs of boundary points for which all or part
of this propagation lies outside of $\volV$ (see Fig.~\ref{FIG:nonconvex}).
On physical grounds, one expects that in the large $k L$ limit the obstacle
between these points would suppress such amplitudes.  Balian and Bloch showed
that this is indeed the case.  To see this, we follow Balian and Bloch and
consider $g^{\rm D}$ for a planar boundary.
Then the MRE~(\ref{EQ:Helmholtz_MRE}) terminates at the second term:
\begin{equation}
g^{\rm D}({\bf x},{\bf x}')=g^{\rm H}({\bf x},{\bf x}')
-2\int_\surS \partial g^{\rm H}({\bf x},\bbox{\alpha})\,
g^{\rm H}(\bbox{\alpha},{\bf x}').
\label{EQ:Helmholtz_planar1}
\end{equation}
On the other hand, by using the method of images one has
\begin{equation}
 g^{\rm D}({\bf x},{\bf x}')=
 g^{\rm H}({\bf x},{\bf x}')
-g^{\rm H}({\bf x},\tilde{\bf x}'),
\label{EQ:Helmholtz_planar2}
\end{equation}
where $\tilde{\bf x}'$ is the mirror image of ${\bf x}'$ with respect to the
planar boundary.  Comparison of Eq.~(\ref{EQ:Helmholtz_planar2}) with Eq.~(\ref{EQ:Helmholtz_planar1}) produces the relation
\begin{equation}
g^{\rm H}({\bf x},\tilde{\bf x}')=
2\int_\surS \partial g^{\rm H}({\bf x},\bbox{\alpha})
\,g^{\rm H}(\bbox{\alpha},{\bf x}').
\label{EQ:cancel1}
\end{equation}
As the length $\vert\bbox{\alpha}-{\bf x}'\vert$ equals the length
$\vert\bbox{\alpha}-\tilde{\bf x}'\vert$, one can
replace ${\bf x}'$ by $\tilde{\bf x}'$ in relation~(\ref{EQ:cancel1})
and obtain a second relation
\begin{equation}
g^{\rm H}({\bf x},\tilde{\bf x}')=
2\int_\surS \partial g^{\rm H}({\bf x},\bbox{\alpha})
\,g^{\rm H}(\bbox{\alpha},\tilde{\bf x}').
\label{EQ:cancel2}
\end{equation}
Now suppose $\surS$ is not planar.  Although
relations~(\ref{EQ:cancel1}) and (\ref{EQ:cancel2}) no longer hold,
in the large $k L$ limit the dominant contribution comes from the
stationary-phase point and, thus,  Eq.~(\ref{EQ:cancel2}) becomes
exact as $k L$ tends to infinity.  That Eq.~(\ref{EQ:cancel1}) is not
exact in this limit is due to the fact that the fluctuation determinant
in this case (viz.~when ${\bf x}$ and ${\bf x}'$ are on the same side
of the surface) depends on the curvature of the surface at the
stationary-phase point.

Now let us return to what happens for nonconvex surfaces.  In particular,
consider a term in MRE in which part of the amplitude has propagation between
boundary points $\bbox{\alpha}$ and $\bbox{\beta}$, where part of the line
joining $\bbox{\alpha}$ and $\bbox{\beta}$ lies outside $\volV$ and
intersects $\surS$ at the points $\bbox{\gamma}_0$ and $\bbox{\delta}_0$
(see Fig.~\ref{FIG:nonconvex}). In relation to this term, consider three
related terms with higher numbers of reflections:
  (i)~the term with one more reflection  near $\bbox{\gamma}_0$,
 (ii)~the term with one more reflection  near $\bbox{\delta}_0$, and
(iii)~the term with two more reflections near $\bbox{\gamma}_0$  and
$\bbox{\delta}_0$.
In the MRE, all four are in the sum.  The sum of the part of the amplitude
involving direct propagation from
$\bbox{\alpha}$ to $\bbox{\beta}$
and the one with one more reflection near $\bbox{\gamma}_0$ is
\begin{equation}
\partial g^{\rm H}(\bbox{\alpha},\bbox{\beta})
-2\int_{\surS}\partial g^{\rm H}(\bbox{\alpha},\bbox{\gamma})\,
\partial g^{\rm H}(\bbox{\gamma},\bbox{\beta}).
\end{equation}
By virtue of Eq.~(\ref{EQ:cancel2}) this sum vanishes in the
$k L\rightarrow \infty$ limit.  The same holds for the sum of remaining
two terms. Thus in the limit $k L\rightarrow \infty$ sum of all terms
involving paths that lie partially outside of $\volV$ vanishes.

These considerations extend readily to Andreev billiards.  In order to see
this, we must extend the identity~(\ref{EQ:cancel2}).  This can be achieved
as follows:
\begin{eqnarray}
2\int_\surS \partial\gf^{\rm N}({\bf x},\bbox{\alpha})\,\psm{3}\,
\gf^{\rm N}(\bbox{\alpha},\tilde{\bf x})
&=&2\int_\surS \pmatrix{\partial g_+({\bf x},\bbox{\alpha})\,g_+(\bbox{\alpha},\tilde{\bf x})&0\cr
                      0&-\partial g_-({\bf x},\bbox{\alpha})\,g_-(\bbox{\alpha},\tilde{\bf x})}
\nonumber \\
\noalign{\medskip}
&=&\pmatrix{g_+({\bf x},\tilde{\bf x})&0\cr
0 & -g_-({\bf x},\tilde{\bf x})}=\gf^{\rm N}({\bf x},\tilde{\bf x})\,,
\label{EQ:cancel_N}
\end{eqnarray}
where, as in the case of Helmholtz Green functions, $\surS$ is a planar surface
and ${\bf x}$ and $\tilde{\bf x}$ lie on different sides of $\surS$, and in
going to the second line we have made use of Eq.~(\ref{EQ:cancel2}).
A similar identity holds for $\gf^{\rm S}$
\begin{equation}
\gf^{\rm S}({\bf x},\tilde{\bf x})=
2\int_\surS \partial\gf^{\rm S}({\bf x},\bbox{\alpha})\,
\psm{3}\,\gf^{\rm S}(\bbox{\alpha},\tilde{\bf x})
=2\int_\surS \gf^{\rm S}(\tilde{\bf x},\bbox{\alpha})\,
\psm{3}\,\delta\gf^{\rm S}(\bbox{\alpha},{\bf x})\,.
\label{EQ:cancel_S}
\end{equation}
This identity can be obtained by performing the matrix multiplication and
using Eq.~(\ref{EQ:cancel2}) in each matrix element.  As in the case of
Helmholtz Green functions, identities~(\ref{EQ:cancel_N}) and
(\ref{EQ:cancel_S}) hold in the large $\Kf L$ limit, even if $\surS$ is
not planar.  As we did for the Helmholtz case, consider a term in MSE in
which part of the amplitude has propagation between boundary points
$\bbox{\alpha}$ and $\bbox{\beta}$, where part of the line joining
$\bbox{\alpha}$ and $\bbox{\beta}$ lies outside $\volV$ and intersects
$\surS$ at the points $\bbox{\gamma}_0$ and $\bbox{\delta}_0$.
However, now we have additional terms owing to the possibility of
propagation between any two boundary points involving either
$\gf^{\rm N}$ or $\gf^{\rm S}$.  By using the
identities~(\ref{EQ:cancel_N}) and~(\ref{EQ:cancel_S}) it is not hard to
see that, in the large $\Kf L$ limit, a propagation from $\bbox{\alpha}$
to $\bbox{\beta}$ is not cancelled only if the line joining these points
lies totally in either $\volV$ or $\notV$, with the propagation
involving the homogeneous Green function appropriate to the corresponding
region.  Thus, in the semiclassical limit the terms that survive in the
MSE consist of pure reflection or pure transmission/tunneling.  For the
example at hand, the surviving term would be the one with normal
propagation from $\bbox{\alpha}$ to $\bbox{\gamma}$, superconducting
propagation from $\bbox{\gamma}$ to $\bbox{\delta}$ (i.e.~tunneling),
and then normal propagation from $\bbox{\delta}$ to $\bbox{\beta}$.



\begin{references}
\bibitem{FT:spinflip}
We shall assume that there are no
single-quasiparticle spin-flip scattering processes operative, so that there
are  no nonzero matrix elements connecting such states with those
involving the addition of spin-down electrons or the removal of
spin-up electrons.
\bibitem{FT:bndryinfty}
The boundary in question at this point is not the billiard boundary but the
boundary at infinity.
\bibitem{FT:CPt_DEF}
In fact, this can be used as the defining property of a critical point
\bibitem{FT:surdisc}
If $\surS$ were infinitely differentiable
except at some isolated points then these points would give rise to critical
points. These critical points correspond to diffraction due to a corner.
\bibitem{FT:dbledrvGF}
Another way to see this
is as follows: if ${\bbox{\alpha}}_{i+1}$ is infinitesimally close to
${\bbox{\alpha}}_i$ then the zeroth-order approximation to the surface around
${\bbox{\alpha}}_i$ is a plane; however, on a plane the quantity
$\partial_{\alpha_i}\,|{\bbox{\alpha}} _i-{\bbox{\alpha}} _{i+1}|^{-1}$ is
identically zero, which shows that the contribution from this singularity
must come from the first-order correction to the local behavior of the surface,
i.e., the curvature correction.
By examining the other possible sources of singular behaviour,
$|{\bbox{\alpha}} _i-{\bbox{\alpha}} _{i+1}|^{-1}$ and
$\partial_{\alpha_i}\,
 \partial_{\alpha_{i+1}}\,
|{\bbox{\alpha}}_i-{\bbox{\alpha}}_{i+1}|^{-1}$,
we see that neither of these singularities vanishes on the plane
and, hence, that both contribute at leading
order
\bibitem{FT:leadingWEYL}
They are the leading order terms
in the Weyl (or Thomas-Fermi) expansion for the Andreev billiard.
\bibitem{FT:mixedAB}
Note that in settings that mix superconducting confinement with
conventional confinement due to a single-particle potential,
it is possible (such as when a chaotic conventional billiard
is in proximity of a superconductor) that the lengthscale of the primitive
periodic orbits be much larger than the linear size of the billiard.
In such cases, Scheme~A breaks down and Scheme~B should be used.
\bibitem{FT:ehsym}
In the context of Andreev billiards
it is usually thought that electron-hole symmetry is broken by the
inclusion of parabolic and higher-order terms in the single-particle
dispersion relation.  However, even if one had a linear dispersion
relation, the difference between electron and hole wavevectors would break
the electron-hole symmetry.  In Scheme~A one uses semiclassical
perturbation theory.  For a discussion of semiclassical perturbation
theory in the context of symmetry breaking, see Refs.~\cite{REF:BBscm}
and~\cite{REF:Creagh}.
\end{references}
\end{document}